\documentclass[12pt]{article}
\usepackage{latexsym}
\usepackage{amsmath}
\usepackage{epsfig,graphics}

\newcommand{\be}{\begin{equation}}
\newcommand{\ee}{\end{equation}}

\newlength{\figsize}
\begin{document}

\begin{titlepage}

\vspace*{-0.5in}
 
\begin{flushright}
DESY 05-021 \\
\end{flushright}

\vspace*{0.5in}
 
\begin{center}
{\large\bf Properties of the deconfining phase
transition \\ in SU(N) gauge theories \\ }
\vspace*{0.5in}
{Biagio Lucini$^{a}$, Michael Teper$^{b}$ and Urs Wenger$^{c}$\\
\vspace*{.3in}
$^{a}$Institute for Theoretical Physics, ETH Z\"urich,\\
CH-8093 Z\"urich, Switzerland\\
\vspace*{.05in}
$^{b}$Theoretical Physics, University of Oxford,\\
1 Keble Road, Oxford OX1 3NP, U.K.\\
\vspace*{.05in}
$^{c}$NIC/DESY Zeuthen, Platanenallee 6, 15738 Zeuthen, Germany
}
\end{center}

\vspace*{0.5in}

\begin{center}
{\bf Abstract}
\end{center}

We extend our earlier investigation of the finite temperature
deconfinement transition in SU($N$) gauge theories, with the
emphasis on what happens as $N\to\infty$. We calculate the 
latent heat, $L_h$, in the continuum limit, and 
find the expected behaviour, $L_h\propto N^2$, at large $N$. 
We confirm that the phase transition, which is second order
for SU(2) and weakly first order for SU(3), becomes robustly 
first order for  $N\geq 4$ and strengthens as $N$ increases. 
As an aside, we explain why the SU(2) specific heat shows no sign 
of any peak as $T$ is varied across what is supposedly 
a second order phase transition. We calculate
the effective string tension and electric gluon masses
at $T\simeq T_c$ confirming the discontinuous nature of the 
transition for $N\geq 3$. We explicitly show that the large-$N$ 
`spatial'  string tension does not vary with $T$ for $T\leq T_c$ 
and that it is discontinuous at $T=T_c$. For $T\geq T_c$ it 
increases $\propto T^2$ to a good approximation, and the
$k$-string tension ratios closely satisfy Casimir Scaling.
Within very small errors, we find a single $T_c$ at which all 
the $k$-strings deconfine, i.e. a step-by-step breaking of the 
relevant centre symmetry does not occur. We calculate the
interface tension but are unable to distinguish between the
$\propto N$ or $\propto N^2$ variations, each of which 
can lead to a striking but different $N=\infty$ deconfinement
scenario. We remark on the location of the bulk phase transition, 
which bounds the range of our large-$N$ calculations on the strong 
coupling side, and within whose hysteresis some of our
larger-$N$ calculations are performed.

\end{titlepage}

\setcounter{page}{1}
\newpage
\pagestyle{plain}

\section{Introduction}
\label{section_intro}

In previous papers,
\cite{oxtemp02,oxtemp03},
we have presented results on some basic properties of the SU($N$)
deconfining phase transition in $3+1$ dimensions, with the
emphasis on the order of the transition and the value of 
the deconfining temperature $T_c$. In the present paper
we shall address a number of more detailed issues such as the
physical latent heat, the mass gap and interface tensions. 
We will also take the opportunity to use some new calculations
to improve upon our earlier calculations of  
$T_c/\surd\sigma$ where $\sigma$ is the confining string tension.
In a companion paper
\cite{oxtempq04}
we have provided a detailed study of the properties of topological 
fluctuations across the phase transition.

In the next Section we provide a background discussion to motivate
the calculations that appear in subsequent sections. We then 
briefly summarise our (entirely conventional) lattice setup,
go on to our improved analysis of $T_c/\surd\sigma$
and then proceed to our calculation of the physical latent heat
and its $N$-dependence. In the following section we present
various mass calculations, both at $T\simeq T_c$ and at
higher $T$. We then address the issue of whether there
are multiple deconfining transitions at larger $N$. We
follow this with an estimate of the $N$-dependence
of the confined-deconfined interface tension. We then
return to SU(2) to examine more carefully why the usual
thermodynamic correlators show no sign of a phase transition
in that case. Finally we summarise some properties of the
bulk phase transition and how it affects our calculations.
In the concluding section we summarise what we have 
learned in this study, with some comments on large-$N$
Master Fields and Eguchi-Kawai space-time reduction.

\section{Deconfinement}
\label{section_deconfinement}

In this Section we shall briefly summarise some theoretical
expectations for the quantities that we shall calculate later 
on in this paper.

We begin with a very basic discussion of deconfinement 
and of the  dynamical significance of the order of the transition.
We then discuss a new phenomenon that may arise for $N>3$: 
deconfinement may occur through several phase transitions
rather than just a single one. Although this seems possible
on general grounds we shall argue that for dynamical
reasons it is very unlikely to occur in the case of SU($N$)
gauge theories. We then discuss the volume and $N$ dependence
of the deconfining transition and how this depends on the 
$N$-dependence of the latent heat and of the confining/deconfining
interface tension. We describe some of the strikingly simple
scenarios that may result at $N=\infty$. We then discuss 
our expectations for the masses which we calculate below, above
and at the phase transition. These include effective string 
tensions, spatial string tensions and the electric Debye mass.
Finally we remind the reader of the fact that not every
aspect of the phase transition on our Euclidean space-time
volumes is directly reflected in the physics of the hot
gauge theory. This will motivate our later investigation of the
SU(2) phase transition where, as is well known (at least
to practitioners), there is no peak visible in the
specific heat despite the apparent second order nature of
the transition.

\subsection{Introduction}
\label{subsection_deconf_intro}

We begin by recalling the following simple argument for deconfinement
\cite{polyTc}.
Consider two static sources in the fundamental representation that
are a distance $r$ apart. Suppose that the theory is linearly 
confining, so that the flux is restricted to a flux tube of 
finite width that ends at the two sources. Clearly such
a flux tube will be of length $l\geq r$. Consider the case
where $l\gg r$. The thermodynamic weight of such a configuration
depends on the string energy $E(l)=\sigma l$, where $\sigma$ is the
string tension, as well as on the density of states, i.e. on the 
number, $n(l)$, of flux tubes of length $l$ with fixed end points:
\begin{equation}
\mathrm{probability} 
\propto
n(l)e^{-\frac{E(l)}{T}}
\propto
e^{cl} e^{-\frac{\sigma l}{T}}.
\label{eqn_prob}
\end{equation}
Here we use the fact that for $l\gg r$ (and ignoring various
inessential complications) $n(l) \sim n(l/2)n(l/2)$ so that 
$n(l) \propto \exp\{cl\}$ up to power corrections. The constant
$c$ will depend on the detailed physics of the flux tube.
Irrespective of its value, we see from eqn(\ref{eqn_prob}) that 
there is a temperature
\begin{equation}
T=T_c=\frac{\sigma}{c}
\label{PolyTc}
\end{equation}
at which the entropy compensates the Boltzman suppression,
so that strings of arbitrary length are  produced thermally.
Clearly at $T=T_c$ the effective string tension,
$\sigma_{\scriptstyle eff}(T) = \sigma -cT$, vanishes 
and the system ceases to be linearly confining.

This simple argument is illuminating in that it makes no reference
to any  underlying microscopic theory, e.g. to gluons,
and shows that any (linearly) confining theory must deconfine at
some finite $T$. Since $\sigma_{\scriptstyle eff}(T) \to 0$
as $T\to T_c$, the corresponding correlation length diverges
and we expect this phase transition to be second order.

Given a phase transition at $T=T_c$ the partition
function 
\begin{equation}
Z 
=
\sum_{states} e^{-\frac{E}{T}} 
\label{eqn_Z}
\end{equation}
should have some singularity there. Since the states contributing
to $Z$ are composed of glueballs, it is interesting to 
ask if there is any argument that $Z$ should be singular 
at precisely the temperature $T=T_c=\sigma/c$  at which confining
strings `condense into the vacuum'. We give the
following simple argument. A natural model for glueballs
is that they are composed of closed loops of confining
flux. (See
\cite{fluxtube}
for explicit models of this kind.) For a loop of length $l$
the mass would be $M\simeq \sigma l$. While this dynamics
need not be convincing for small $l$ and $M$ (where the size 
of the state would be comparable to the width of the flux
tube) it is hard to imagine excluding its contribution
to the very massive glueballs that correspond to large $l$.
Now for large $l$ the number of states should be proportional
to the number of different closed loops of length $l$.
This number is clearly $\propto \exp\{cl\}$, with the same
exponent $c$ as in eqn(\ref{eqn_prob}). Thus $Z$ will
be singular at precisely  $T=T_c=\sigma/c$ with a Hagedorn-style
divergence. (This classical number of states counting 
should be reliable for the relevant very heavy glueballs 
on the basis of the standard  classical-quantum correspondence
for large quantum numbers.)
Note that the fact that the lightest glueball is very much
heavier than $T_c$, as are the splittings between the
lightest glueballs, should not be regarded as paradoxical
(see e.g. 
\cite{Tcpuzzle});
what matters is the exponential growth of the number of
glueball `resonances' and the scale of this growth is,
as we have just seen, precisely the same as in the earlier
argument for deconfinement through string condensation. 

The above discussion has been formulated entirely within  
the effective low-energy confining theory. This means it
does not depend on details, and so possesses an
attractive universality, but it also means that it is 
necessarily incomplete. In particular it ignores
the fact that in addition to the energy eigenstates composed
of colour singlet glueballs, there are also states composed
of gluons where the gluons are densely packed throughout 
the volume so that they avoid the constraints of 
confinement (in the same way that dense quark matter
avoids these constraints). Neglecting any ambiguity in the 
separation between these two classes of states, we can write  
the partition function as
\begin{equation}
Z 
\simeq
\sum_{glueballs} e^{-\frac{E}{T}} 
+
\sum_{gluons} e^{-\frac{E}{T}}
=
e^{-\frac{F_G}{T}}
+
e^{-\frac{F_g}{T}}
\label{eqn_ZGg}
\end{equation}
where the $F_G,F_g$ are the corresponding free energies. Gluons 
clearly have an entropy $S_g\propto O(N^2)$ in contrast to the 
$S_G\propto O(N^0)$ entropy of the colourless glueball states.
(This is not quite correct because
highly excited glueballs can be composed not only of closed
loops of fundamental flux, but also out of closed $k$-strings
with $k\leq N/2$. But this is not important to the present
argument.) This means that $F_g=\overline{E_g}-TS_g$ will become equal 
to  $F_G=\overline{E_G}-TS_G$ at some $T=T_c$ and beyond $T_c$ the
gluon states will be overwhelmingly more probable -- 
we will be in the `gluon plasma' phase.
If this happens at a lower value of $T$ than the 
`string condensation' temperature discussed earlier, the
deconfining phase transition will in fact go directly from
a phase with a finite effective string tension to the 
non-confining `gluon plasma' --  and so we would expect
it to  be first order.

The reason that the confining phase exists at all is that
the confining vacuum energy is below that of the `perturbative'
vacuum (which one can roughly think of as the vacuum for
the states that form the gluon plasma). This difference
is often parameterised in terms of the gluon condensate and 
is usually expected to be $\sim -O(N^2)$. Thus low-lying
hadron states are lighter by $O(N^2)V$ than low-lying
gluon states. Since at low $T$ energy counts for everything and 
entropy for nothing, the gluon states are clearly irrelevant there.
At large $N$ the value of $F_G$ will be dominated by this
vacuum energy at all $T$, since all other pieces are $O(N^0)$.
So it is the competition between this and the $-O(N^2)T$
contribution to the gluon free energy $F_g$ that will
determine $T_c$. Since the vacuum energy density
is on the scale of $-O(N^2\sigma^2)$ we would expect
\begin{equation}
\lim_{N\to\infty}T_c 
=
O(\surd{\sigma})
\label{eqn_TcNc2}
\end{equation}
and the fact that this is what one observes
\cite{oxtemp02,oxtemp03}
may thus be regarded as evidence that the confining vacuum 
energy density is indeed  $-O(N^2)$.

We know that for SU($N\ge 3$) gauge theories the phase transition 
is first order
\cite{oxtemp02,oxtemp03}
and becomes more strongly so as $N$ increases. 
This tells us that for these cases deconfinement occurs  
before the second order string condensation transition has had an 
opportunity to occur. For SU(2), however, the transition is 
indeed second order and the effective string tension vanishes as 
$T\to T^-_c$. It is therefore possible that here the phase transition 
proceeds by string condensation. The same is true of both
SU(2) and SU(3) in $D=2+1$. In all these cases the
deconfined phase immediately above $T_c$ might 
exhibit properties very different from those of a
gluon plasma and more characteristic of what one would
expect in the deconfined phase of a fundamental
string theory -- an interesting possibility whose investigation
lies beyond the scope of the present paper. We note that
if we assume the simplest effective string action, the
Nambu-Gotto action, and if we assume it to be valid at all scales, 
then the deconfinement temperature occurs at
\cite{ArvisLW}
\begin{equation}
\frac{T_c}{\surd\sigma}
=
\sqrt{\frac{3}{\pi (D-2)}}
=
\left\{ \begin{array}{ll}
0.691 & \ \ \ {\mathrm{D=3+1}} \\
0.977 & \ \ \ {\mathrm{D=2+1}} 
\end{array}
\right. 
\label{eqn_NGTc}
\end{equation}
which we can compare to the lattice values
\cite{oxtemp03,mtd3su2,legeland}:
\begin{equation}
\frac{T_c}{\surd\sigma}
=
\left\{ \begin{array}{ll}
0.709(4)   & \ \ \ {\mathrm{SU(2),\, D=3+1}} \\
1.121(8)   & \ \ \ {\mathrm{SU(2),\, D=2+1}} \\
0.985(12)  & \ \ \ {\mathrm{SU(3),\, D=2+1}} \ . 
\end{array}
\right. 
\label{eqn_latTc}
\end{equation}
These predicted and lattice values are remarkably close (as has been
previously observed
\cite{legeland}),
simultaneously suggesting that 
the Nambu-Gotto action provides a surprisingly accurate representation 
of the dynamics of the confining flux tube all the way down to scales
of $O(T_c)$ and that these phase transitions are indeed driven by
`string condensation'.

\subsection{Multiple transitions?}
\label{subsection_deconf_mult}

We have assumed in the above discussion that there is a single 
deconfinement transition. However this need not be the
case. As is well known, deconfinement is associated 
with the complete spontaneous breaking of a $Z_N$ symmetry, 
and for $N\geq 4$ this symmetry could be broken in more than one
step. In the case of SU(4), for example, one could imagine a 
breaking $Z_4 \to Z_2$ at $T=T_c$ followed by a breaking of the
remaining $Z_2$ at $T=T_{c^\prime} >T_c$.  
Such a pattern of breakings occurs in certain spin models
\cite{CKA}
and so is not entirely idle speculation.

Physically such a step-by-step breaking of the $Z_N$ symmetry
can be related to the deconfinement of different $k$-strings at
different temperatures $T_c(k)$. Recall (see e.g.
\cite{blmt-string})
that at low $T$ a source that is like a local clump of $k$ 
fundamental charges will be joined to a distant conjugate source
by a stable flux tube that is called a $k$-string 
(for $k\leq N/2$). One can think of a $k$-string as a
bound state of $k$ fundamental strings. The tension
of such a string is found to lie somewhere between the 
Casimir scaling
\cite{CS} 
and `MQCD' 
\cite{MQCD} 
conjectures
\cite{blmt-string,oxglue04}:
\begin{equation}
\frac{k(N-k)}{N-1}
\leq
\frac{\sigma_k}{\sigma} 
\leq
\frac{\sin{\frac{k\pi}{N}}}{\sin{\frac{\pi}{N}}}
\label{eqn_kstring}.
\end{equation}
The two-step deconfinement that we outlined in the previous
paragraph would correspond to $k=2$ charges becoming
deconfined at $T=T_c$ while $k=1$ charges remain confined
until a phase transition at a higher temperature $T=T_{c^\prime}$.
This can be generalised in an obvious way to larger $N$ and $k$.

When deconfinement proceeds via string condensation, it is easy 
to see what is needed for the above scenario to occur. The entropy 
factor in eqn(\ref{eqn_prob}) will in general depend on $k$, 
$n_k(l) \propto \exp c_k l$, so that $k$-strings will condense at
\begin{equation}
T_c(k)
=
\frac{\sigma_k}{c_k}.
\label{eqn_Tck}
\end{equation}
If we had no knowledge of the detailed dynamics we could
easily imagine that the $c_k$ might be such that a 
$T_c(k=2) < T_c(k=1)$. (Note that the reverse is not
possible because once $k=1$ sources deconfine, all
$k>1$ sources will also deconfine.) However the effective
string theory arises from a gauge theory and we will now
argue that in a gauge theory this probably cannot occur.

First we note that $c_k$ must contain a factor $1/l_k$
that provides a scale for the $l$ in the exponent.
This scale $l_k$ will be given, in the simplest case, 
by the radius, $r_k$,  of the $k$-string. This assumes
that different flux tubes are geometrically much the same
apart from their radii. Now, a flux tube carries
non-Abelian electric flux and if we assume that the flux is 
distributed uniformly across the cross-section of the flux tube,
a simple classical calculation would predict
\begin{equation}
\sigma_k
\propto
\int_0^{r_k} rdr \text{Tr}_k \frac{E}{r^2_k}\frac{E}{r^2_k}
\propto
\frac{k(N-k)(N+1)}{2N} \times \frac{1}{r^2_k}
\label{eqn_rsigk}
\end{equation}
where we have used the fact that the trace is the quadratic
Casimir and we have inserted the value appropriate to a
$k$-source in the totally antisymmetric representation.
(This classical calculation is just like the old bag model
\cite{bagTH}
but without the variational determination of the radius.)  
Inserting this and  $c_k=c_0/r_k$ into eqn(\ref{eqn_Tck})
we find
\begin{equation}
\frac{T_c(k)}{T_c(k=1)}
=
\sqrt{
\frac{\sigma_{k}}{\sigma_{k=1}}
\frac{k(N-k)}{N-1}
}
> 1.
\label{eqn_Tckr}
\end{equation}
Although the argument is clearly too simple, the ratio is
so much larger than unity that it seems very plausible that 
fundamental sources must deconfine at the lowest temperature,
leading to a simultaneous deconfinement of all other sources.

For $N\geq 3$ and for fundamental sources the phase transition
is first order, passing directly to a `gluon plasma' without 
string condensation. This dynamics will presumably break 
the centre symmetry completely so that there is a single
$T_c$. The possibility that there might be some earlier
second order phase transition corresponding to the 
condensation of higher $k$-strings is unlikely given
our arguments above coupled to the fact that the first order
transition temperature is less than the string condensation
temperature of fundamental strings.

Although our conclusion is that there is almost certainly
a single deconfinement transition, the possibility that things
are otherwise is too interesting to ignore. We therefore
perform a numerical investigation in Section \ref{section_Tck}, 
although without unearthing any surprises.

\subsection{V and N dependence}
\label{subsection_deconf_Ndep}

When the deconfining phase transition is first order, one
determines $T_c$ as lying in the range of $T$ where the
system undergoes  back-and-forth tunnelling between the
confined and deconfined phases. This tunnelling necessarily
proceeds through intermediate field configurations containing 
two interfaces that divide the periodic spatial volume into roughly 
two equal parts. Thus, at $T=T_c$ where the two phases have the same 
free energies, the tunnelling probability will be suppressed as
\begin{equation}
P_{tunnel}
\propto 
\exp\{-2\sigma_{cd}A/T_c\}
\label{eqn_ptunnel}
\end{equation}
where $\sigma_{cd}$ is the surface tension of the interface and
$A$ is the (smallest) spatial cross-section, e.g. $A = V^\frac{2}{3}$ 
for a symmetric volume $V$. So to be able to locate the actual phase
transition we need to work on a small enough volume for this tunnelling 
to be non-negligible. In that case, such back-and-forth tunnelling 
will occur for a range  $\Delta T$ of temperatures around $T_c$ 
for which  the difference in the free energies  $\Delta F/T$ is 
$O(1)$. (Once $\Delta F/T \gg 1$ the transition will effectively just
go one way.) For $T$ close to $T_c$ we can use the approximation 
$\Delta F \sim L_h V (1-T/T_c)$
where $L_h$ is the latent heat (per unit volume) of the transition,
and we see that $\Delta T$ and hence the uncertainty in locating
$T_c$ on the finite volume, is given by  
\begin{equation}
\Delta T \propto \frac{T^2_c}{L_h V}
\label{eqn_DT}
\end{equation}

Thus we see from eqn(\ref{eqn_DT}) that the latent heat determines
the uncertainty in determining $T_c$ on a volume $V$, while
eqn(\ref{eqn_ptunnel}) tells us that
the interface tension determines how large a $V$ one can consider 
and still see the transition at $T=T_c$. Of course, even if $V$ is 
so large that no  back-and-forth tunnelling can be seen at $T=T_c$, 
if we move $T$ away from $T_c$ then the growing free energy difference 
will eventually drive the tunnelling in one direction.
The tunnelling will proceed by the formation of a small
bubble that grows, improbably, to a certain critical size 
beyond which the free energy difference between the two phases 
outweighs the cost of the bubble interface free energy.
At this point the bubble will rapidly grow to fill the available
volume. (Such a continuous description is appropriate since 
the lattice Monte Carlo proceeds by changing one link matrix
at a time, and so is effectively a continuous deformation of the fields.)
It is clear that as the bubble grows, the cost of the bubble will
first grow and then decrease. If we estimate the tunnelling
probability by the thermodynamic suppression of the most
suppressed bubble in this process, we find that for
\begin{equation}
(T-T_c)^2 > c\frac{\sigma^3_{cd}}{L^2_h}T_c
\label{eqn_DThys1}
\end{equation}
($c$ is some numerical constant) one-way tunnelling will occur however
large is the volume. This gives the width of the hysteresis
and provides a finite uncertainty on our estimate of $T_c$ if
we work on very large volumes. To calculate $T_c$ one 
therefore works on volumes
that are small enough to possess tunnelling right across the
transition and one then extrapolates to infinite volume. 
On the other hand, if we choose to work with $V$ so large that no 
tunnelling occurs in a (realistic) simulation at $T\simeq T_c$, 
we will be able to calculate properties
of the confining (or deconfined) phase at $T=T_c$ --
and in its neighbourhood. We shall make use of this strategy for
$N\geq 4$ where the transition is strongly enough first order
for this to be practical.

In an earlier paper 
\cite{oxtemp03}
we performed a calculation of the latent heat at a fixed
value of the lattice spacing, $a\simeq 1/5T_c$, and found that
\begin{equation}
L_h 
\stackrel{N\to\infty}{\propto}
N^2,
\label{eqn_LhN}
\end{equation}
just as one expects from the
usual large-$N$ counting arguments. In the present paper we
will calculate the continuum limit of $L_h$ for 
$N=4,6,8$ and so determine not only the variation with $N$ 
but also its value in physical units. In
\cite{oxtemp03}
we also showed that the interface tension grows with $N$.
In the present paper we attempt to distinguish 
between $\sigma_{cd} \propto N$ and $\sigma_{cd} \propto N^2$
using a different method, but unfortunately our accuracy
will not allow us to answer this question convincingly.
(There is a simple theoretical motivation for 
$\sigma_{cd} \propto N^2$ that has been given in
\cite{oxtemp03}.)
Each of these two possibilities leads to a particularly
simple picture of the phase transition as $N\to\infty$.
We first note from eqn(\ref{eqn_ptunnel}) that for either power
of $N$ the back-and-forth tunnelling when $T=T_c$ and $\Delta F=0$
will be exponentially suppressed in $N$ (or in $N^2$) however
small we make $V$. However we also see from  eqn(\ref{eqn_LhN}) that 
if $\sigma_{cd} \propto N$ then 
the width of the hysteresis will be no more than 
$\Delta T \propto 1/N$. Thus as $N\to\infty$ on a fixed 
volume $V$, however small or large, the phase transition will become
completely sharp.
If, on the other hand, $\sigma_{cd} \propto N^2$ then
this is no longer the case. In this case eqn(\ref{eqn_DThys1})
tells us that on a large volume the hysteresis grows
as $\Delta T \propto N$ i.e. becomes infinite at $N=\infty$.
In this limit of infinite hysteresis the confined and deconfined 
phases exist at all $T$ with no tunnelling between them.
(If $V$ is smaller, below the critical bubble size, the
contribution of the interface tension always dominates
as the bubble grows, so this infinite hysteresis
scenario would in fact occur on all volumes.)
Of course our argument for eqn(\ref{eqn_DThys1}) breaks down
once $\Delta T$ becomes large, and in any case the string condensation 
transition will, presumably, intervene somewhere above $T_c$.

These two ideal scenarios -- zero or `infinite' hysteresis
on any volume $V$ -- provide one motivation
for being interested in the value and $N$-dependence of
the interface tension $\sigma_{cd}$. Unfortunately the
numerical results we present in Section~\ref{section_interface}
will not be conclusive. All this is also reason to be
interested in the latent heat although, since that is
a fundamental quantity characterising the phase transition,
we would be interested in it in any case. Here our
numerical results, in
\cite{oxtemp03}
and Section~\ref{section_Lh} are more precise.

\subsection{Masses}
\label{subsection_deconf_masses}

The size of the latent heat (rescaled by $N^2$) tells
us how strongly first order the phase transition is.
Another measure of this strength is the mass gap in
each of the phases as $T\to T_c$. For a second order
transition the mass gap vanishes; for a weakly first
order transition it becomes small but non-zero; for a normal 
first order transition it should be on the order of the
typical dynamical length scale of the theory, e.g.
$T_c$ or the $T=0$ string tension. If
the mass gap increases with $N$ it provides us with 
additional evidence that the transition is becoming more
strongly first order. We provide some calculations of this
mass gap in Section~\ref{section_masses}.
 
Since the transition is to do with deconfinement, the
natural mass to look at is the effective string tension
at finite $T$, $\sigma_{eff}(T)$. As discussed, for example,
in Section 3.1 of
\cite{oxtemp03}
this is given by the mass $m_t(l_t)$ of the lightest flux loop 
that winds around the timelike torus: 
\begin{equation}
\sigma_{eff}(T)
=
\frac{1}{l_t} m_t(l_t)
=
T m_t(l_t=\frac{1}{T}).
\label{eqn_sigT}
\end{equation}
Thus at $T=T_c$ it will be this mass,
$m_t(l_t=1/T_c)$, that will vanish for a second order
transition, and should be $\sim T_c$ for a normal
first order transition. 

Note also that eqn(\ref{eqn_sigT}) allows
us to derive the dependence of  $\sigma_{eff}(T)$ on $T$
for low temperatures, $T\sim 0$, in a very simple way. 
We recall that
the leading $l_t$ correction to the linear dependence of 
$m_t(l_t=1/T)$ is provided by the universal bosonic 
string correction, $\pi/3l_t$, so that
eqn(\ref{eqn_sigT}) implies
\begin{equation}
\sigma_{eff}(T)
\stackrel{T\to 0}{=}
\sigma - \frac{\pi}{3} T^2 + O(T^4).
\label{eqn_sigTb}
\end{equation}
A parallel argument holds for $k$-strings and their associated
string tensions, at any fixed $N$. If we take $N\to\infty$
at fixed $k$, the $k$-string becomes an ever more loosely
bound state of $k$ fundamental strings and,  for reasons
discussed in Section 3.1.2 of
\cite{oxglue04}, 
this implies that the leading correction in eqn(\ref{eqn_sigTb})
will hold in an ever decreasing interval around $T=0$.

This winding flux loop is the lightest state coupling to
timelike Wilson lines (Polyakov loops). In the deconfined
phase this operator acquires a non-zero vacuum expectation 
value. At high $T$ it can be expanded and the first non-trivial
term is $\propto Tr\{A^2_0\}$ showing that the strongest
coupling of the vacuum subtracted operator will be to two 
(electric) Debye-screened gluons. To leading order in the
't Hooft coupling, the Debye mass must clearly be 
$m^2_D \propto g^2(T) N T^2$ on dimensional grounds, and
a one-loop calculation
\cite{GPY}
shows that in fact
\begin{equation}
m_t(T)
=
2m_D
=
2\sqrt{\frac{g^2(T)N}{3}} T .
\label{eqn_mDT}
\end{equation}
Near $T=T_c$ this perturbative calculation and interpretation is no 
longer compelling but we shall use the same language for simplicity.

A quite different quantity from the flux loop that winds around
the timelike torus is the one that winds around a spatial torus. 
As $T\to 0$ this difference disappears
of course and either will provide a measure of the $T=0$
string tension. At high $T$ the mass, $m_s(l_s)$, of this 
spacelike flux loop and its associated string tension, 
$\sigma_s$, will vary as
\begin{equation}
\sigma_{s}(l_s;T)
=
\frac{1}{l_s} 
m_s(l_s)
\propto
T^2
\label{eqn_msT}
\end{equation}
on dimensional grounds. The `spacelike' string tension, 
$\sigma_s$, becomes the string tension of the linearly
confining D=2+1
gauge-scalar theory to which the D=3+1 SU($N$) gauge
theory is dimensionally reduced at high enough $T$.
In principle the spatial string tensions in the
confined and deconfined phases at $T\simeq T_c$ could
be the same, and this has often been 
inferred from earlier calculations in SU(3). However 
when one works at higher $N$ where the phase transition 
is quite strongly first order, so that the separation
into the two phases can be easily made unambiguous,
one can show, as we shall in Section~\ref{section_masses}, 
that there is in fact a significant 
discontinuity in $\sigma_s$ across $T_c$.

\subsection{Thermal average or Euclidean artifact?}
\label{subsection_deconf_artifact}

In using the Euclidean time framework for calculating
thermal averages at a temperature $T$ it is important to
be aware that dynamical features of the 4 dimensional
Euclidean fields will only be features of the finite
temperature field theory if they are encoded as such in
the (equal time) thermal averages. 

A cautionary example concerns the well-known $Z_N$ spontaneous 
symmetry breaking of SU($N$) gauge theories for $T > T_c$. 
It is certainly the case that as the Euclidean
time extent, $l_t$, of our 4-volume decreases through the 
critical value $l_t=1/T_c$, the 4-dimensional system
undergoes a phase transition to one of $N$ degenerate vacua.
The timelike Wilson line acquires a value that is proportional
to one of the $N$'th roots of unity and this serves as an order 
parameter distinguishing these various phases. There are interfaces
between these vacua and these are perturbatively calculable at high 
enough $T$
\cite{ckaDW,oxDW96}. 
Indeed we note that if we relabel co-ordinates 
$(x,y,z,t)\to (t,x,y,z)$ so that the short torus is now
in the $z$-direction and the Euclidean-time torus is 
now long, so that the theory is effectively at zero
temperature, then this phase transition becomes a real finite
volume phase transition, obtained by reducing $l_z$ while keeping
the other three lengths fixed (and large). All the properties 
listed above (suitably relabelled) become physically observable.
Nonetheless, despite appearances, it is now widely
accepted that all this does not mean that the deconfined phase
of SU($N$) gauge theories possesses such a symmetry breaking
with, for example, bubbles of different $Z_N$ vacua in the
high-$T$ gluon plasma. At least, no-one has succeeded in showing
that such a structure is encoded in an appropriate fashion
in equal-time thermal averages. (The timelike Polyakov loop
clearly cannot serve such a purpose.) 

It is therefore always important to consider if what one is seeing
might be a Euclidean artifact rather than a genuine property of
the gauge theory at finite $T$. This applies to the details of the
phase transition itself. Of course the first order transitions
for SU($N\geq 3$) are reflected in discontinuities in thermal 
averages and there is no ambiguity. However the second order
SU(2) transition is more problematic. In particular the
diverging correlation length -- easily seen in the correlation
of timelike Wilson lines -- is in practice not visible in equal-time
correlators and one finds no growing peak in the specific heat.
It is therefore interesting to ask in what way this transition
manifests itself in the high-$T$ gauge theory, and this we do
in Section~\ref{section_su2}.

\section{The lattice setup}
\label{section_lattice}

Our Euclidean space-time is discretised to a hypercubic, periodic 
lattice of size $L_s^3 L_t$ and lattice spacing $a$,
with lengths denoted by $L$ when there is no ambiguity. We assign
SU($N$) matrices, $U_l$, to the links $l$ (also denoted by $U_\mu(n)$ 
for the link emanating in the $\mu$ direction from site $n$). 
We use the standard plaquette action  
\begin{equation}
S = \beta \sum_{p}\{1-{\frac{1}{N}}{\mathrm {ReTr}} U_p\},
\label{B1}
\end{equation}
where $U_p$ is  the ordered product of the SU($N$) matrices 
around the boundary of the plaquette $p$. (We shall often
use the shorthand $u_p = {\mathrm {ReTr}} U_p/N$.)
$S$ appears in the Euclidean Path Integral as 
\begin{equation}
Z = \int \prod_l dU_l e^{-S}
\label{eqn_latZ}
\end{equation}
and becomes, in the continuum limit, the usual Yang-Mills action with
\begin{equation}
\beta = \frac{2N}{g^2}.
\label{B2}
\end{equation}
Since the theory is asymptotically free, we approach the continuum 
limit by reducing the bare coupling, 
$g^2(a)\stackrel{a\to 0}{\longrightarrow} 0$, and so $\beta\to\infty$. 
Our simulations in this paper are performed with a combination of 
heat bath and over-relaxation updates in which we update all the 
$N(N-1)/2$ SU(2) subgroups of the SU($N$) link matrices, as described 
in 
\cite{blmt-glue,oxglue04}.

The thermodynamic partition function of the gauge theory 
at a temperature $T$ is identical to the Euclidean Feynman Path 
Integral with temporal periodicity of $1/T$. So by calculating
appropriate expectation values on an $L_s^3L_t$ lattice with
$L_s\gg L_t$ we are calculating a lattice approximation to the
corresponding thermal average at
\begin{equation}
T = \frac{1}{a(\beta)L_t},
\label{B4}
\end{equation}
which becomes exact when we extrapolate to the continuum
limit at fixed $T$.

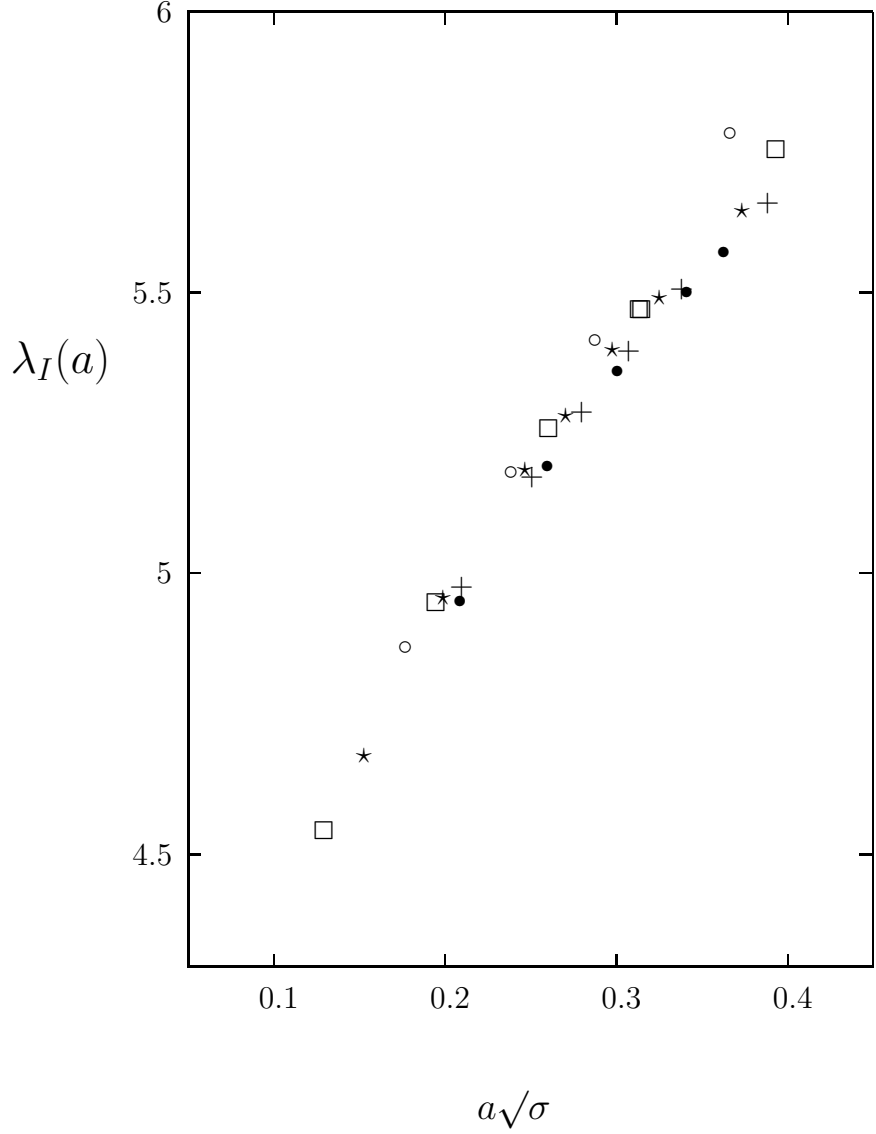
\begin	{figure}[p]
\begin	{center}
\leavevmode
\setlength{\unitlength}{0.240900pt}
\ifx\plotpoint\undefined\newsavebox{\plotpoint}\fi
\sbox{\plotpoint}{\rule[-0.200pt]{0.400pt}{0.400pt}}%
\begin{picture}(1500,1800)(0,0)
\font\gnuplot=cmr10 at 12pt
\gnuplot
\sbox{\plotpoint}{\rule[-0.200pt]{0.400pt}{0.400pt}}%
\put(350.0,426.0){\rule[-0.200pt]{4.818pt}{0.400pt}}
\put(325,426){\makebox(0,0)[r]{\ \ {$4.5$}}}
\put(1405.0,426.0){\rule[-0.200pt]{4.818pt}{0.400pt}}
\put(350.0,868.0){\rule[-0.200pt]{4.818pt}{0.400pt}}
\put(325,868){\makebox(0,0)[r]{\ \ {$5$}}}
\put(1405.0,868.0){\rule[-0.200pt]{4.818pt}{0.400pt}}
\put(350.0,1309.0){\rule[-0.200pt]{4.818pt}{0.400pt}}
\put(325,1309){\makebox(0,0)[r]{\ \ {$5.5$}}}
\put(1405.0,1309.0){\rule[-0.200pt]{4.818pt}{0.400pt}}
\put(350.0,1750.0){\rule[-0.200pt]{4.818pt}{0.400pt}}
\put(325,1750){\makebox(0,0)[r]{\ \ {$6$}}}
\put(1405.0,1750.0){\rule[-0.200pt]{4.818pt}{0.400pt}}
\put(484.0,250.0){\rule[-0.200pt]{0.400pt}{4.818pt}}
\put(484,200){\makebox(0,0){\ {$0.1$}}}
\put(484.0,1730.0){\rule[-0.200pt]{0.400pt}{4.818pt}}
\put(753.0,250.0){\rule[-0.200pt]{0.400pt}{4.818pt}}
\put(753,200){\makebox(0,0){\ {$0.2$}}}
\put(753.0,1730.0){\rule[-0.200pt]{0.400pt}{4.818pt}}
\put(1022.0,250.0){\rule[-0.200pt]{0.400pt}{4.818pt}}
\put(1022,200){\makebox(0,0){\ {$0.3$}}}
\put(1022.0,1730.0){\rule[-0.200pt]{0.400pt}{4.818pt}}
\put(1291.0,250.0){\rule[-0.200pt]{0.400pt}{4.818pt}}
\put(1291,200){\makebox(0,0){\ {$0.4$}}}
\put(1291.0,1730.0){\rule[-0.200pt]{0.400pt}{4.818pt}}
\put(350.0,250.0){\rule[-0.200pt]{258.967pt}{0.400pt}}
\put(1425.0,250.0){\rule[-0.200pt]{0.400pt}{361.350pt}}
\put(350.0,1750.0){\rule[-0.200pt]{258.967pt}{0.400pt}}
\put(150,1200){\makebox(0,0){\Large{$\lambda_I(a)$}}}
\put(862,25){\makebox(0,0){\large{$a\surd\sigma$}}}
\put(350.0,250.0){\rule[-0.200pt]{0.400pt}{361.350pt}}
\put(1191,1373){\circle*{18}}
\put(1133,1310){\circle*{18}}
\put(1024,1185){\circle*{18}}
\put(914,1036){\circle*{18}}
\put(777,825){\circle*{18}}
\put(1260,1449){\makebox(0,0){$+$}}
\put(1125,1314){\makebox(0,0){$+$}}
\put(1042,1217){\makebox(0,0){$+$}}
\put(968,1121){\makebox(0,0){$+$}}
\put(890,1019){\makebox(0,0){$+$}}
\put(780,846){\makebox(0,0){$+$}}
\put(1220,1437){\makebox(0,0){$\star$}}
\put(1090,1301){\makebox(0,0){$\star$}}
\put(1016,1219){\makebox(0,0){$\star$}}
\put(943,1115){\makebox(0,0){$\star$}}
\put(879,1030){\makebox(0,0){$\star$}}
\put(750,830){\makebox(0,0){$\star$}}
\put(626,581){\makebox(0,0){$\star$}}
\put(1273,1532){\raisebox{-.8pt}{\makebox(0,0){$\Box$}}}
\put(1062,1281){\raisebox{-.8pt}{\makebox(0,0){$\Box$}}}
\put(1058,1280){\raisebox{-.8pt}{\makebox(0,0){$\Box$}}}
\put(916,1093){\raisebox{-.8pt}{\makebox(0,0){$\Box$}}}
\put(739,820){\raisebox{-.8pt}{\makebox(0,0){$\Box$}}}
\put(563,462){\raisebox{-.8pt}{\makebox(0,0){$\Box$}}}
\put(1201,1559){\circle{18}}
\put(989,1234){\circle{18}}
\put(857,1027){\circle{18}}
\put(691,752){\circle{18}}
\end{picture}
\end	{center}
\vskip 0.15in
\caption{The value of the `t Hooft coupling on the scale $a$,
as obtained from $\beta$ in eqn(\ref{eqn_ggNI}),
for $N=2(\circ),3(\Box),4(\star),6(+),8(\bullet)$, plotted against the
values of $a$ expressed in physical units.}
\label{fig_ggNI}
\end 	{figure}

When we vary $N$ we expect 
\cite{largeN,lattice-thooft}
that we will need to keep constant the 't Hooft coupling, $\lambda$,
\begin{equation}
\lambda \equiv g^2 N  
\label{B3}
\end{equation}
to obtain a smooth large-$N$ limit. Non-perturbative calculations
for $2\leq N \leq 5$ have supported this expectation
\cite{blmt-glue}.
Using the string tension calculations in our recent paper
\cite{oxglue04}
we can extend this study to  $2\leq N \leq 8$. Following 
\cite{blmt-glue}
we define a (lattice improved) 't Hooft coupling on the scale $a$ by
\begin{equation}
\lambda_I(a) = g_I(a)^2 N
=\frac{2N^2}{\beta \langle u_p \rangle}
\label{eqn_ggNI}
\end{equation}
and plot in Fig.\ref{fig_ggNI} the result. We observe very nice
evidence that the running 't Hooft coupling approaches a 
limiting $N$-independent function of the length scale 
$l=a\surd\sigma$ as $N\to\infty$, and that it does so very
rapidly, once we are away from the coarsest values of $a$ that
are close to the bulk transition/cross-over 
(see Section~\ref{section_bulk}).

Finally two comments for the general reader.

Our Monte Carlo generates a sequence of fields by changing the 
value of one link matrix at a time, with the local action density
controlling the size of the change. Thus the Monte Carlo 
proceeds by what is essentially a local deformation of the
fields and so it mimics features of a real phase transition
in a statistical mechanics system. That is to say, a first order 
transition can be described as a continuous tunnelling process,
in terms of growing bubbles of one phase within another phase,
and if at $T=T_c$ that probability is low enough then we will
observe a hysteresis effect. 

Our second comment is that the deconfining phase transition
occurs only on physical length scales. Thus in lattice units
the latent heat will be 
$a^4 L_h\propto a^4 \stackrel{a\to 0}{\longrightarrow} 0$
and the mass gap at the transition will also satisfy 
$am_g \stackrel{a\to 0}{\longrightarrow} 0$.
So from the point of view of the lattice, the transition
is very weakly first order and becomes second order as $a\to 0$. 
This is however an illusion, merely reflecting the fact that
the continuum limit is a second order transition of the
lattice theory. When we express the mass gap and latent heat 
in physical units, e.g. as $L_h/T^4_c$ and  $m_g/T_c$,
it becomes clear that the transition is robustly first
order for $N\geq 4$.

\section{The deconfining temperature}
\label{section_Tc}

Since presenting our results on $T_c/\surd\sigma$ in
\cite{oxtemp03}
we have significantly improved our SU(8) calculations.
We now have a greater range of volumes at $L_t=5$ and
this allows us to obtain a more reliable estimate of the
coefficient $h$ of the leading $1/V$ correction to $\beta_c$:
\begin{equation}
\beta_c(V) 
\stackrel{V\to\infty}{=} 
\beta_c(\infty) - h\frac{L^3_t}{L^3_s}.
\label{Tc1}
\end{equation}
We also have better calculations of the SU(8) lattice string tensions
\cite{oxglue04}
and this allows us to perform a more accurate continuum extrapolation
of  $T_c/\surd\sigma$. 

We recall
\cite{oxtemp03}
that our strategy is to use the value of $h$ calculated at 
$a=1/5T_c$ to perform $V\to\infty$ extrapolations at the smaller
values of $a$ at which we usually have performed calculations
on only one volume. In 
\cite{oxtemp03}
we did not attempt to compute the $a$-dependence of $h$ and
we simply doubled the statistical error on $h$ in the hope
that this would encompass any variation of $h$ with $a(\beta)$. 
However the form of this $a$ dependence of $h$ is known
\cite{oxtemp03}
\begin{equation}
h(a) 
\stackrel{a\to 0}{=}
\frac{h_0}{\frac{d}{d\beta}\ln a(\beta=\beta_c)}
+
O(a^2)
\label{Tc2}
\end{equation}
and so we can use the values of  $d\ln a/d\beta$ that we calculate later
on in this paper to rescale $h$ at the $\beta$ values of interest.
We shall do this not just for $N=8$, but for $N=4,6$ as well.
We then increase the error on $h$ by $15\%$ in order to cover the 
remaining $O(a^2)$ corrections which, given what we observe with
other dimensionless quantities, we expect to be small.

In Tables~\ref{table_su4lt} -- \ref{table_su8lt} we list our
calculated values of the critical $\beta$, obtained by extrapolating
to $V=\infty$ using eqns(\ref{Tc1},\ref{Tc2}) with the values of $h$
as shown. We calculate the string tension at each $\beta$
using the values in
\cite{oxglue04}
and this provides us with the listed values of $T_c/\surd\sigma$.
Comparing with Table 5 of
\cite{oxtemp03}
we see that the SU(4) and SU(6) values have changed very little,
while the SU(8) values, where we employ better calculations
of $\sigma$, have changed more substantially.

We extrapolate to the continuum limit using an $a^2\sigma$
correction, as described in 
\cite{oxtemp03},
and list the resulting continuum values  of $T_c/\surd\sigma$
in Table~\ref{table_suN}.
For completeness we have included the SU(2) and SU(3) values although
their analysis is unchanged from 
\cite{oxtemp03}.
We note that the fits all have an 
acceptable $\chi^2$ per degree of freedom except for SU(4) where 
the fit is very poor. This promises to create a problem for any
statistical analysis that involves the SU(4) value, as for example
in the extrapolation to $N=\infty$ to which we now turn.

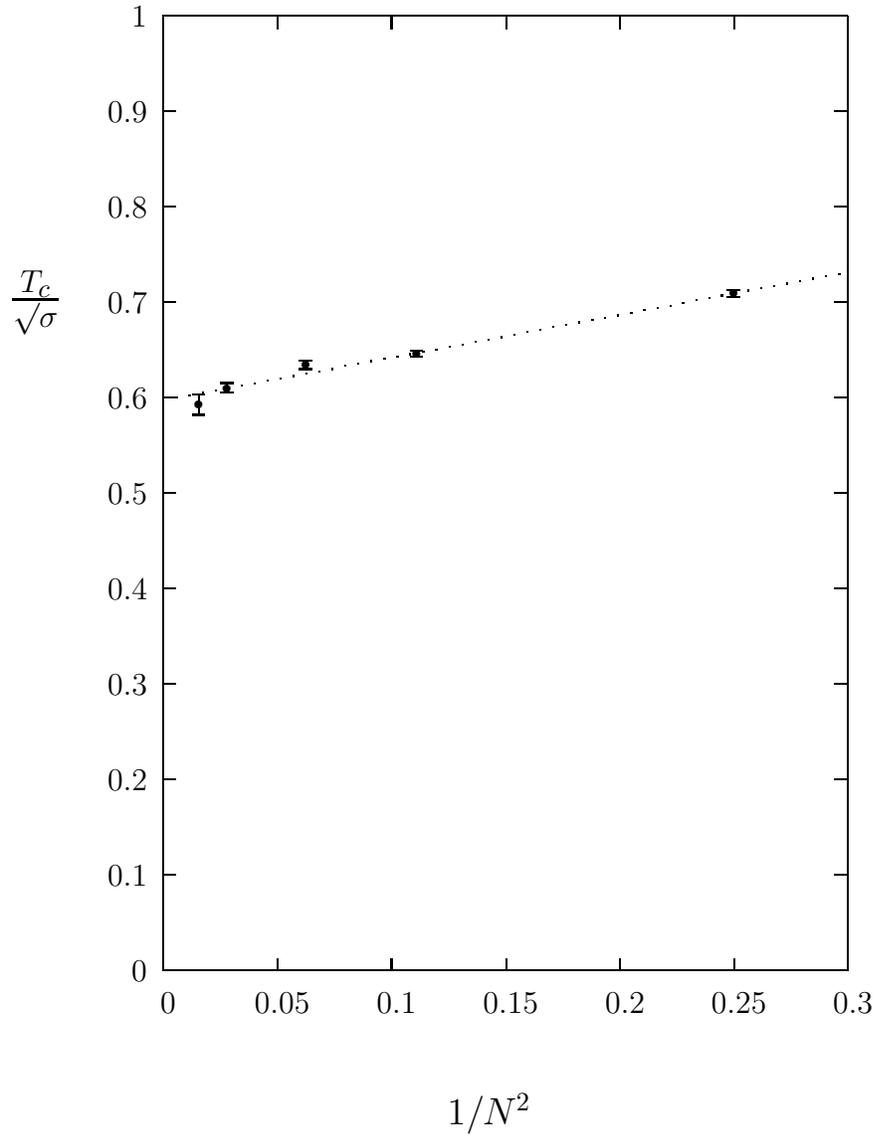
\begin	{figure}[p]
\begin	{center}
\leavevmode
\setlength{\unitlength}{0.240900pt}
\ifx\plotpoint\undefined\newsavebox{\plotpoint}\fi
\sbox{\plotpoint}{\rule[-0.200pt]{0.400pt}{0.400pt}}%
\begin{picture}(1500,1800)(0,0)
\font\gnuplot=cmr10 at 12pt
\gnuplot
\sbox{\plotpoint}{\rule[-0.200pt]{0.400pt}{0.400pt}}%
\put(350.0,250.0){\rule[-0.200pt]{4.818pt}{0.400pt}}
\put(325,250){\makebox(0,0)[r]{\ \ {$0$}}}
\put(1405.0,250.0){\rule[-0.200pt]{4.818pt}{0.400pt}}
\put(350.0,400.0){\rule[-0.200pt]{4.818pt}{0.400pt}}
\put(325,400){\makebox(0,0)[r]{\ \ {$0.1$}}}
\put(1405.0,400.0){\rule[-0.200pt]{4.818pt}{0.400pt}}
\put(350.0,550.0){\rule[-0.200pt]{4.818pt}{0.400pt}}
\put(325,550){\makebox(0,0)[r]{\ \ {$0.2$}}}
\put(1405.0,550.0){\rule[-0.200pt]{4.818pt}{0.400pt}}
\put(350.0,700.0){\rule[-0.200pt]{4.818pt}{0.400pt}}
\put(325,700){\makebox(0,0)[r]{\ \ {$0.3$}}}
\put(1405.0,700.0){\rule[-0.200pt]{4.818pt}{0.400pt}}
\put(350.0,850.0){\rule[-0.200pt]{4.818pt}{0.400pt}}
\put(325,850){\makebox(0,0)[r]{\ \ {$0.4$}}}
\put(1405.0,850.0){\rule[-0.200pt]{4.818pt}{0.400pt}}
\put(350.0,1000.0){\rule[-0.200pt]{4.818pt}{0.400pt}}
\put(325,1000){\makebox(0,0)[r]{\ \ {$0.5$}}}
\put(1405.0,1000.0){\rule[-0.200pt]{4.818pt}{0.400pt}}
\put(350.0,1150.0){\rule[-0.200pt]{4.818pt}{0.400pt}}
\put(325,1150){\makebox(0,0)[r]{\ \ {$0.6$}}}
\put(1405.0,1150.0){\rule[-0.200pt]{4.818pt}{0.400pt}}
\put(350.0,1300.0){\rule[-0.200pt]{4.818pt}{0.400pt}}
\put(325,1300){\makebox(0,0)[r]{\ \ {$0.7$}}}
\put(1405.0,1300.0){\rule[-0.200pt]{4.818pt}{0.400pt}}
\put(350.0,1450.0){\rule[-0.200pt]{4.818pt}{0.400pt}}
\put(325,1450){\makebox(0,0)[r]{\ \ {$0.8$}}}
\put(1405.0,1450.0){\rule[-0.200pt]{4.818pt}{0.400pt}}
\put(350.0,1600.0){\rule[-0.200pt]{4.818pt}{0.400pt}}
\put(325,1600){\makebox(0,0)[r]{\ \ {$0.9$}}}
\put(1405.0,1600.0){\rule[-0.200pt]{4.818pt}{0.400pt}}
\put(350.0,1750.0){\rule[-0.200pt]{4.818pt}{0.400pt}}
\put(325,1750){\makebox(0,0)[r]{\ \ {$1$}}}
\put(1405.0,1750.0){\rule[-0.200pt]{4.818pt}{0.400pt}}
\put(350.0,250.0){\rule[-0.200pt]{0.400pt}{4.818pt}}
\put(350,200){\makebox(0,0){\ {$0$}}}
\put(350.0,1730.0){\rule[-0.200pt]{0.400pt}{4.818pt}}
\put(529.0,250.0){\rule[-0.200pt]{0.400pt}{4.818pt}}
\put(529,200){\makebox(0,0){\ {$0.05$}}}
\put(529.0,1730.0){\rule[-0.200pt]{0.400pt}{4.818pt}}
\put(708.0,250.0){\rule[-0.200pt]{0.400pt}{4.818pt}}
\put(708,200){\makebox(0,0){\ {$0.1$}}}
\put(708.0,1730.0){\rule[-0.200pt]{0.400pt}{4.818pt}}
\put(888.0,250.0){\rule[-0.200pt]{0.400pt}{4.818pt}}
\put(888,200){\makebox(0,0){\ {$0.15$}}}
\put(888.0,1730.0){\rule[-0.200pt]{0.400pt}{4.818pt}}
\put(1067.0,250.0){\rule[-0.200pt]{0.400pt}{4.818pt}}
\put(1067,200){\makebox(0,0){\ {$0.2$}}}
\put(1067.0,1730.0){\rule[-0.200pt]{0.400pt}{4.818pt}}
\put(1246.0,250.0){\rule[-0.200pt]{0.400pt}{4.818pt}}
\put(1246,200){\makebox(0,0){\ {$0.25$}}}
\put(1246.0,1730.0){\rule[-0.200pt]{0.400pt}{4.818pt}}
\put(1425.0,250.0){\rule[-0.200pt]{0.400pt}{4.818pt}}
\put(1425,200){\makebox(0,0){\ {$0.3$}}}
\put(1425.0,1730.0){\rule[-0.200pt]{0.400pt}{4.818pt}}
\put(350.0,250.0){\rule[-0.200pt]{258.967pt}{0.400pt}}
\put(1425.0,250.0){\rule[-0.200pt]{0.400pt}{361.350pt}}
\put(350.0,1750.0){\rule[-0.200pt]{258.967pt}{0.400pt}}
\put(150,1300){\makebox(0,0){\Large{${{T_c}\over{\surd\sigma}}$}}}
\put(862,25){\makebox(0,0){\large{$1/N^2$}}}
\put(350.0,250.0){\rule[-0.200pt]{0.400pt}{361.350pt}}
\put(1246.0,1308.0){\rule[-0.200pt]{0.400pt}{2.650pt}}
\put(1236.0,1308.0){\rule[-0.200pt]{4.818pt}{0.400pt}}
\put(1236.0,1319.0){\rule[-0.200pt]{4.818pt}{0.400pt}}
\put(748.0,1215.0){\rule[-0.200pt]{0.400pt}{2.168pt}}
\put(738.0,1215.0){\rule[-0.200pt]{4.818pt}{0.400pt}}
\put(738.0,1224.0){\rule[-0.200pt]{4.818pt}{0.400pt}}
\put(574.0,1195.0){\rule[-0.200pt]{0.400pt}{3.132pt}}
\put(564.0,1195.0){\rule[-0.200pt]{4.818pt}{0.400pt}}
\put(564.0,1208.0){\rule[-0.200pt]{4.818pt}{0.400pt}}
\put(450.0,1158.0){\rule[-0.200pt]{0.400pt}{3.613pt}}
\put(440.0,1158.0){\rule[-0.200pt]{4.818pt}{0.400pt}}
\put(440.0,1173.0){\rule[-0.200pt]{4.818pt}{0.400pt}}
\put(406.0,1123.0){\rule[-0.200pt]{0.400pt}{7.709pt}}
\put(396.0,1123.0){\rule[-0.200pt]{4.818pt}{0.400pt}}
\put(1246,1314){\circle*{12}}
\put(748,1219){\circle*{12}}
\put(574,1202){\circle*{12}}
\put(450,1165){\circle*{12}}
\put(406,1139){\circle*{12}}
\put(396.0,1155.0){\rule[-0.200pt]{4.818pt}{0.400pt}}
\put(350,1146){\usebox{\plotpoint}}
\put(350.00,1146.00){\usebox{\plotpoint}}
\put(370.42,1149.71){\usebox{\plotpoint}}
\put(390.82,1153.56){\usebox{\plotpoint}}
\put(411.23,1157.31){\usebox{\plotpoint}}
\put(431.65,1161.03){\usebox{\plotpoint}}
\put(452.07,1164.74){\usebox{\plotpoint}}
\put(472.46,1168.63){\usebox{\plotpoint}}
\put(492.88,1172.34){\usebox{\plotpoint}}
\put(513.30,1176.05){\usebox{\plotpoint}}
\put(533.72,1179.77){\usebox{\plotpoint}}
\put(554.11,1183.66){\usebox{\plotpoint}}
\put(574.53,1187.37){\usebox{\plotpoint}}
\put(594.95,1191.08){\usebox{\plotpoint}}
\put(615.35,1194.87){\usebox{\plotpoint}}
\put(635.54,1199.64){\usebox{\plotpoint}}
\put(655.96,1203.36){\usebox{\plotpoint}}
\put(676.38,1207.07){\usebox{\plotpoint}}
\put(696.77,1210.95){\usebox{\plotpoint}}
\put(717.19,1214.67){\usebox{\plotpoint}}
\put(737.61,1218.38){\usebox{\plotpoint}}
\put(758.03,1222.10){\usebox{\plotpoint}}
\put(778.42,1225.98){\usebox{\plotpoint}}
\put(798.84,1229.70){\usebox{\plotpoint}}
\put(819.26,1233.41){\usebox{\plotpoint}}
\put(839.68,1237.14){\usebox{\plotpoint}}
\put(860.07,1241.01){\usebox{\plotpoint}}
\put(880.49,1244.72){\usebox{\plotpoint}}
\put(900.91,1248.44){\usebox{\plotpoint}}
\put(921.11,1253.11){\usebox{\plotpoint}}
\put(941.50,1257.00){\usebox{\plotpoint}}
\put(961.92,1260.71){\usebox{\plotpoint}}
\put(982.34,1264.43){\usebox{\plotpoint}}
\put(1002.76,1268.15){\usebox{\plotpoint}}
\put(1023.15,1272.03){\usebox{\plotpoint}}
\put(1043.57,1275.74){\usebox{\plotpoint}}
\put(1063.99,1279.45){\usebox{\plotpoint}}
\put(1084.39,1283.28){\usebox{\plotpoint}}
\put(1104.80,1287.05){\usebox{\plotpoint}}
\put(1125.22,1290.77){\usebox{\plotpoint}}
\put(1145.64,1294.48){\usebox{\plotpoint}}
\put(1166.02,1298.37){\usebox{\plotpoint}}
\put(1186.23,1303.04){\usebox{\plotpoint}}
\put(1206.65,1306.75){\usebox{\plotpoint}}
\put(1227.07,1310.47){\usebox{\plotpoint}}
\put(1247.46,1314.36){\usebox{\plotpoint}}
\put(1267.88,1318.07){\usebox{\plotpoint}}
\put(1288.30,1321.78){\usebox{\plotpoint}}
\put(1308.71,1325.54){\usebox{\plotpoint}}
\put(1329.10,1329.38){\usebox{\plotpoint}}
\put(1349.52,1333.10){\usebox{\plotpoint}}
\put(1369.95,1336.81){\usebox{\plotpoint}}
\put(1390.34,1340.67){\usebox{\plotpoint}}
\put(1410.75,1344.41){\usebox{\plotpoint}}
\put(1425,1347){\usebox{\plotpoint}}
\end{picture}
\end	{center}
\vskip 0.15in
\caption{The SU($N$) continuum deconfining temperature in units of 
the string tension, with an extrapolation to $N=\infty$ using
a leading $O(1/N^2)$ correction.}
\label{fig_tcN}
\end 	{figure}

In Fig.\ref{fig_tcN} we plot our continuum values of  $T_c/\surd\sigma$
against $1/N^2$. The usual large-$N$ counting tells us that 
the leading large-$N$ correction should be $O(1/N^2)$, and this
corresponds to a straight line on this plot. While it is clear
that our results do fall approximately onto a straight line,
the best fit $\chi^2$ is in fact very poor. It is reasonable to 
suppose that this is due to our SU(4) value being off perhaps
by several standard deviations. To try and deal with this
we rescale the errors on the SU(4) lattice calculations 
by almost a factor of two, so that
the continuum extrapolation has a $\chi^2$ per degree of freedom 
of unity. This increases the error on the continuum extrapolation
to the value shown in square brackets in Table~\ref{table_suN}.
Performing a fit with this modified SU(4) error, we obtain
\begin{equation}
\frac{T_c}{\surd\sigma}
=
0.5970(38) + \frac{0.449(29)}{N^2}
\label{Tc3}
\end{equation}
with an acceptable $\chi^2$ per degree of freedom of about unity. 
This fit is plotted in Fig.\ref{fig_tcN}. As we have noted before
\cite{oxtemp02,oxtemp03}
it is remarkable and perhaps puzzling that the leading correction 
suffices all the way down to SU(2) despite the fact that the 
transition changes from first to second order between $N=3$
and $N=2$.

In performing our large-$N$ extrapolation we assume a
conventional $O(1/N^2)$ correction. Since this expectation
is based on diagrams (albeit to all orders) while 
$T_c/\surd\sigma$ is a non-perturbative quantity, it is 
interesting to see whether our results are accurate
enough to constrain the power of the leading correction.
We therefore perform fits to a $1/N^\alpha$ correction.
We find that the $\chi^2$ per degree of freedom of the
best fit is  
\begin{equation}
\frac{\chi^2}{n_{df}}
=
\left\{ \begin{array}{r@{\quad:\quad}r}
3.5  &   \alpha = 1 \\ 0.8  &   \alpha = 2 \\ 3.1  &   \alpha = 3  
\end{array} \right.
\label{power}
\end{equation}
which provides some evidence that the leading correction is
indeed $O(1/N^2)$.

\section{Latent heat}
\label{section_Lh}

\subsection{Background}
\label{subsection_Lhbackground}

The free energy, $F$, and the free energy per unit volume, $f$, 
are defined to be
\begin{equation}
Z 
\equiv
\sum_{states} e^{-\frac{E}{T}} 
=
e^{-\frac{F}{T}}
=
e^{-\frac{fV}{T}}
\label{eqn_fZ}
\end{equation}
where $V$ is the spatial volume and $Z$ is the standard  
finite temperature partition function. The internal energy
density, $\epsilon$, is the average energy per unit volume
\begin{equation}
\epsilon
\equiv
\frac{\overline{E}}{V}
=
\frac{T^2}{V} \frac{\partial}{\partial T}\ln Z.
\label{eqn_EZ}
\end{equation}

Suppose we are on a $L_s^3L_t$ lattice that remains fixed in
lattice units. The Euclidean Feynman Path integral provides a
definition of a thermodynamic partition function of the
lattice system at $T=1/aL_t$ which will tend to the desired 
continuum partition function as we reduce $a$ (as long as the 
spatial volume is sufficiently large). $a$ and hence $T$ are 
varied by varying $\beta$, so eqn(\ref{eqn_EZ}) becomes
\begin{equation}
\epsilon
=
\frac{T^2}{V}  \frac{\partial\beta}{\partial T}
\frac{\partial}{\partial\beta}\ln Z
=
\frac{T^2}{V} \Bigl\{-L_ta^2 \frac{\partial\beta}{\partial a} \Bigr\}
 \Bigl\{ 6 L^3_s L_t\langle u_p \rangle  \Bigr\}
=
- 6\frac{\langle u_p \rangle}{a^4} a \frac{\partial\beta}{\partial a}
\label{eqn_EZb}
\end{equation}
where $\langle u_p \rangle$ is the average plaquette.
We thus see that the latent heat in units of $T_c$ is given by
\begin{equation}
\frac{L_h}{T^4_c}
\equiv
\frac{\Delta\epsilon}{T^4_c}
=
-L_t^4 a \frac{\partial\beta}{\partial a} 
 6 \Delta\langle u_p \rangle.
\label{eqn_Lhplaq}
\end{equation}
Note that if we wished to calculate the internal energy 
or the free energy, we would need to be concerned about
regularising it, for example by subtracting from 
$\langle u_p \rangle$ in eqn(\ref{eqn_EZb}) the
value of the plaquette at some reference value of
$T$ (but at the same value of $a$). Since we are
only interested in the difference in internal energies,
such issues need not concern us here.

The finite $T$ phase transition is clearly also a phase transition 
of a statistical system in 4 space dimensions whose classical 
partition function is our Euclidean Feynman path integral.
In this reinterpretation, $\beta$ is the inverse temperature and 
the action $S$ is the energy. In this case the latent heat is simply 
the jump in the action i.e. it is just what we find in
eqn(\ref{eqn_Lhplaq}) but without the factor of  
$a {\partial\beta}/{\partial a}$. This latter factor is,
however, crucial if we want to obtain an estimate
of the real deconfining latent heat of the gauge theory
in 3 space dimensions.

We can calculate $a {\partial\beta}/{\partial a}$
using
\begin{equation}
a \frac{\partial\beta}{\partial a} 
=
a\mu \frac{\partial\beta}{\partial (a\mu)}
\label{eqn_abeta}
\end{equation}
where $\mu$ is some physical quantity with dimensions of mass,
e.g. $\mu=\surd\sigma$ or $\mu = T_c$. Such different
choices will differ by $O(a^2)$ lattice corrections
in ${\Delta\epsilon}/{T^4_c}$. We shall choose to use 
$\mu=\surd\sigma$. To obtain the value of 
${\partial (a\surd\sigma)}/{\partial\beta}$ at the
values of $\beta$ that correspond to $T=T_c$ for
various values of $L_t$ we interpolate between our
calculated values of $a\surd\sigma$ (as listed in Tables 5-9 of
\cite{oxglue04})
using a fitting function
\begin{equation}
a\sqrt{\sigma(\beta)}
=
c_0 \exp\Bigl\{ -\frac{12\pi^2}{11}\frac{(\beta-\beta_0)}{N^2}
+ c_1 x + c_2 x^2 + c_3 x^3 \Bigr\}
\ \ \ \ ; \ \ 
x \equiv N^2 (\frac{1}{\beta}-\frac{1}{\beta_0}).
\label{eqn_Kfit}
\end{equation}
The value of $\beta_0$ is chosen to coincide with one of the
calculations near the middle of the interesting range of $\beta$.
(Given that we fit $c_0$, this is not an extra parameter.)
Where we can obtain  a good fit without the $O(x^3)$
term we do so. Note that although the fitting function is designed 
so that it automatically tends to one-loop scaling at
large enough $\beta$, the particular form of the fit does
not really matter as long as we intend to use it to interpolate 
rather than to extrapolate in $\beta$. (Including a $\ln \beta$
term corresponding to 2-loop scaling does not improve the fits
and does not even appear to improve the rate at which scaling is 
approached when the fits are extrapolated to larger $\beta$.)
In Table~\ref{table_Kfit}
we list the values of the parameters of the best fit for each $N$
together with the range of $\beta$ interpolated (and the $\chi^2$
per degree of freedom of the best fit).

\subsection{Results}
\label{subsection_Lhresults}

In
\cite{oxtemp03}
we calculated the value of $\Delta\langle u_p \rangle$
from the $V\to\infty$ limit of the specific heat peak. 
(Note that in
\cite{oxtemp03}
we referred to this as $L_h$, and the latent heat
was defined as the jump in the average `energy' per
plaquette of the four dimensional lattice system. This differs
from the physical latent heat of the three dimensional
thermal gauge theory, as discussed above.) This calculation
requires a detailed finite volume study, which we have only
carried out at $a\simeq 1/5T_c$, with the results shown
in Table~\ref{table_lhCmax}, and so it is only at this
fixed lattice spacing that we have been able to compare
latent heats for different $N$. What we would now like to
obtain is the comparison in the continuum limit, and for
that we need a different technique.

We begin by observing that on our very largest lattice volumes at 
$a\simeq 1/5T_c$ there is no tunnelling for $N\geq 4$ at 
$T\simeq T_c$ in our simulations. 
We can therefore produce ensembles of lattice fields that are
entirely confining or deconfining by obtaining the starting
lattice field from $T\to T^-_c$ and  $T\to T^+_c$ respectively
(and subsequently thermalising at $T=T_c$). From these two
ensembles we can calculate $\langle u_p \rangle_c$ and  
$\langle u_p \rangle_d$, respectively, and hence  
$\Delta\langle u_p \rangle$. The values of the latter are 
listed in the $L_t=5$ rows of Table~\ref{table_lh}. 
They are all compatible with the values obtained from
the specific heat extrapolation, as listed in Table~\ref{table_lhCmax},
reinforcing our confidence in both methods. 

We can extend this method to smaller volumes where the
tunnelling is present but is relatively rare. We can use
the mean Polyakov loop as an order parameter to categorise
subsequences in the Monte Carlo generated sequence of
lattice fields as being confined, deconfined or 
(much more rarely) tunnelling. There is clearly some
ambiguity in this separation, but this ambiguity 
disappears as the tunnelling becomes less frequent. 
We try to incorporate this ambiguity into the error
estimate by using various different cuts on the order
parameter and by averaging the order parameter over 
subsequences of lattice fields of different length.
However once the spatial volume becomes sufficiently
small in physical units -- as measured by $L_s/L_t$ --
all these remedies become inadequate and our values
will be increasingly unreliable. A test of this method
for the case of SU(4) can be found in Fig.5 of
\cite{oxtemp02} 
where the $L=32$ lattice has no tunnelling and provides
the benchmark for the comparison. We observe that the
error becomes rapidly larger as we decrease $V$, as it
should, but that the value of $\Delta\langle u_p \rangle$
remains consistent with the $V\to\infty$ value, even on the
smallest volumes where some separation between phases
is apparent. This reassures us that this method can be
used at smaller values of $a$ where we do not have calculations
on volumes that are very large in physical units.

Our results are listed in Table~\ref{table_lh}. For $N\geq 4$ 
we have three values of $L_t$. Since our calculations are
at a fixed temperature, $T\simeq T_c$, we have $a\propto 1/L_t$
and this will allow us to perform a continuum extrapolation.
At each $N$ and $L_t$ we have used only the largest lattices.
We show the value of $\beta$ at which 
$\Delta\langle u_p \rangle$ is calculated. In fact around
each of these values of $\beta$ there are usually two or three 
nearby values which exhibit tunnelling and are useful. We
transport these values to the indicated value, varying
$a^4$ using the listed values of $\partial\beta/\partial\ln a$.
We then transport from these reference values of $\beta$
to the values of $\beta_c(V=\infty;L_t)$ listed in 
Tables~\ref{table_su4lt}-\ref{table_su8lt} and using
eqn(\ref{eqn_Lhplaq}) we obtain the latent heat at $T=T_c$ in
physical units, as shown. We now discuss these results.

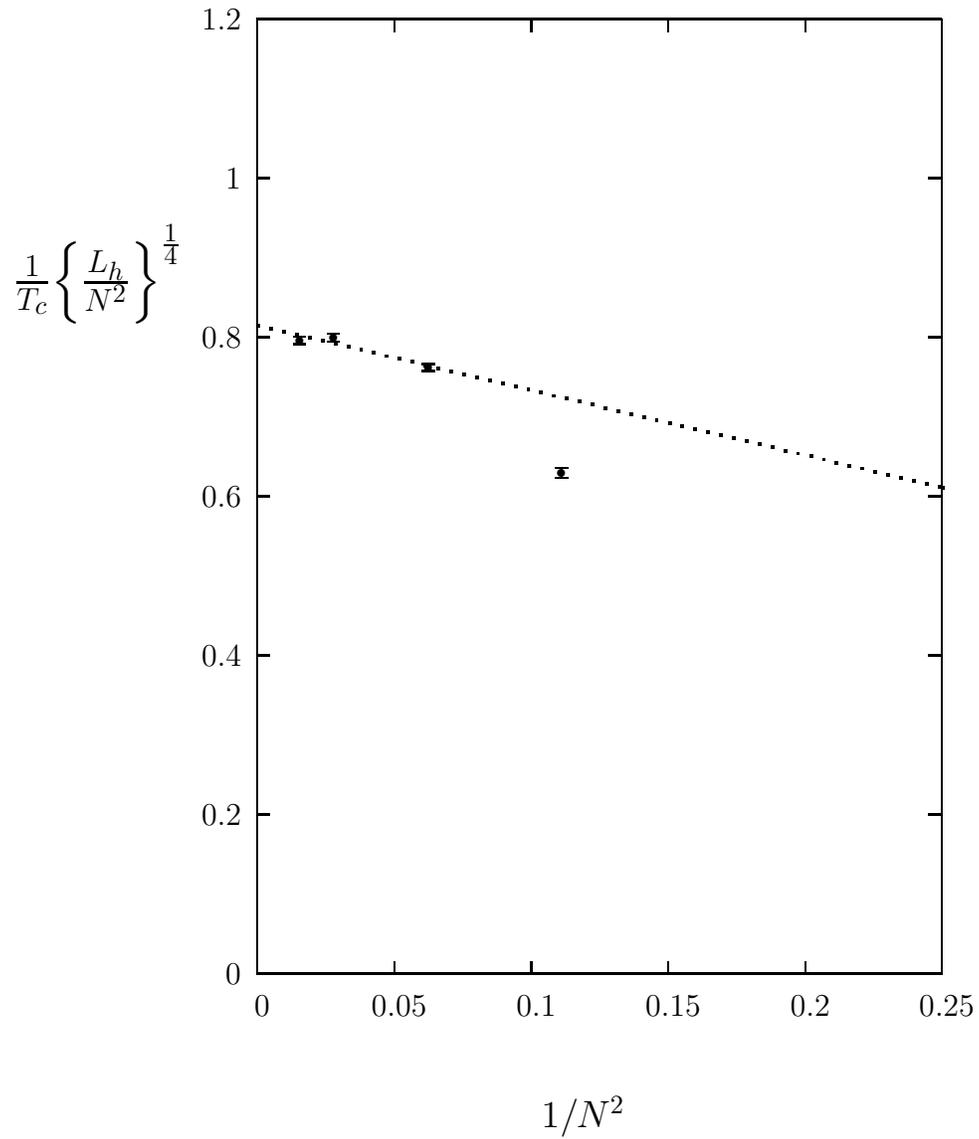
\begin	{figure}[p]
\begin	{center}
\leavevmode
\setlength{\unitlength}{0.240900pt}
\ifx\plotpoint\undefined\newsavebox{\plotpoint}\fi
\sbox{\plotpoint}{\rule[-0.200pt]{0.400pt}{0.400pt}}%
\begin{picture}(1500,1800)(0,0)
\font\gnuplot=cmr10 at 12pt
\gnuplot
\sbox{\plotpoint}{\rule[-0.200pt]{0.400pt}{0.400pt}}%
\put(350.0,250.0){\rule[-0.200pt]{4.818pt}{0.400pt}}
\put(325,250){\makebox(0,0)[r]{\ \ {$0$}}}
\put(1405.0,250.0){\rule[-0.200pt]{4.818pt}{0.400pt}}
\put(350.0,500.0){\rule[-0.200pt]{4.818pt}{0.400pt}}
\put(325,500){\makebox(0,0)[r]{\ \ {$0.2$}}}
\put(1405.0,500.0){\rule[-0.200pt]{4.818pt}{0.400pt}}
\put(350.0,750.0){\rule[-0.200pt]{4.818pt}{0.400pt}}
\put(325,750){\makebox(0,0)[r]{\ \ {$0.4$}}}
\put(1405.0,750.0){\rule[-0.200pt]{4.818pt}{0.400pt}}
\put(350.0,1000.0){\rule[-0.200pt]{4.818pt}{0.400pt}}
\put(325,1000){\makebox(0,0)[r]{\ \ {$0.6$}}}
\put(1405.0,1000.0){\rule[-0.200pt]{4.818pt}{0.400pt}}
\put(350.0,1250.0){\rule[-0.200pt]{4.818pt}{0.400pt}}
\put(325,1250){\makebox(0,0)[r]{\ \ {$0.8$}}}
\put(1405.0,1250.0){\rule[-0.200pt]{4.818pt}{0.400pt}}
\put(350.0,1500.0){\rule[-0.200pt]{4.818pt}{0.400pt}}
\put(325,1500){\makebox(0,0)[r]{\ \ {$1$}}}
\put(1405.0,1500.0){\rule[-0.200pt]{4.818pt}{0.400pt}}
\put(350.0,1750.0){\rule[-0.200pt]{4.818pt}{0.400pt}}
\put(325,1750){\makebox(0,0)[r]{\ \ {$1.2$}}}
\put(1405.0,1750.0){\rule[-0.200pt]{4.818pt}{0.400pt}}
\put(350.0,250.0){\rule[-0.200pt]{0.400pt}{4.818pt}}
\put(350,200){\makebox(0,0){\ {$0$}}}
\put(350.0,1730.0){\rule[-0.200pt]{0.400pt}{4.818pt}}
\put(565.0,250.0){\rule[-0.200pt]{0.400pt}{4.818pt}}
\put(565,200){\makebox(0,0){\ {$0.05$}}}
\put(565.0,1730.0){\rule[-0.200pt]{0.400pt}{4.818pt}}
\put(780.0,250.0){\rule[-0.200pt]{0.400pt}{4.818pt}}
\put(780,200){\makebox(0,0){\ {$0.1$}}}
\put(780.0,1730.0){\rule[-0.200pt]{0.400pt}{4.818pt}}
\put(995.0,250.0){\rule[-0.200pt]{0.400pt}{4.818pt}}
\put(995,200){\makebox(0,0){\ {$0.15$}}}
\put(995.0,1730.0){\rule[-0.200pt]{0.400pt}{4.818pt}}
\put(1210.0,250.0){\rule[-0.200pt]{0.400pt}{4.818pt}}
\put(1210,200){\makebox(0,0){\ {$0.2$}}}
\put(1210.0,1730.0){\rule[-0.200pt]{0.400pt}{4.818pt}}
\put(1425.0,250.0){\rule[-0.200pt]{0.400pt}{4.818pt}}
\put(1425,200){\makebox(0,0){\ {$0.25$}}}
\put(1425.0,1730.0){\rule[-0.200pt]{0.400pt}{4.818pt}}
\put(350.0,250.0){\rule[-0.200pt]{258.967pt}{0.400pt}}
\put(1425.0,250.0){\rule[-0.200pt]{0.400pt}{361.350pt}}
\put(350.0,1750.0){\rule[-0.200pt]{258.967pt}{0.400pt}}
\put(100,1350){\makebox(0,0){\Large{$\frac{1}{T_c}\Bigl\{\frac{L_h}{N^2}\Bigr\}^\frac{1}{4}$}}}
\put(862,25){\makebox(0,0){\large{$1/N^2$}}}
\put(350.0,250.0){\rule[-0.200pt]{0.400pt}{361.350pt}}
\put(828.0,1029.0){\rule[-0.200pt]{0.400pt}{3.613pt}}
\put(818.0,1029.0){\rule[-0.200pt]{4.818pt}{0.400pt}}
\put(818.0,1044.0){\rule[-0.200pt]{4.818pt}{0.400pt}}
\put(619.0,1197.0){\rule[-0.200pt]{0.400pt}{2.650pt}}
\put(609.0,1197.0){\rule[-0.200pt]{4.818pt}{0.400pt}}
\put(609.0,1208.0){\rule[-0.200pt]{4.818pt}{0.400pt}}
\put(470.0,1243.0){\rule[-0.200pt]{0.400pt}{2.891pt}}
\put(460.0,1243.0){\rule[-0.200pt]{4.818pt}{0.400pt}}
\put(460.0,1255.0){\rule[-0.200pt]{4.818pt}{0.400pt}}
\put(417.0,1239.0){\rule[-0.200pt]{0.400pt}{2.891pt}}
\put(407.0,1239.0){\rule[-0.200pt]{4.818pt}{0.400pt}}
\put(828,1037){\circle*{12}}
\put(619,1202){\circle*{12}}
\put(470,1249){\circle*{12}}
\put(417,1245){\circle*{12}}
\put(407.0,1251.0){\rule[-0.200pt]{4.818pt}{0.400pt}}
\sbox{\plotpoint}{\rule[-0.500pt]{1.000pt}{1.000pt}}%
\put(350,1268){\usebox{\plotpoint}}
\put(350.00,1268.00){\usebox{\plotpoint}}
\put(370.24,1263.48){\usebox{\plotpoint}}
\put(390.42,1258.77){\usebox{\plotpoint}}
\put(410.55,1253.81){\usebox{\plotpoint}}
\put(430.76,1249.14){\usebox{\plotpoint}}
\put(450.90,1244.21){\usebox{\plotpoint}}
\put(471.09,1239.43){\usebox{\plotpoint}}
\put(491.33,1234.91){\usebox{\plotpoint}}
\put(511.54,1230.27){\usebox{\plotpoint}}
\put(531.59,1224.93){\usebox{\plotpoint}}
\put(551.77,1220.15){\usebox{\plotpoint}}
\put(572.01,1215.63){\usebox{\plotpoint}}
\put(592.25,1211.11){\usebox{\plotpoint}}
\put(612.29,1205.74){\usebox{\plotpoint}}
\put(632.47,1200.91){\usebox{\plotpoint}}
\put(652.70,1196.36){\usebox{\plotpoint}}
\put(672.93,1191.84){\usebox{\plotpoint}}
\put(693.06,1186.79){\usebox{\plotpoint}}
\put(713.25,1182.05){\usebox{\plotpoint}}
\put(733.45,1177.37){\usebox{\plotpoint}}
\put(753.62,1172.56){\usebox{\plotpoint}}
\put(773.80,1167.78){\usebox{\plotpoint}}
\put(794.02,1163.18){\usebox{\plotpoint}}
\put(814.07,1157.80){\usebox{\plotpoint}}
\put(834.30,1153.28){\usebox{\plotpoint}}
\put(854.49,1148.50){\usebox{\plotpoint}}
\put(874.73,1143.98){\usebox{\plotpoint}}
\put(894.79,1138.68){\usebox{\plotpoint}}
\put(914.99,1134.00){\usebox{\plotpoint}}
\put(935.16,1129.25){\usebox{\plotpoint}}
\put(955.39,1124.71){\usebox{\plotpoint}}
\put(975.54,1119.81){\usebox{\plotpoint}}
\put(995.75,1115.14){\usebox{\plotpoint}}
\put(1015.90,1110.29){\usebox{\plotpoint}}
\put(1036.06,1105.44){\usebox{\plotpoint}}
\put(1056.30,1100.92){\usebox{\plotpoint}}
\put(1076.50,1096.27){\usebox{\plotpoint}}
\put(1096.65,1091.43){\usebox{\plotpoint}}
\put(1116.72,1086.17){\usebox{\plotpoint}}
\put(1136.96,1081.65){\usebox{\plotpoint}}
\put(1157.18,1077.05){\usebox{\plotpoint}}
\put(1177.19,1071.60){\usebox{\plotpoint}}
\put(1197.40,1066.93){\usebox{\plotpoint}}
\put(1217.63,1062.37){\usebox{\plotpoint}}
\put(1237.81,1057.66){\usebox{\plotpoint}}
\put(1257.95,1052.74){\usebox{\plotpoint}}
\put(1278.15,1048.06){\usebox{\plotpoint}}
\put(1298.36,1043.39){\usebox{\plotpoint}}
\put(1318.46,1038.33){\usebox{\plotpoint}}
\put(1338.69,1033.81){\usebox{\plotpoint}}
\put(1358.91,1029.20){\usebox{\plotpoint}}
\put(1379.11,1024.52){\usebox{\plotpoint}}
\put(1399.12,1019.06){\usebox{\plotpoint}}
\put(1419.36,1014.54){\usebox{\plotpoint}}
\put(1425,1013){\usebox{\plotpoint}}
\end{picture}
\end	{center}
\vskip 0.15in
\caption{The latent heat calculated at $a=1/5T_c$ for various
SU($N$) groups.}
\label{fig_lht5N}
\end 	{figure}

Our best values of $L_h$, in the sense of possessing the
smallest statistical and systematic errors, are those
for $L_t = 5$. In Fig.\ref{fig_lht5N} we plot these
latent heats against $1/N^2$ and show a fit of the form
\begin{equation}
\frac{1}{T_c}\Bigl\{\frac{L_h}{N^2}\Bigr\}^\frac{1}{4}
=
c_0 + \frac{c_1}{N^2}.
\label{eqn_lhlt5N}
\end{equation}
This form is expected because the average energy of the
gluon plasma should be $O(N^2)$ and the leading large-$N$
correction should be $O(1/N^2)$. We see that $L_h/N^2$
does indeed appear to go to a finite limit and that the
SU(3) value is well below the fit, in line with the
usual belief that it is weakly first order. We note
that $L_h/N^2$ increases to an $N=\infty$ limit that is
$O(1)$ in units of $T_c$. That is to say, 
the transition is robustly first-order in that limit. 

\begin	{figure}[p]
\begin	{center}
\leavevmode
\input	{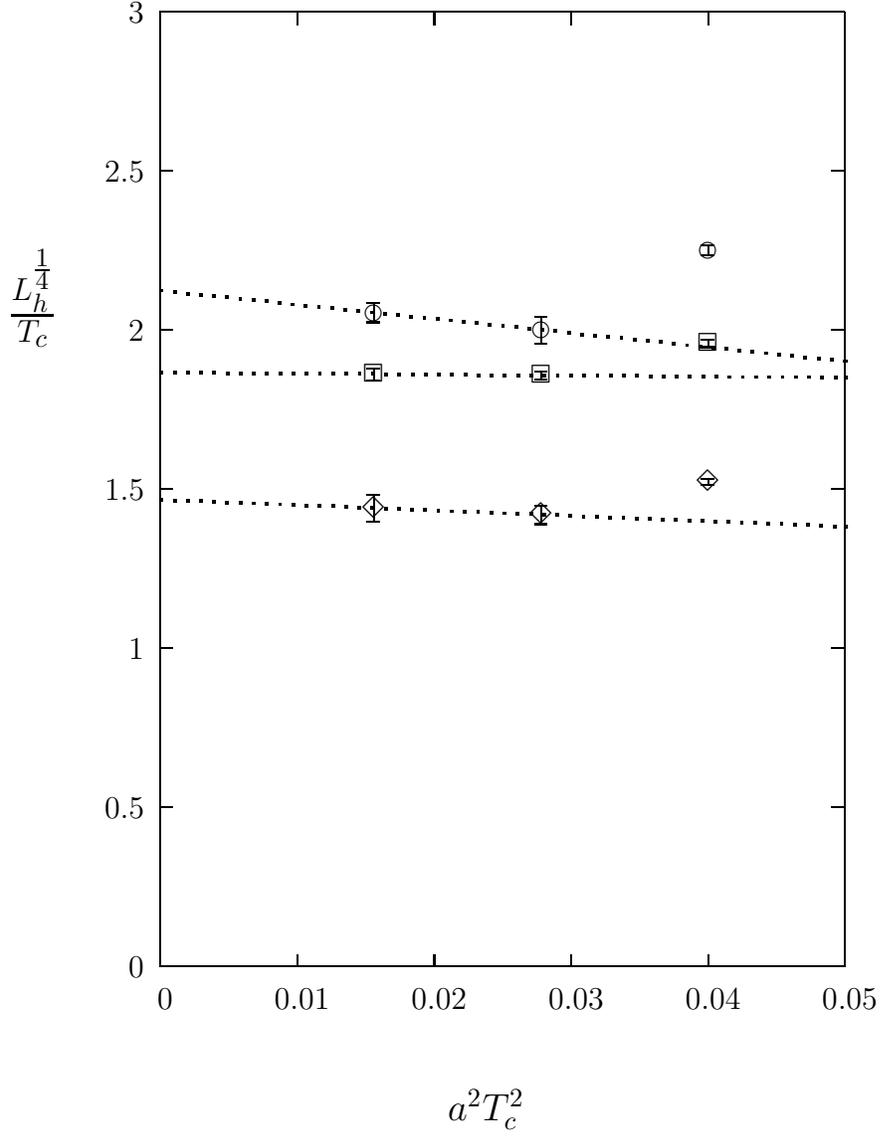}
\end	{center}
\vskip 0.15in
\caption{The deconfining latent heat in SU(4), $\diamond$, 
SU(6), $\Box$ and SU(8), $\circ$ gauge theories at various $a$,
with extrapolations to the continuum limit using a $O(a^2)$ 
correction.}
\label{fig_lhsuN}
\end 	{figure}

In Fig.\ref{fig_lhsuN} we plot the dimensionless ratio 
$L^\frac{1}{4}_h/T_c$ against $a^2T^2_c$. For the plaquette
action the leading lattice correction is known to be $O(a^2)$ 
so we expect to be able to extrapolate to the continuum 
limit using
\begin{equation}
\frac{L^\frac{1}{4}_h(a)}{T_c(a)}
=
\frac{L^\frac{1}{4}_h(a=0)}{T_c(a=0)} + c a^2T^2_c(a).
\label{eqn_lhcont}
\end{equation}
Unfortunately it is clear that such straight-line fits will be 
very poor and indeed they lead to $\chi^2/n_{df}\simeq 3,9,11$
for $N=4,6,8$ respectively, which is unacceptable. All
the plots show the similar feature that the $L_t=5$ value
is much too high to lie on the fit. Although in some cases
$L_s/L_t$ is smaller for $L_t=6$ than for $L_t=8$, in other
cases it is larger, so we do not attribute the problem to
large finite volume corrections afflicting all the $L_t=6$ 
calculations. We therefore take this to be a real
effect rather than a manifestation of some unexpectedly 
large systematic error. That means that we must extrapolate
using just the two values, at $L_t=6$ and  at $L_t=8$. This
means we have no evidence that including just the $O(a^2)$
correction is sufficient. Nonetheless, it is plausible to
assume that it is since for other quantities, such as glueball masses,
the deconfining temperature, etc., we know that it is in this
range of $a$. Making this assumption we obtain the continuum
values listed in Table~\ref{table_lhcont}, using the 
extrapolations shown in Fig.\ref{fig_lhsuN}. 

We can now take these continuum values of $L^\frac{1}{4}_h/T_c$
and see whether they determine the $N=\infty$ value. We find
that they are indeed well fit by the expected functional form
\begin{equation}
\frac{L^\frac{1}{4}_h}{N^\frac{1}{2}T_c}
=
0.766(40) -\frac{0.34(1.60)}{N^2} 
\quad : \quad \chi^2/n_{df} = 0.3,
\label{eqn_lhcontN}
\end{equation}
although the errors are rather large because of the fact that
we could not use our most accurate data, which was at $L_t=5$.

As we shall discuss in Section~\ref{section_bulk}, the value
of $\beta$ corresponding to $a=1/5T_c$ is very close to the
bulk transition. This is a rapid crossover for $N=4$, and becomes
a strong first order transition for $N\geq 6$. Indeed, for $N=8$,
$a=1/5T_c$ is very close to the lower $\beta$ edge of the
bulk hysteresis curve. At this first order bulk transition the 
jump in the plaquette, $\Delta\langle u_p \rangle$, is about 50 times 
larger than it is for the deconfining transition on the $L_t=5$ 
lattice system. It is therefore plausible that the whole
effective potential is becoming deformed at this value of
$a$ and that this is causing the anomalously large value for
the latent heat at $L_t=5$. The fact that the effect is larger
for $N=6,8$ than for $N=4$, where the transition is only a crossover,
supports this hypothesis. This transition can be naturally regarded 
as a finite $N$ manifestation of the D=1+1 Gross-Witten transition
\cite{gross-witten,mtN04}
and the plaquette is the natural order parameter. It is therefore
perhaps plausible that it should be the average plaquette
that is most sensitive to the nearby presence of the bulk phase.

\section{Masses}
\label{section_masses}

At $T=T_c$ we can perform calculations separately
in the confined and deconfined phases by working
with volumes large enough for there to be no sign
of any tunnelling (or attempted tunnelling). We have
performed such calculations on $32^3 5$, $16^3 5$ and
$12^3 5$ lattices for SU(4), SU(6) and SU(8) respectively.
In SU(3) we use a $64^3 5$ lattice, on which there is
tunnelling, but it is very rare and we can 
work with long subsequences 
of field configurations that are in a single phase.
(The reason that we are able to work on smaller spatial 
volumes, $V=l_s^3$, as 
$N$ increases is because in the tunnelling probability, 
$\propto \exp\{-2\sigma_{cd}l_s^2/T\}$,
the interface tension, $\sigma_{cd}$, increases with $N$
-- as discussed in Section~\ref{section_interface}.)
Although the calculations in this Section are performed at
a fixed value of the lattice spacing, this value is small
enough that the results should be representative of what one
would obtain in the continuum limit.

\subsection{Mass gap at ${\bf T=T_c}$}
\label{subsection_mgatTc}

We calculate the lightest mass that couples to the
Polyakov loop which winds around the time-torus. We do
this for different $k$-loops (see 
Section~\ref{subsection_deconf_mult}) at each value of $N$ and
we present the masses for the confined and deconfined phases
in Tables~\ref{table_mtCTc} and \ref{table_mtDTc} respectively. 
(In the deconfined phase we use the vacuum subtracted loop.)

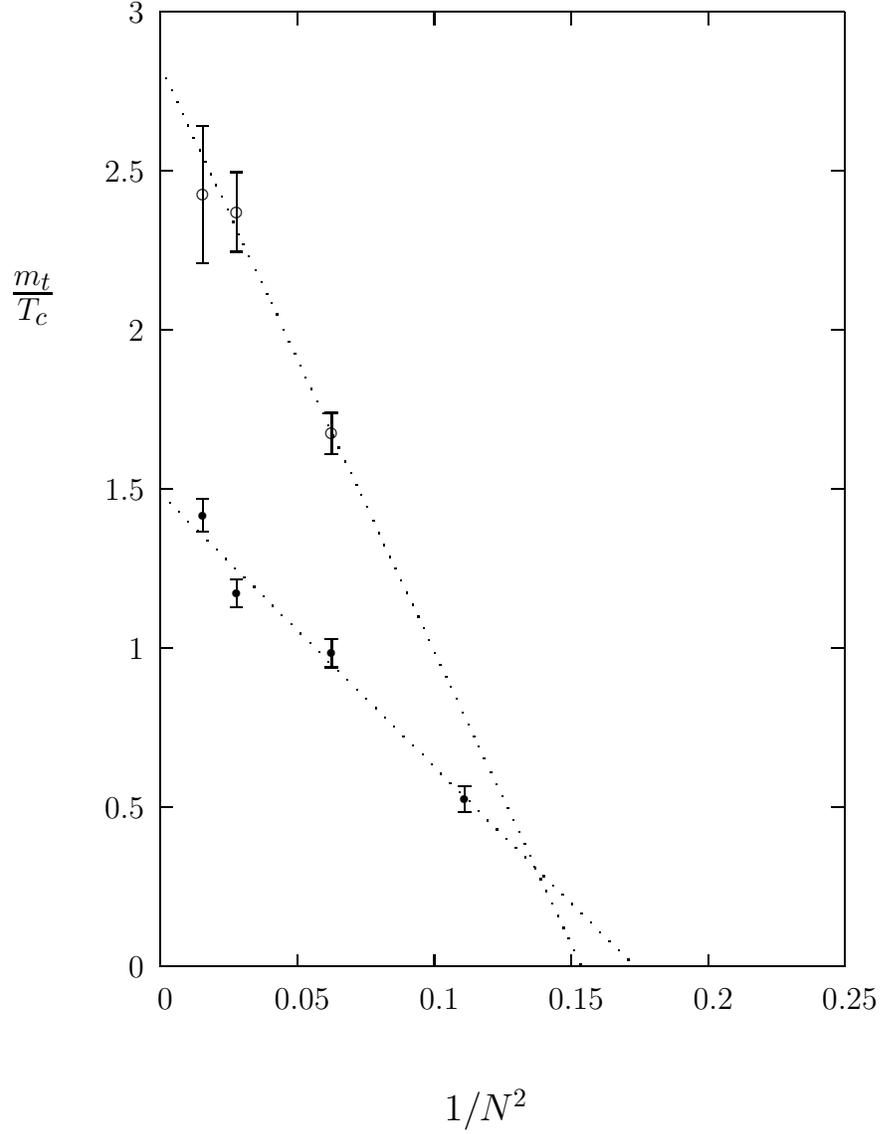
\begin	{figure}[p]
\begin	{center}
\leavevmode
\setlength{\unitlength}{0.240900pt}
\ifx\plotpoint\undefined\newsavebox{\plotpoint}\fi
\sbox{\plotpoint}{\rule[-0.200pt]{0.400pt}{0.400pt}}%
\begin{picture}(1500,1800)(0,0)
\font\gnuplot=cmr10 at 12pt
\gnuplot
\sbox{\plotpoint}{\rule[-0.200pt]{0.400pt}{0.400pt}}%
\put(350.0,250.0){\rule[-0.200pt]{4.818pt}{0.400pt}}
\put(325,250){\makebox(0,0)[r]{\ \ {$0$}}}
\put(1405.0,250.0){\rule[-0.200pt]{4.818pt}{0.400pt}}
\put(350.0,500.0){\rule[-0.200pt]{4.818pt}{0.400pt}}
\put(325,500){\makebox(0,0)[r]{\ \ {$0.5$}}}
\put(1405.0,500.0){\rule[-0.200pt]{4.818pt}{0.400pt}}
\put(350.0,750.0){\rule[-0.200pt]{4.818pt}{0.400pt}}
\put(325,750){\makebox(0,0)[r]{\ \ {$1$}}}
\put(1405.0,750.0){\rule[-0.200pt]{4.818pt}{0.400pt}}
\put(350.0,1000.0){\rule[-0.200pt]{4.818pt}{0.400pt}}
\put(325,1000){\makebox(0,0)[r]{\ \ {$1.5$}}}
\put(1405.0,1000.0){\rule[-0.200pt]{4.818pt}{0.400pt}}
\put(350.0,1250.0){\rule[-0.200pt]{4.818pt}{0.400pt}}
\put(325,1250){\makebox(0,0)[r]{\ \ {$2$}}}
\put(1405.0,1250.0){\rule[-0.200pt]{4.818pt}{0.400pt}}
\put(350.0,1500.0){\rule[-0.200pt]{4.818pt}{0.400pt}}
\put(325,1500){\makebox(0,0)[r]{\ \ {$2.5$}}}
\put(1405.0,1500.0){\rule[-0.200pt]{4.818pt}{0.400pt}}
\put(350.0,1750.0){\rule[-0.200pt]{4.818pt}{0.400pt}}
\put(325,1750){\makebox(0,0)[r]{\ \ {$3$}}}
\put(1405.0,1750.0){\rule[-0.200pt]{4.818pt}{0.400pt}}
\put(350.0,250.0){\rule[-0.200pt]{0.400pt}{4.818pt}}
\put(350,200){\makebox(0,0){\ {$0$}}}
\put(350.0,1730.0){\rule[-0.200pt]{0.400pt}{4.818pt}}
\put(565.0,250.0){\rule[-0.200pt]{0.400pt}{4.818pt}}
\put(565,200){\makebox(0,0){\ {$0.05$}}}
\put(565.0,1730.0){\rule[-0.200pt]{0.400pt}{4.818pt}}
\put(780.0,250.0){\rule[-0.200pt]{0.400pt}{4.818pt}}
\put(780,200){\makebox(0,0){\ {$0.1$}}}
\put(780.0,1730.0){\rule[-0.200pt]{0.400pt}{4.818pt}}
\put(995.0,250.0){\rule[-0.200pt]{0.400pt}{4.818pt}}
\put(995,200){\makebox(0,0){\ {$0.15$}}}
\put(995.0,1730.0){\rule[-0.200pt]{0.400pt}{4.818pt}}
\put(1210.0,250.0){\rule[-0.200pt]{0.400pt}{4.818pt}}
\put(1210,200){\makebox(0,0){\ {$0.2$}}}
\put(1210.0,1730.0){\rule[-0.200pt]{0.400pt}{4.818pt}}
\put(1425.0,250.0){\rule[-0.200pt]{0.400pt}{4.818pt}}
\put(1425,200){\makebox(0,0){\ {$0.25$}}}
\put(1425.0,1730.0){\rule[-0.200pt]{0.400pt}{4.818pt}}
\put(350.0,250.0){\rule[-0.200pt]{258.967pt}{0.400pt}}
\put(1425.0,250.0){\rule[-0.200pt]{0.400pt}{361.350pt}}
\put(350.0,1750.0){\rule[-0.200pt]{258.967pt}{0.400pt}}
\put(150,1300){\makebox(0,0){\Large{${{m_t}\over{T_c}}$}}}
\put(862,25){\makebox(0,0){\large{$1/N^2$}}}
\put(350.0,250.0){\rule[-0.200pt]{0.400pt}{361.350pt}}
\put(828.0,493.0){\rule[-0.200pt]{0.400pt}{9.636pt}}
\put(818.0,493.0){\rule[-0.200pt]{4.818pt}{0.400pt}}
\put(818.0,533.0){\rule[-0.200pt]{4.818pt}{0.400pt}}
\put(619.0,720.0){\rule[-0.200pt]{0.400pt}{10.600pt}}
\put(609.0,720.0){\rule[-0.200pt]{4.818pt}{0.400pt}}
\put(609.0,764.0){\rule[-0.200pt]{4.818pt}{0.400pt}}
\put(470.0,814.0){\rule[-0.200pt]{0.400pt}{10.600pt}}
\put(460.0,814.0){\rule[-0.200pt]{4.818pt}{0.400pt}}
\put(460.0,858.0){\rule[-0.200pt]{4.818pt}{0.400pt}}
\put(417.0,933.0){\rule[-0.200pt]{0.400pt}{12.527pt}}
\put(407.0,933.0){\rule[-0.200pt]{4.818pt}{0.400pt}}
\put(828,513){\circle*{12}}
\put(619,742){\circle*{12}}
\put(470,836){\circle*{12}}
\put(417,959){\circle*{12}}
\put(407.0,985.0){\rule[-0.200pt]{4.818pt}{0.400pt}}
\put(619.0,1055.0){\rule[-0.200pt]{0.400pt}{15.658pt}}
\put(609.0,1055.0){\rule[-0.200pt]{4.818pt}{0.400pt}}
\put(609.0,1120.0){\rule[-0.200pt]{4.818pt}{0.400pt}}
\put(470.0,1373.0){\rule[-0.200pt]{0.400pt}{30.112pt}}
\put(460.0,1373.0){\rule[-0.200pt]{4.818pt}{0.400pt}}
\put(460.0,1498.0){\rule[-0.200pt]{4.818pt}{0.400pt}}
\put(417.0,1355.0){\rule[-0.200pt]{0.400pt}{51.793pt}}
\put(407.0,1355.0){\rule[-0.200pt]{4.818pt}{0.400pt}}
\put(619,1088){\circle{18}}
\put(470,1435){\circle{18}}
\put(417,1463){\circle{18}}
\put(407.0,1570.0){\rule[-0.200pt]{4.818pt}{0.400pt}}
\put(350,993){\usebox{\plotpoint}}
\put(350.00,993.00){\usebox{\plotpoint}}
\put(364.68,978.32){\usebox{\plotpoint}}
\put(379.35,963.65){\usebox{\plotpoint}}
\put(393.52,948.48){\usebox{\plotpoint}}
\put(408.19,933.81){\usebox{\plotpoint}}
\put(423.24,919.51){\usebox{\plotpoint}}
\put(438.03,904.97){\usebox{\plotpoint}}
\put(452.71,890.29){\usebox{\plotpoint}}
\put(466.98,875.22){\usebox{\plotpoint}}
\put(481.62,860.52){\usebox{\plotpoint}}
\put(496.72,846.28){\usebox{\plotpoint}}
\put(511.39,831.61){\usebox{\plotpoint}}
\put(526.07,816.93){\usebox{\plotpoint}}
\put(540.47,801.99){\usebox{\plotpoint}}
\put(555.37,787.57){\usebox{\plotpoint}}
\put(570.07,772.93){\usebox{\plotpoint}}
\put(584.75,758.25){\usebox{\plotpoint}}
\put(599.43,743.57){\usebox{\plotpoint}}
\put(614.10,728.90){\usebox{\plotpoint}}
\put(628.78,714.22){\usebox{\plotpoint}}
\put(643.46,699.54){\usebox{\plotpoint}}
\put(658.13,684.87){\usebox{\plotpoint}}
\put(672.81,670.19){\usebox{\plotpoint}}
\put(687.93,655.98){\usebox{\plotpoint}}
\put(702.14,640.86){\usebox{\plotpoint}}
\put(716.81,626.19){\usebox{\plotpoint}}
\put(731.49,611.51){\usebox{\plotpoint}}
\put(746.41,597.09){\usebox{\plotpoint}}
\put(761.33,582.67){\usebox{\plotpoint}}
\put(775.50,567.50){\usebox{\plotpoint}}
\put(790.17,552.83){\usebox{\plotpoint}}
\put(804.85,538.15){\usebox{\plotpoint}}
\put(820.01,523.99){\usebox{\plotpoint}}
\put(834.69,509.31){\usebox{\plotpoint}}
\put(848.86,494.15){\usebox{\plotpoint}}
\put(863.53,479.47){\usebox{\plotpoint}}
\put(878.54,465.15){\usebox{\plotpoint}}
\put(893.37,450.63){\usebox{\plotpoint}}
\put(908.05,435.95){\usebox{\plotpoint}}
\put(922.72,421.28){\usebox{\plotpoint}}
\put(936.93,406.16){\usebox{\plotpoint}}
\put(952.05,391.95){\usebox{\plotpoint}}
\put(966.73,377.27){\usebox{\plotpoint}}
\put(981.40,362.60){\usebox{\plotpoint}}
\put(996.08,347.92){\usebox{\plotpoint}}
\put(1010.76,333.24){\usebox{\plotpoint}}
\put(1025.43,318.57){\usebox{\plotpoint}}
\put(1040.11,303.89){\usebox{\plotpoint}}
\put(1054.79,289.21){\usebox{\plotpoint}}
\put(1069.58,274.66){\usebox{\plotpoint}}
\put(1084.30,260.07){\usebox{\plotpoint}}
\put(1094,250){\usebox{\plotpoint}}
\put(350,1664){\usebox{\plotpoint}}
\multiput(350,1664)(8.955,-18.724){2}{\usebox{\plotpoint}}
\put(367.91,1626.55){\usebox{\plotpoint}}
\put(376.70,1607.75){\usebox{\plotpoint}}
\put(385.25,1588.84){\usebox{\plotpoint}}
\multiput(393,1571)(8.955,-18.724){2}{\usebox{\plotpoint}}
\put(411.47,1532.37){\usebox{\plotpoint}}
\put(420.24,1513.56){\usebox{\plotpoint}}
\put(428.99,1494.74){\usebox{\plotpoint}}
\multiput(437,1478)(8.955,-18.724){2}{\usebox{\plotpoint}}
\put(455.86,1438.57){\usebox{\plotpoint}}
\put(464.37,1419.64){\usebox{\plotpoint}}
\put(472.81,1400.68){\usebox{\plotpoint}}
\multiput(480,1385)(8.955,-18.724){2}{\usebox{\plotpoint}}
\put(499.42,1344.39){\usebox{\plotpoint}}
\put(508.38,1325.67){\usebox{\plotpoint}}
\put(517.33,1306.94){\usebox{\plotpoint}}
\multiput(524,1293)(8.648,-18.868){2}{\usebox{\plotpoint}}
\put(543.14,1250.28){\usebox{\plotpoint}}
\put(551.94,1231.49){\usebox{\plotpoint}}
\put(560.90,1212.76){\usebox{\plotpoint}}
\put(569.75,1193.99){\usebox{\plotpoint}}
\multiput(578,1176)(8.955,-18.724){2}{\usebox{\plotpoint}}
\put(596.33,1137.68){\usebox{\plotpoint}}
\put(605.28,1118.96){\usebox{\plotpoint}}
\put(613.99,1100.12){\usebox{\plotpoint}}
\multiput(621,1084)(8.648,-18.868){2}{\usebox{\plotpoint}}
\put(639.89,1043.50){\usebox{\plotpoint}}
\put(648.84,1024.78){\usebox{\plotpoint}}
\put(657.80,1006.06){\usebox{\plotpoint}}
\multiput(665,991)(8.648,-18.868){2}{\usebox{\plotpoint}}
\put(684.27,949.70){\usebox{\plotpoint}}
\put(692.76,930.76){\usebox{\plotpoint}}
\put(701.36,911.88){\usebox{\plotpoint}}
\put(710.32,893.15){\usebox{\plotpoint}}
\multiput(719,875)(8.648,-18.868){2}{\usebox{\plotpoint}}
\put(736.79,836.80){\usebox{\plotpoint}}
\put(745.75,818.07){\usebox{\plotpoint}}
\put(754.70,799.35){\usebox{\plotpoint}}
\multiput(763,782)(7.983,-19.159){2}{\usebox{\plotpoint}}
\put(780.35,742.63){\usebox{\plotpoint}}
\put(789.31,723.91){\usebox{\plotpoint}}
\put(798.26,705.18){\usebox{\plotpoint}}
\multiput(806,689)(8.955,-18.724){2}{\usebox{\plotpoint}}
\put(824.85,648.88){\usebox{\plotpoint}}
\put(833.69,630.10){\usebox{\plotpoint}}
\put(842.37,611.25){\usebox{\plotpoint}}
\multiput(849,596)(8.955,-18.724){2}{\usebox{\plotpoint}}
\put(868.69,554.83){\usebox{\plotpoint}}
\put(877.42,536.00){\usebox{\plotpoint}}
\put(886.21,517.20){\usebox{\plotpoint}}
\put(895.16,498.47){\usebox{\plotpoint}}
\multiput(904,480)(8.955,-18.724){2}{\usebox{\plotpoint}}
\put(921.79,442.19){\usebox{\plotpoint}}
\put(930.25,423.24){\usebox{\plotpoint}}
\put(938.73,404.30){\usebox{\plotpoint}}
\multiput(947,387)(8.955,-18.724){2}{\usebox{\plotpoint}}
\put(965.59,348.12){\usebox{\plotpoint}}
\put(974.36,329.31){\usebox{\plotpoint}}
\put(983.11,310.49){\usebox{\plotpoint}}
\multiput(991,294)(8.955,-18.724){2}{\usebox{\plotpoint}}
\put(1009.28,254.00){\usebox{\plotpoint}}
\put(1011,250){\usebox{\plotpoint}}
\end{picture}
\end	{center}
\vskip 0.15in
\caption{The masses of the lightest $k=1$ ($\bullet$) and
$k=2$ ($\circ$) flux loops that wind around the time-torus,
in the confining phase at $T=T_c$, $a=1/5T_c$ for various
SU($N$) gauge theories.}
\label{fig_mtCN}
\end 	{figure}

We observe that in the confined phase the higher $k$ loops
are more massive, just as they are at $T=0$. 
If we take the mass of the SU(8) loop
in the confined phase, and apply the leading string correction
({\it a priori} inadequate for such a short string) so as to extract 
a string tension ${\sigma} = m/l_t + \pi/3l^2_t$, we obtain,
for SU(8), a value, $a\surd{\sigma} \simeq 0.314(3)$ that is
quite close to the $T=0$ value $a\surd{\sigma} \simeq 0.346$
\cite{oxglue04}.
This suggests that this first order transition is quite strong. 
If we express the loop mass in the natural units of $T_c$ and 
plot it, in Fig.\ref{fig_mtCN}, against $1/N^2$ we see that this
mass gap tends to a finite $N=\infty$ limit, albeit with
a large $O(1/N^2)$ correction. Indeed
we see that for SU(3) the correlation length, $\xi_t=1/m_t$,
is quite large indicating a transition that is quite weakly 
first order.

The $k=2$ mass gap in  Fig.\ref{fig_mtCN} also appears to
tend to a finite  $N=\infty$ limit. Although we have used
a leading $O(1/N^2)$ correction, this is not guaranteed
to be correct for $k > 1$ strings. The alternative is
an $O(1/N)$ correction which arises in, for example,
the Casimir Scaling conjecture
\cite{CS}.
It is interesting to note that the string tension ratio
\begin{equation}
\frac{\sigma_{eff}(k=2)}{\sigma_{eff}(k=1)}
=
\frac{m_t(k=2)}{m_t(k=1)}
\stackrel{N\to\infty}{\longrightarrow}
\left\{ \begin{array}{ll}
1.91 \pm 0.13 & \ \ \ {\delta\sigma_{k=2} = O(1/N^2)} \\
2.36 \pm 0.21 & \ \ \ {\delta\sigma_{k=2} = O(1/N)}
\end{array}
\right. 
\label{eqn_ratkTc}
\end{equation}
is compatible with the value of 2 that one expects from 
factorisation in the $N=\infty$ limit (even at finite $T$)
for both forms of the correction.

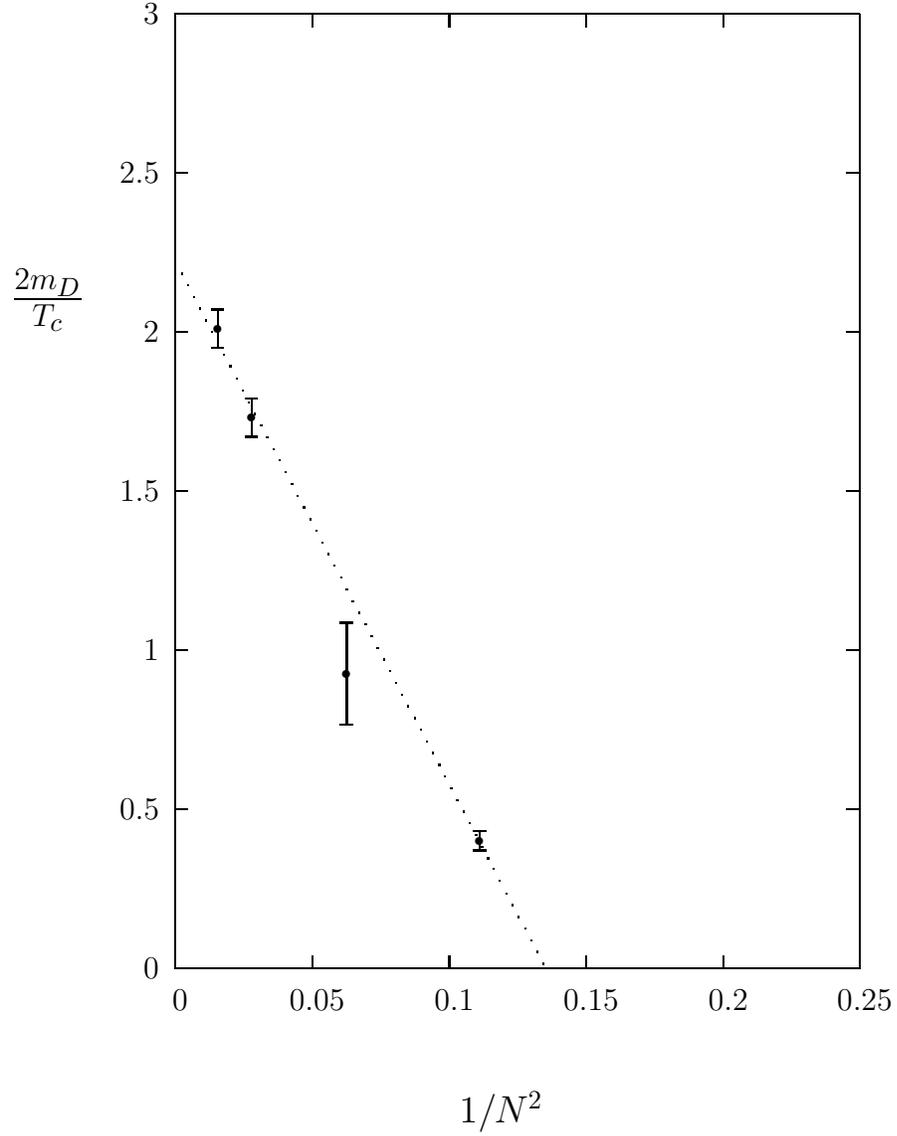
\begin	{figure}[p]
\begin	{center}
\leavevmode
\setlength{\unitlength}{0.240900pt}
\ifx\plotpoint\undefined\newsavebox{\plotpoint}\fi
\sbox{\plotpoint}{\rule[-0.200pt]{0.400pt}{0.400pt}}%
\begin{picture}(1500,1800)(0,0)
\font\gnuplot=cmr10 at 12pt
\gnuplot
\sbox{\plotpoint}{\rule[-0.200pt]{0.400pt}{0.400pt}}%
\put(350.0,250.0){\rule[-0.200pt]{4.818pt}{0.400pt}}
\put(325,250){\makebox(0,0)[r]{\ \ {$0$}}}
\put(1405.0,250.0){\rule[-0.200pt]{4.818pt}{0.400pt}}
\put(350.0,500.0){\rule[-0.200pt]{4.818pt}{0.400pt}}
\put(325,500){\makebox(0,0)[r]{\ \ {$0.5$}}}
\put(1405.0,500.0){\rule[-0.200pt]{4.818pt}{0.400pt}}
\put(350.0,750.0){\rule[-0.200pt]{4.818pt}{0.400pt}}
\put(325,750){\makebox(0,0)[r]{\ \ {$1$}}}
\put(1405.0,750.0){\rule[-0.200pt]{4.818pt}{0.400pt}}
\put(350.0,1000.0){\rule[-0.200pt]{4.818pt}{0.400pt}}
\put(325,1000){\makebox(0,0)[r]{\ \ {$1.5$}}}
\put(1405.0,1000.0){\rule[-0.200pt]{4.818pt}{0.400pt}}
\put(350.0,1250.0){\rule[-0.200pt]{4.818pt}{0.400pt}}
\put(325,1250){\makebox(0,0)[r]{\ \ {$2$}}}
\put(1405.0,1250.0){\rule[-0.200pt]{4.818pt}{0.400pt}}
\put(350.0,1500.0){\rule[-0.200pt]{4.818pt}{0.400pt}}
\put(325,1500){\makebox(0,0)[r]{\ \ {$2.5$}}}
\put(1405.0,1500.0){\rule[-0.200pt]{4.818pt}{0.400pt}}
\put(350.0,1750.0){\rule[-0.200pt]{4.818pt}{0.400pt}}
\put(325,1750){\makebox(0,0)[r]{\ \ {$3$}}}
\put(1405.0,1750.0){\rule[-0.200pt]{4.818pt}{0.400pt}}
\put(350.0,250.0){\rule[-0.200pt]{0.400pt}{4.818pt}}
\put(350,200){\makebox(0,0){\ {$0$}}}
\put(350.0,1730.0){\rule[-0.200pt]{0.400pt}{4.818pt}}
\put(565.0,250.0){\rule[-0.200pt]{0.400pt}{4.818pt}}
\put(565,200){\makebox(0,0){\ {$0.05$}}}
\put(565.0,1730.0){\rule[-0.200pt]{0.400pt}{4.818pt}}
\put(780.0,250.0){\rule[-0.200pt]{0.400pt}{4.818pt}}
\put(780,200){\makebox(0,0){\ {$0.1$}}}
\put(780.0,1730.0){\rule[-0.200pt]{0.400pt}{4.818pt}}
\put(995.0,250.0){\rule[-0.200pt]{0.400pt}{4.818pt}}
\put(995,200){\makebox(0,0){\ {$0.15$}}}
\put(995.0,1730.0){\rule[-0.200pt]{0.400pt}{4.818pt}}
\put(1210.0,250.0){\rule[-0.200pt]{0.400pt}{4.818pt}}
\put(1210,200){\makebox(0,0){\ {$0.2$}}}
\put(1210.0,1730.0){\rule[-0.200pt]{0.400pt}{4.818pt}}
\put(1425.0,250.0){\rule[-0.200pt]{0.400pt}{4.818pt}}
\put(1425,200){\makebox(0,0){\ {$0.25$}}}
\put(1425.0,1730.0){\rule[-0.200pt]{0.400pt}{4.818pt}}
\put(350.0,250.0){\rule[-0.200pt]{258.967pt}{0.400pt}}
\put(1425.0,250.0){\rule[-0.200pt]{0.400pt}{361.350pt}}
\put(350.0,1750.0){\rule[-0.200pt]{258.967pt}{0.400pt}}
\put(150,1300){\makebox(0,0){\Large{${{2m_D}\over{T_c}}$}}}
\put(862,25){\makebox(0,0){\large{$1/N^2$}}}
\put(350.0,250.0){\rule[-0.200pt]{0.400pt}{361.350pt}}
\put(828.0,435.0){\rule[-0.200pt]{0.400pt}{7.227pt}}
\put(818.0,435.0){\rule[-0.200pt]{4.818pt}{0.400pt}}
\put(818.0,465.0){\rule[-0.200pt]{4.818pt}{0.400pt}}
\put(619.0,633.0){\rule[-0.200pt]{0.400pt}{38.544pt}}
\put(609.0,633.0){\rule[-0.200pt]{4.818pt}{0.400pt}}
\put(609.0,793.0){\rule[-0.200pt]{4.818pt}{0.400pt}}
\put(470.0,1085.0){\rule[-0.200pt]{0.400pt}{14.454pt}}
\put(460.0,1085.0){\rule[-0.200pt]{4.818pt}{0.400pt}}
\put(460.0,1145.0){\rule[-0.200pt]{4.818pt}{0.400pt}}
\put(417.0,1225.0){\rule[-0.200pt]{0.400pt}{14.454pt}}
\put(407.0,1225.0){\rule[-0.200pt]{4.818pt}{0.400pt}}
\put(828,450){\circle*{12}}
\put(619,713){\circle*{12}}
\put(470,1115){\circle*{12}}
\put(417,1255){\circle*{12}}
\put(407.0,1285.0){\rule[-0.200pt]{4.818pt}{0.400pt}}
\put(350,1360){\usebox{\plotpoint}}
\multiput(350,1360)(9.631,-18.386){2}{\usebox{\plotpoint}}
\put(369.26,1323.23){\usebox{\plotpoint}}
\put(378.89,1304.84){\usebox{\plotpoint}}
\put(388.32,1286.35){\usebox{\plotpoint}}
\put(397.78,1267.88){\usebox{\plotpoint}}
\put(407.41,1249.49){\usebox{\plotpoint}}
\put(417.04,1231.11){\usebox{\plotpoint}}
\multiput(426,1214)(10.002,-18.186){2}{\usebox{\plotpoint}}
\put(446.34,1176.17){\usebox{\plotpoint}}
\put(455.97,1157.78){\usebox{\plotpoint}}
\put(465.12,1139.15){\usebox{\plotpoint}}
\put(474.65,1120.73){\usebox{\plotpoint}}
\put(484.48,1102.45){\usebox{\plotpoint}}
\put(494.11,1084.06){\usebox{\plotpoint}}
\put(503.74,1065.68){\usebox{\plotpoint}}
\multiput(513,1048)(10.002,-18.186){2}{\usebox{\plotpoint}}
\put(533.04,1010.74){\usebox{\plotpoint}}
\put(542.11,992.07){\usebox{\plotpoint}}
\put(551.76,973.71){\usebox{\plotpoint}}
\put(561.55,955.40){\usebox{\plotpoint}}
\put(571.18,937.02){\usebox{\plotpoint}}
\put(580.81,918.63){\usebox{\plotpoint}}
\put(590.50,900.28){\usebox{\plotpoint}}
\multiput(600,883)(9.631,-18.386){2}{\usebox{\plotpoint}}
\put(619.10,844.99){\usebox{\plotpoint}}
\put(628.58,826.53){\usebox{\plotpoint}}
\put(638.45,808.27){\usebox{\plotpoint}}
\put(648.25,789.97){\usebox{\plotpoint}}
\put(657.88,771.59){\usebox{\plotpoint}}
\put(667.51,753.20){\usebox{\plotpoint}}
\put(677.19,734.84){\usebox{\plotpoint}}
\multiput(687,717)(8.923,-18.739){2}{\usebox{\plotpoint}}
\put(705.65,679.48){\usebox{\plotpoint}}
\put(715.56,661.25){\usebox{\plotpoint}}
\put(725.32,642.93){\usebox{\plotpoint}}
\put(734.95,624.54){\usebox{\plotpoint}}
\put(744.58,606.16){\usebox{\plotpoint}}
\put(754.30,587.82){\usebox{\plotpoint}}
\put(764.16,569.56){\usebox{\plotpoint}}
\multiput(773,551)(9.631,-18.386){2}{\usebox{\plotpoint}}
\put(792.35,514.05){\usebox{\plotpoint}}
\put(802.25,495.81){\usebox{\plotpoint}}
\put(812.02,477.50){\usebox{\plotpoint}}
\put(821.65,459.11){\usebox{\plotpoint}}
\put(831.28,440.73){\usebox{\plotpoint}}
\put(840.85,422.31){\usebox{\plotpoint}}
\multiput(849,406)(9.631,-18.386){2}{\usebox{\plotpoint}}
\put(869.43,366.99){\usebox{\plotpoint}}
\put(879.37,348.77){\usebox{\plotpoint}}
\put(889.10,330.44){\usebox{\plotpoint}}
\put(898.73,312.06){\usebox{\plotpoint}}
\put(908.36,293.67){\usebox{\plotpoint}}
\put(918.11,275.35){\usebox{\plotpoint}}
\put(927.81,257.01){\usebox{\plotpoint}}
\put(931,250){\usebox{\plotpoint}}
\end{picture}
\end	{center}
\vskip 0.15in
\caption{The lightest mass that couples to the vacuum subtracted 
Wilson line winding  around the time-torus,
in the deconfined phase at $T=T_c$ and $a=1/5T_c$ for various
SU($N$) gauge theories.}
\label{fig_mtDN}
\end 	{figure}

In the deconfined phase all the $k$-loops couple to the same
object (the centre symmetry is spontaneously broken) and,
as we see in Table~\ref{table_mtDTc}, their masses are the same.
(It is interesting to note that the operator with the 
largest $k$ appears to have the best overlap.)  
Writing this mass as twice the Debye mass, we plot its value against 
$1/N^2$ in Fig.\ref{fig_mtDN}. Once again we see a finite
large-$N$ limit with a large finite-$N$ correction that leads to
a small SU(3) mass, confirming its weakly first order character.

It is of course quite possible  that some of our calculated masses 
suffer significant finite volume corrections. To partially address
this question we show in Table~\ref{table_mtTcN4} the masses one
gets in SU(4) on smaller lattices at the same value of $a$. The
smaller the lattice the greater the tunnelling and the more
difficult it is to distinguish the two phases, hence the increasing
errors. To partially control this we also show some calculations
performed at slightly different values of $\beta$ where there
is very little tunnelling. The conclusion of this comparison
is that the confined $k=1$ mass has at most modest corrections but
that they may be more substantial for the $k=2$ mass. In the deconfined
phase, any finite $V$ correction to the $L_s\geq 20$ calculations
look small. Since finite $V$ corrections are likely to be
functions of the lattice size in units of the mass,
the above suggests that our SU(6) and SU(8) lattices are
large enough for the confined $k=1$ masses and probably 
large enough for the deconfined masses,
but that the $k>1$ confined masses may well suffer substantial
corrections. (See also
\cite{hmmt-string04}.)
\subsection{Debye mass at ${\bf T > T_c}$}
\label{subsection_mdatTc}

We see from eqn(\ref{eqn_mDT}) that at leading order in
$g^2(T)N$ the Debye electric screening mass  $m_D$
is independent of $N$ and grows linearly in 
$T$ up to the weakly varying coupling factor. 

We extract this mass from correlations of the vacuum-subtracted  
timelike  Polyakov loop on $L_s=8$ lattices with $L_t=2,3,4$ at 
a value of $\beta$ where $L_t=5$ corresponds to $T=T_c$ and
list the results in Table~\ref{table_mtDT}.
Thus our lattice calculations cover the range 
$1.25 \leq T/T_c \leq 2.5$. In the case of SU(8) the tunnelling
on an $8^3 5$ lattice is sufficiently well-defined that we are
able to separate fields in the confined and deconfined phase
and so calculate $m_D$ down to $T=T_c$. For $T=T_c$ we also
have calculations on the larger volumes listed in
Table~\ref{table_mtDTc}. We see that at $T=T_c$, where the
masses are smallest, the finite $V$ shift is at the level of
$\sim 20\%$. At the highest $T$, where the masses are largest,
we have also performed calculations on $6^3 2$ lattices,
for SU(4) and SU(8), and there is no finite $V$ shift
visible within quite small errors.

\begin	{figure}[p]
\begin	{center}
\leavevmode
\setlength{\unitlength}{0.240900pt}
\ifx\plotpoint\undefined\newsavebox{\plotpoint}\fi
\sbox{\plotpoint}{\rule[-0.200pt]{0.400pt}{0.400pt}}%
\begin{picture}(1500,1800)(0,0)
\font\gnuplot=cmr10 at 12pt
\gnuplot
\sbox{\plotpoint}{\rule[-0.200pt]{0.400pt}{0.400pt}}%
\put(350.0,250.0){\rule[-0.200pt]{4.818pt}{0.400pt}}
\put(325,250){\makebox(0,0)[r]{\ \ {$0$}}}
\put(1405.0,250.0){\rule[-0.200pt]{4.818pt}{0.400pt}}
\put(350.0,625.0){\rule[-0.200pt]{4.818pt}{0.400pt}}
\put(325,625){\makebox(0,0)[r]{\ \ {$0.5$}}}
\put(1405.0,625.0){\rule[-0.200pt]{4.818pt}{0.400pt}}
\put(350.0,1000.0){\rule[-0.200pt]{4.818pt}{0.400pt}}
\put(325,1000){\makebox(0,0)[r]{\ \ {$1$}}}
\put(1405.0,1000.0){\rule[-0.200pt]{4.818pt}{0.400pt}}
\put(350.0,1375.0){\rule[-0.200pt]{4.818pt}{0.400pt}}
\put(325,1375){\makebox(0,0)[r]{\ \ {$1.5$}}}
\put(1405.0,1375.0){\rule[-0.200pt]{4.818pt}{0.400pt}}
\put(350.0,1750.0){\rule[-0.200pt]{4.818pt}{0.400pt}}
\put(325,1750){\makebox(0,0)[r]{\ \ {$2$}}}
\put(1405.0,1750.0){\rule[-0.200pt]{4.818pt}{0.400pt}}
\put(350.0,250.0){\rule[-0.200pt]{0.400pt}{4.818pt}}
\put(350,200){\makebox(0,0){\ {$0$}}}
\put(350.0,1730.0){\rule[-0.200pt]{0.400pt}{4.818pt}}
\put(529.0,250.0){\rule[-0.200pt]{0.400pt}{4.818pt}}
\put(529,200){\makebox(0,0){\ {$0.5$}}}
\put(529.0,1730.0){\rule[-0.200pt]{0.400pt}{4.818pt}}
\put(708.0,250.0){\rule[-0.200pt]{0.400pt}{4.818pt}}
\put(708,200){\makebox(0,0){\ {$1$}}}
\put(708.0,1730.0){\rule[-0.200pt]{0.400pt}{4.818pt}}
\put(888.0,250.0){\rule[-0.200pt]{0.400pt}{4.818pt}}
\put(888,200){\makebox(0,0){\ {$1.5$}}}
\put(888.0,1730.0){\rule[-0.200pt]{0.400pt}{4.818pt}}
\put(1067.0,250.0){\rule[-0.200pt]{0.400pt}{4.818pt}}
\put(1067,200){\makebox(0,0){\ {$2$}}}
\put(1067.0,1730.0){\rule[-0.200pt]{0.400pt}{4.818pt}}
\put(1246.0,250.0){\rule[-0.200pt]{0.400pt}{4.818pt}}
\put(1246,200){\makebox(0,0){\ {$2.5$}}}
\put(1246.0,1730.0){\rule[-0.200pt]{0.400pt}{4.818pt}}
\put(1425.0,250.0){\rule[-0.200pt]{0.400pt}{4.818pt}}
\put(1425,200){\makebox(0,0){\ {$3$}}}
\put(1425.0,1730.0){\rule[-0.200pt]{0.400pt}{4.818pt}}
\put(350.0,250.0){\rule[-0.200pt]{258.967pt}{0.400pt}}
\put(1425.0,250.0){\rule[-0.200pt]{0.400pt}{361.350pt}}
\put(350.0,1750.0){\rule[-0.200pt]{258.967pt}{0.400pt}}
\put(150,1300){\makebox(0,0){\Large{$2am_D$}}}
\put(862,25){\makebox(0,0){\large{$T/T_c$}}}
\put(350.0,250.0){\rule[-0.200pt]{0.400pt}{361.350pt}}
\put(708.0,543.0){\rule[-0.200pt]{0.400pt}{4.336pt}}
\put(698.0,543.0){\rule[-0.200pt]{4.818pt}{0.400pt}}
\put(698.0,561.0){\rule[-0.200pt]{4.818pt}{0.400pt}}
\put(798.0,744.0){\rule[-0.200pt]{0.400pt}{4.336pt}}
\put(788.0,744.0){\rule[-0.200pt]{4.818pt}{0.400pt}}
\put(788.0,762.0){\rule[-0.200pt]{4.818pt}{0.400pt}}
\put(944.0,1026.0){\rule[-0.200pt]{0.400pt}{7.227pt}}
\put(934.0,1026.0){\rule[-0.200pt]{4.818pt}{0.400pt}}
\put(934.0,1056.0){\rule[-0.200pt]{4.818pt}{0.400pt}}
\put(1242.0,1415.0){\rule[-0.200pt]{0.400pt}{16.622pt}}
\put(1232.0,1415.0){\rule[-0.200pt]{4.818pt}{0.400pt}}
\put(708,552){\circle*{18}}
\put(798,753){\circle*{18}}
\put(944,1041){\circle*{18}}
\put(1242,1449){\circle*{18}}
\put(1232.0,1484.0){\rule[-0.200pt]{4.818pt}{0.400pt}}
\put(708.0,306.0){\rule[-0.200pt]{0.400pt}{2.168pt}}
\put(698.0,306.0){\rule[-0.200pt]{4.818pt}{0.400pt}}
\put(698.0,315.0){\rule[-0.200pt]{4.818pt}{0.400pt}}
\put(798.0,696.0){\rule[-0.200pt]{0.400pt}{7.468pt}}
\put(788.0,696.0){\rule[-0.200pt]{4.818pt}{0.400pt}}
\put(788.0,727.0){\rule[-0.200pt]{4.818pt}{0.400pt}}
\put(951.0,1019.0){\rule[-0.200pt]{0.400pt}{5.300pt}}
\put(941.0,1019.0){\rule[-0.200pt]{4.818pt}{0.400pt}}
\put(941.0,1041.0){\rule[-0.200pt]{4.818pt}{0.400pt}}
\put(1249.0,1417.0){\rule[-0.200pt]{0.400pt}{14.454pt}}
\put(1239.0,1417.0){\rule[-0.200pt]{4.818pt}{0.400pt}}
\put(708,310){\circle{18}}
\put(798,711){\circle{18}}
\put(951,1030){\circle{18}}
\put(1249,1447){\circle{18}}
\put(1239.0,1477.0){\rule[-0.200pt]{4.818pt}{0.400pt}}
\put(350,250){\usebox{\plotpoint}}
\put(350.00,250.00){\usebox{\plotpoint}}
\put(362.75,266.38){\usebox{\plotpoint}}
\put(375.15,283.01){\usebox{\plotpoint}}
\put(387.68,299.56){\usebox{\plotpoint}}
\put(399.86,316.36){\usebox{\plotpoint}}
\put(412.50,332.82){\usebox{\plotpoint}}
\put(424.88,349.48){\usebox{\plotpoint}}
\put(437.66,365.84){\usebox{\plotpoint}}
\put(450.37,382.24){\usebox{\plotpoint}}
\put(462.59,399.02){\usebox{\plotpoint}}
\put(475.00,415.64){\usebox{\plotpoint}}
\put(487.49,432.22){\usebox{\plotpoint}}
\put(500.16,448.66){\usebox{\plotpoint}}
\put(512.98,464.98){\usebox{\plotpoint}}
\put(525.31,481.67){\usebox{\plotpoint}}
\put(537.82,498.22){\usebox{\plotpoint}}
\put(549.82,515.14){\usebox{\plotpoint}}
\put(562.64,531.46){\usebox{\plotpoint}}
\put(575.11,548.05){\usebox{\plotpoint}}
\put(587.80,564.47){\usebox{\plotpoint}}
\put(600.60,580.81){\usebox{\plotpoint}}
\put(612.84,597.57){\usebox{\plotpoint}}
\put(625.15,614.28){\usebox{\plotpoint}}
\put(637.71,630.79){\usebox{\plotpoint}}
\put(650.30,647.29){\usebox{\plotpoint}}
\put(662.73,663.91){\usebox{\plotpoint}}
\put(675.46,680.31){\usebox{\plotpoint}}
\put(688.15,696.72){\usebox{\plotpoint}}
\put(699.96,713.77){\usebox{\plotpoint}}
\put(712.79,730.09){\usebox{\plotpoint}}
\put(725.33,746.63){\usebox{\plotpoint}}
\put(737.94,763.11){\usebox{\plotpoint}}
\put(750.76,779.43){\usebox{\plotpoint}}
\put(763.09,796.13){\usebox{\plotpoint}}
\put(775.19,812.99){\usebox{\plotpoint}}
\put(787.62,829.61){\usebox{\plotpoint}}
\put(800.44,845.93){\usebox{\plotpoint}}
\put(812.96,862.48){\usebox{\plotpoint}}
\put(825.60,878.94){\usebox{\plotpoint}}
\put(838.42,895.26){\usebox{\plotpoint}}
\put(850.11,912.41){\usebox{\plotpoint}}
\put(862.93,928.73){\usebox{\plotpoint}}
\put(875.55,945.20){\usebox{\plotpoint}}
\put(888.08,961.74){\usebox{\plotpoint}}
\put(900.57,978.32){\usebox{\plotpoint}}
\put(913.24,994.76){\usebox{\plotpoint}}
\put(926.05,1011.08){\usebox{\plotpoint}}
\put(937.75,1028.22){\usebox{\plotpoint}}
\put(950.57,1044.54){\usebox{\plotpoint}}
\put(963.16,1061.04){\usebox{\plotpoint}}
\put(975.72,1077.56){\usebox{\plotpoint}}
\put(988.55,1093.88){\usebox{\plotpoint}}
\put(1000.93,1110.54){\usebox{\plotpoint}}
\put(1013.03,1127.40){\usebox{\plotpoint}}
\put(1025.40,1144.06){\usebox{\plotpoint}}
\put(1038.23,1160.38){\usebox{\plotpoint}}
\put(1050.79,1176.90){\usebox{\plotpoint}}
\put(1063.38,1193.39){\usebox{\plotpoint}}
\put(1076.20,1209.71){\usebox{\plotpoint}}
\put(1087.90,1226.85){\usebox{\plotpoint}}
\put(1100.71,1243.18){\usebox{\plotpoint}}
\put(1113.38,1259.61){\usebox{\plotpoint}}
\put(1125.87,1276.19){\usebox{\plotpoint}}
\put(1138.40,1292.73){\usebox{\plotpoint}}
\put(1151.02,1309.21){\usebox{\plotpoint}}
\put(1163.26,1325.97){\usebox{\plotpoint}}
\put(1175.55,1342.69){\usebox{\plotpoint}}
\put(1188.37,1359.01){\usebox{\plotpoint}}
\put(1201.01,1375.47){\usebox{\plotpoint}}
\put(1213.52,1392.03){\usebox{\plotpoint}}
\put(1226.35,1408.35){\usebox{\plotpoint}}
\put(1238.23,1425.35){\usebox{\plotpoint}}
\put(1250.85,1441.82){\usebox{\plotpoint}}
\put(1263.19,1458.51){\usebox{\plotpoint}}
\put(1276.01,1474.83){\usebox{\plotpoint}}
\put(1288.63,1491.31){\usebox{\plotpoint}}
\put(1301.16,1507.84){\usebox{\plotpoint}}
\put(1313.51,1524.52){\usebox{\plotpoint}}
\put(1325.74,1541.29){\usebox{\plotpoint}}
\put(1338.51,1557.65){\usebox{\plotpoint}}
\put(1351.23,1574.05){\usebox{\plotpoint}}
\put(1363.67,1590.67){\usebox{\plotpoint}}
\put(1376.25,1607.16){\usebox{\plotpoint}}
\put(1388.42,1623.98){\usebox{\plotpoint}}
\put(1401.01,1640.47){\usebox{\plotpoint}}
\put(1413.37,1657.14){\usebox{\plotpoint}}
\put(1425,1672){\usebox{\plotpoint}}
\end{picture}
\end	{center}
\vskip 0.15in
\caption{The electric screening mass plotted against
$T/T_c$ for SU(3) ($\circ$) and SU(8) ($\bullet$).}
\label{fig_mtDT}
\end 	{figure}
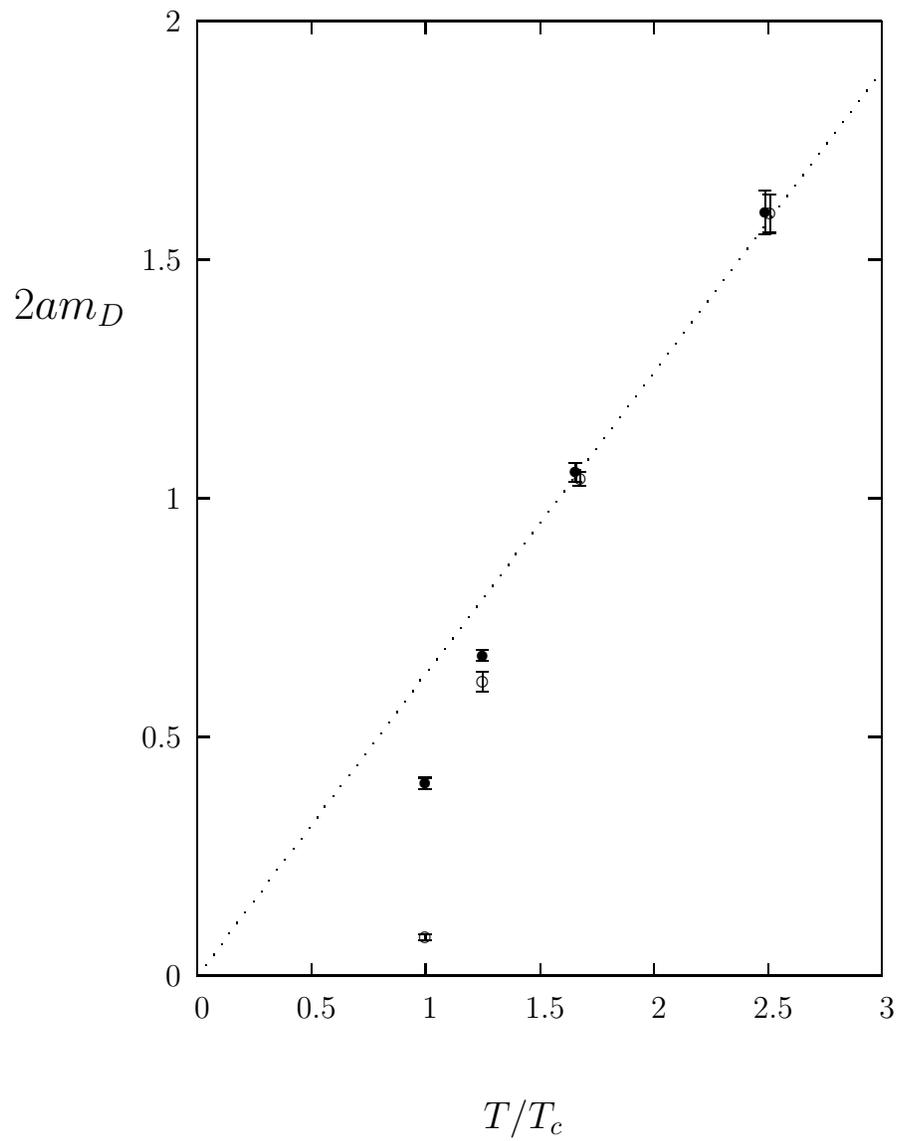

In Fig.\ref{fig_mtDT} we plot $m_t=2m_D$ against $T/T_c$
for both SU(8) and SU(3). We observe that for $T\geq 5T_c/3$ the 
mass is independent of $N$ and appears to increase $\propto T$.
There are only small deviations from this behaviour at
$T=5T_c/4$ and it is only at $T\simeq T_c$ that we see any
large deviation. 

Thus the asymptotic $T\to\infty$ behaviour sets in very close to 
$T_c$, especially at larger $N$. This is remarkable since for
$T \sim T_c$ the coupling will surely not be small. Indeed, if we
apply eqn(\ref{eqn_mDT}) to our calculated values, we find
\begin{equation}
\frac{g^2(T\simeq 1.25T_c)N}{4\pi}
\simeq 
0.43 .
\label{eqn_alpha}
\end{equation}
Such a substantial coupling also means that the lowest order
calculation is probably unreliable (as may be the interpretation of 
$m_t$ as $2m_D$). Indeed  eqn(\ref{eqn_mDT}) would predict that
the approach to the linear behaviour at high $T$ should be
from above, since $g^2(T)$ decreases with increasing $T$, whereas
what we observe in Fig.\ref{fig_mtDT} is that it does so from
below.

\subsection{Spatial string tensions at all ${\bf T}$}
\label{section_masses_sigT}

The expectation values of 
large spatial Wilson loops are known to decay exponentially with 
their area at all $T$ and the coefficient of the area is the
`spatial string tension'. 
At high $T$ this corresponds to the string tension
of the dimensionally reduced $D=2+1$ SU($N$) gauge-Higgs theory.
It also corresponds to the string tension in a $L^2_s L_t$
spatial volume at a temperature $aT=1/L_s$, for sources separated 
along one of the long directions. (This follows from  
a trivial relabelling of the Euclidean co-ordinates.)

The spatial string tension, $\sigma_s(T)$, can be obtained 
from the correlation of Polyakov loops that wind around a spatial 
torus. From this we obtain the `mass', $m_s$, of the lightest flux 
loop that winds around that torus and applying the usual string
correction, which should be adequate since $L_s$ is large,
we obtain the spatial string tension:
\begin{equation}
\frac{1}{L_s} am_s(L_s;T)
=
a^2 \sigma_s(T)
-
\frac{\pi D_{\perp}}{6}\frac{1}{L^2_s} .
\label{eqn_sigsT}
\end{equation}
Here $D_{\perp}$ is the number of dimensions transverse to
the Wilson loop. For D=3+1 and low $T$ we have $D_{\perp}=D-2=2$
and we would expect this to remain the case for any $T$ in
the confining phase. It is less clear what one should use
in the deconfined phase: at very large $T$ one presumably has
$D_{\perp} = 1$. Since  eqn(\ref{eqn_msT}) tells us that  
$\sigma_s(T) \propto T^2$, this correction will be $O(1/T^2)$
and so unimportant at high $T$ if $L_s$ is kept fixed.

Below $T_c$  $\sigma_s(T)$ is more-or-less constant and
above $T_c$ it grows with $T$. Is the change smooth
or discontinuous? Given that the transition is first-order
a discontinuity would be natural but the existing folklore,
based on SU(3) calculations, is that it is continuous.
To address this question we show in 
Table~\ref{table_msCD} the values of $am_s$ as obtained
on sequences of $12^3 5$ lattices at $T=T_c$ in SU(8), that
are entirely in confined and deconfined phases respectively.
(These volumes are large enough to suppress any sign 
of tunnelling.) What we actually show in the Table are
effective masses obtained from local cosh fits to the
correlation function. They are labelled by the distance
$n_t$ at which they are calculated. Although the errors
become large for $n_t\geq 3$ (particularly in the confined
phase) it is quite clear that there is a discontinuity
in the value of $m_s$ as we cross from the confined to
the deconfined phase. (The string operators we use have
excellent overlaps onto the lightest string states, and
we expect
\cite{oxglue04}
the $n_t = 2$ effective mass to provide an excellent 
approximation to the true mass.) Since the transition is strongly 
first order for SU(8), perhaps the real surprise is that this
jump (decrease) in the spatial string tension is only 
about 15\%.

From Table~\ref{table_msCD} we estimate $am_s$ in the confined
phase to be $am_s=1.377(40)$. Using  eqn(\ref{eqn_sigsT})
with $D_{\perp}=2$ we then obtain $a\surd\sigma_s(T=T_c) = 0.349(5)$.
We can compare this to the `$T=0$' value by interpolating
to $\beta=43.965$ the values in Table 9 of 
\cite{oxglue04}
giving  $a\surd\sigma_s(T=0) = 0.346(2)$. Thus we see that to a
good approximation the spatial string tension does not
change at all when $T$ increases from $0$ to $T_c$
-- at least for large enough $N$.

\begin	{figure}[p]
\begin	{center}
\leavevmode
\input	{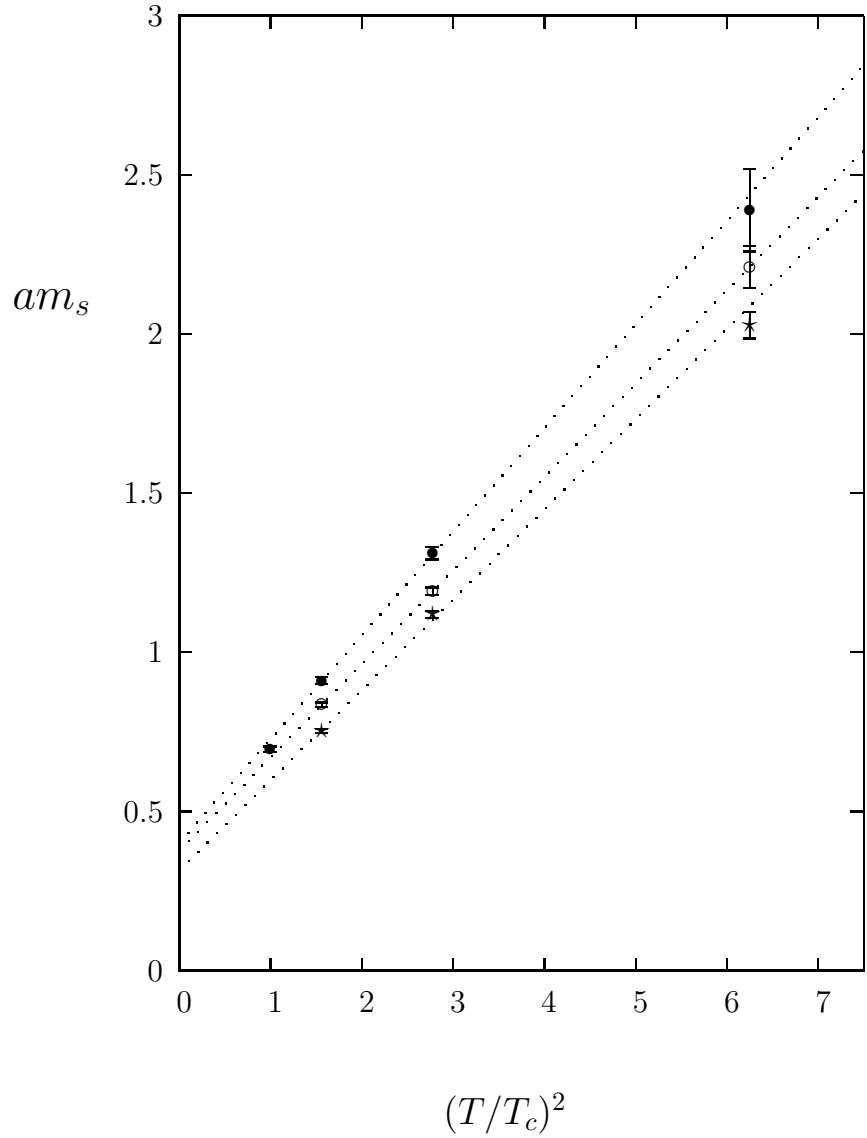}
\end	{center}
\vskip 0.15in
\caption{The mass of the lightest $k=1$ flux loop that
winds around a spatial torus. Plotted against
$T^2$ for SU(3) ($\star$), SU(4) ($\circ$) and SU(8) ($\bullet$),
all at $a\simeq 1/5T_c$.}
\label{fig_msDT}
\end 	{figure}

We have also calculated $am_s$ on the $L_s=8$ and  $L_s=6$
lattices that were used in Section~\ref{subsection_mdatTc}
to calculate the $T$-dependence of the electric gluon mass 
in the deconfined phase. The extracted loop masses are listed
in Table~\ref{table_msDT}. In Fig.\ref{fig_msDT} we plot the 
$k=1$ spatial loop mass against $(T/T_c)^2$, since that is
the expected behaviour at high $T$ (see eqn(\ref{eqn_msT})).
We plot the best linear fits to the data and we clearly
observe good evidence for the early onset of the quadratic 
behaviour. We note that the $T=0$ intercept  corresponds
to relative  corrections that are $O(1/T^2)$, and which 
include the string correction. We note some variation of
the slope with $N$, but this is close to what we have observed
\cite{oxtemp03}
in the variation of $T_c/\surd\sigma$ with $N$ and so it is 
presumably an $O(1/N^2)$ correction.

\begin	{figure}[p]
\begin	{center}
\leavevmode
\input	{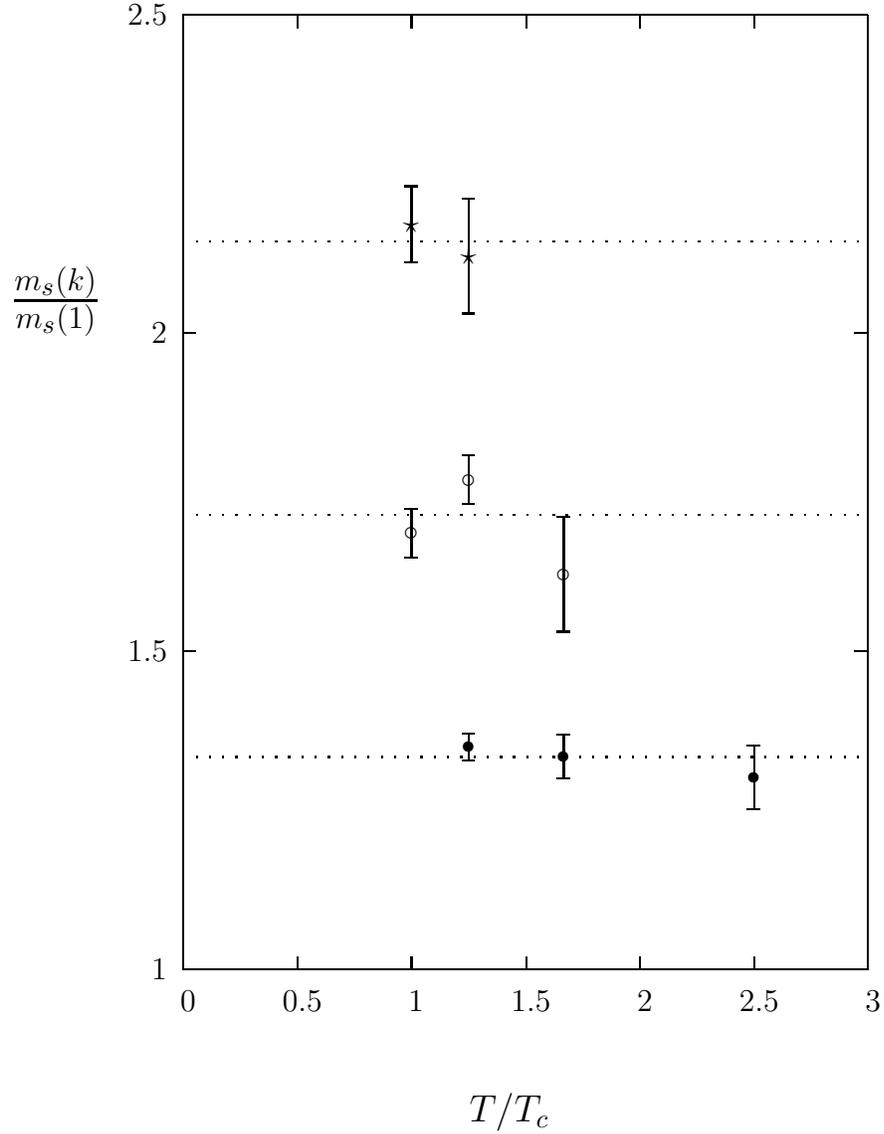}
\end	{center}
\vskip 0.15in
\caption{The mass ratio of the $k=2$ to $k=1$ spatial loops in
SU(4) ($\bullet$) and in 
SU(8) ($\circ$), and of the $k=3$ to $k=1$ loop in
SU(8) ($\star$). All in the deconfined phase
and for  $a\simeq 1/5T_c$. High $T$
Casimir scaling predictions are shown for comparison.}
\label{fig_mskDT}
\end 	{figure}

In Fig.\ref{fig_mskDT} we plot the ratios of the $k$-loop masses 
to the fundamental ($k=1$) loop masses. At higher $T$, where string
corrections become negligible, this just becomes the ratio of
the correponding spatial $k$-string tensions. For comparison
we show the predictions of the Casimir scaling conjecture 
\cite{CS}
in eqn(\ref{eqn_kstring}). (This also arises naturally in a 
model where the high-$T$ plasma contains a gas of adjoint 
magnetic pseudo-particles   
\cite{CKAmon}.)
It is clear from Fig.\ref{fig_mskDT} that we have quite good 
agreement with this conjecture. By contrast, the trigonometric 
`MQCD'  conjecture in  eqn(\ref{eqn_kstring}) is strongly
disfavoured. The inclusion of a string correction
strengthens this conclusion.

\section{${\bf T_c(k)}$}
\label{section_Tck}

We pointed out in Section~\ref{subsection_deconf_mult}
that deconfinement may proceed in several discrete steps,
if the relevant $Z_N$ centre symmetry breaks through
a sequence of subgroups. In that case different $k$-string 
operators will acquire non-zero expectation values at the 
corresponding temperatures, $T_c(k)$. As we remarked in
Section~\ref{subsection_deconf_mult} one expects, on general grounds, 
that $T_c(k)\leq T_c(k=1)$. Thus the temperature we have
identified as $T_c$ using the $k=1$ Polyakov loop as the order 
parameter, will be the highest of these temperatures. 
If, then, we calculate the
masses of the various $k$-strings in the confined phase at 
$T=T_c(k=1)$, any such expectation values will become manifest,
and we will have direct evidence for the existence of such
multiple phase transitions.

We have performed such mass calculations for many
of the values of $N$, lattice sizes and $\beta$ values
discussed in Section~\ref{section_masses}. 
In  Table~\ref{table_mtCTc} we summarised some
of our `cleanest' results. These calculations were performed 
without any vacuum subtraction so that a non-zero
vacuum expectation value would manifest itself in a (nearly) 
vanishing mass gap. It is clear that none of the masses are 
near zero, demonstrating  that in all
these case the $k$-string operators have not developed 
non-zero expectation values for $T< T_c(k=1)$. This provides
our first, quite convincing evidence that deconfinement
proceeds through a single phase transition.

Of course the reason the mass calculations in Table~\ref{table_mtCTc}
are `clean' is that the spatial volumes used are large enough for 
the  confined-deconfined  tunnelling  to be negligible.
One might therefore worry that even if there was an
earlier phase transition, the corresponding tunnelling would 
be improbable. This is clearly not going to be the case
unless the $T_c(k)$ are sufficiently close, since otherwise 
the free energy difference between the relevant phases will 
promote the tunnelling probability at $T_c(k=1)$. Nonetheless
this uncertainty suggests that we should also look at smaller
volumes where there is significant tunnelling. We have indeed
done so and find no suggestion for non-zero expectation
values for any $k>1$ string operators. Such a study does however
suffer from some small ambiguity to do with separating the
confined and deconfined phases on smaller volumes.

As an explicit illustration, we consider SU(4). We take the
largest lattice, $20^3 5$, on which we see any tunnelling
(within our statistics) and a value of $T$ slightly below $T_c$;
roughly $T\simeq 0.996 T_c$. In Fig.\ref{fig_histk1}
we plot a histogram of the values taken by $|{\bar l_{k=1}}|$, 
the modulus of the Polyakov loop averaged over a single lattice
field configuration. We see the very large peak at
low $|{\bar l}|$ corresponding to the confined phase
($l$ is complex hence the suppression at $l=0$) and a small
peak at larger $|{\bar l}|$ corresponding to fields in
the deconfined phase. If there was a separate $k=2$
transition then we would expect that some of the fields in the
low $|{\bar l}_{k=1}|$ peak in Fig.\ref{fig_histk1} would
be confined for $k=2$ strings
and some would be deconfined. To test whether this is
so we have taken separately all the fields with   
$|{\bar l_{k=1}}| \leq 0.05$ and $|{\bar l_{k=1}}| > 0.05$
(i.e. mid-way between the confined and deconfined peaks)
and for each of these two ensembles we plot separately the 
distribution of $|{\bar l}_{k=2}|$ (in the totally 
antisymmetric representation).
These are plotted in  Fig.\ref{fig_histk2n} and
we see that there is no evidence at all for a $k=2$
deconfined phase for which $k=1$ strings are still
confined. We infer that there is a single deconfining
transition at any $N$.

\section{Interface tensions}
\label{section_interface}

We saw in Section~\ref{subsection_deconf_Ndep} that
at $T=T_c$ the interface tension, $\sigma_{cd}$, 
determines the rate of tunnelling
between the confined and deconfined phases through 
eqn(\ref{eqn_ptunnel}). In this Section we shall 
use the observed rate of tunnelling to estimate the 
$N$-dependence of $\sigma_{cd}$, borrowing standard methods 
that have been used in earlier work 
\cite{interface}
in the case of SU(3). (For a different approach,
which appeared as this paper was being written, see
\cite{phdf04}.)
We shall start 
by formulating in more detail the expression for tunnelling,
with the focus on making explicit all the important $N$-dependence.

We use the average Polyakov loop on each lattice field, 
$\bar l$, as our order parameter. (In practice we will
use its modulus, which leads to some minor changes
as described in
\cite{interface}.)
Let $i=c,d$ label the confined and deconfined phases and
let $P_i({\bar l})$ be the normalised probability distribution. 
If $\xi_i$ is the length scale on which the fluctuations
of Polyakov loops decorrelate, and if $V$ is large enough,
then we can expect that making the usual assumption of a 
single-scale Gaussian distribution
\begin{equation}
P_i({\bar l})
=
c_i
e^{-\frac{({\bar l}-{\bar l}_i)^2}{d^2_i}}
\label{eqn_probli}
\end{equation}
where 
\begin{equation}
\frac{1}{d^2_i} 
\propto
N^2 \frac{V}{\xi_i^3}
\hspace{0.5cm}
;
\hspace{0.5cm}
c_i 
\propto
\frac{1}{d_i}
\label{eqNdi}
\end{equation}
is a harmless approximation for our purposes.
The factor of $N^2$ in eqn(\ref{eqNdi}) follows from the
usual large-$N$ factorisation arguments for gauge invariant
operators, which imply that
$\langle {\bar l}^2 \rangle = \langle{\bar l}\rangle^2
(1+O(1/N^2))$ in both phases.

At $T=T_c$ we have tunnelling and both Gaussians peaks are present. 
The probability that a value of ${\bar l}$ mid-way between the peaks
comes from one of these Gaussians is $\propto \exp\{- b_i N^2 V\}$,
where the $b_i$ are some constants, and this
is much less than the tunnelling probability, 
$\propto \exp\{- 2 \sigma_{cd} A/T\}$, for large enough $V$ 
and/or $N$. So in the latter limit, on an $L^3_s L_t$ lattice,
the observed probability of such a value of ${\bar l}$ should 
reflect the tunnelling probability
\begin{equation}
p_t
\propto
e^{-2\frac{a^2\sigma_{cd} L_s^2}{T}}
=
e^{-2 a^3\sigma_{cd} L_s^2L_t}
\label{eqn_probt}
\end{equation}
in lattice units.

Let us assume that we have defined $T_c$ on a finite volume
as the temperature at which the probability of the
confined phase equals that of the deconfined phase. (The 
maximum of the specific heat or of the loop susceptibility 
will correspond to this definition at large $V$.) We 
see from the above that if we consider the overall probability
distribution for ${\bar l}$ then the value at the minimum $p_{min}$,
between the two Gaussian peaks, is related to the value at one of 
these peaks, $p_{max}$, by
\begin{equation}
\frac{p_{min}}{p_{max}}
 \propto 
\frac
{ L^2_s e^{-2 a^3\sigma_{cd} L_s^2 L_t} }
{ \sqrt{N^2 V} } 
 = 
\frac{L^\frac{1}{2}_s}{N}
e^{-2 a^3\sigma_{cd} L_s^2 L_t}.
\label{eqn_pminmax}
\end{equation}
The factor of $L^2_s$ comes from translations of the two walls
(and being careful with the way this affects the measure).
The fact that there are $N$ deconfined phases in the Euclidean
system does not alter this result. Note that
we have neglected the contribution from the interiors of the
domain walls and various factors that have non-zero large-$N$
limits.

Apart from the obvious caveats, an additional uncertainty in  
the above analysis is that it ignores the contribution
of tunnellings between the $N$ different deconfined phases. 
If we label the deconfined vacuum by $k$, corresponding to the
$k$'th element of $Z_N$, then one can see that a tunnelling between 
the $k$ and $k^\prime$ vacua will pass through fields with 
\begin{equation}
|{\bar l}|
\simeq 
|l_d| \cdot |\cos \{ \pi\frac{k-k^\prime}{N} \}| \ .
\label{eqn_lbarddb}
\end{equation}
For $k-k^\prime \sim O(N/2)$ the value of $|{\bar l}|$ will
be partway between $|l_c|$ and $|l_d|$, i.e. precisely where
we have applied  eqn(\ref{eqn_pminmax}). Since we expect
(at least at high $T$ where it is calculable
\cite{ckaDW})
that $\sigma_{kk^\prime}\sim O(N^2)$ when $k-k^\prime \sim O(N/2)$,
such tunnellings are strongly suppressed, and will only
be important for $N,V\to\infty$ if  $\sigma_{cd}\sim O(N^2)$
so that the deconfining tunnellings are similarly suppressed.
In principle one can identify and exclude these 
deconfined-deconfined tunnellings. In practice this requires 
large enough $V$ for the transitions to be very well separated.
This represents a potentially significant systematic error. 

As an aside, we remark 
that the above problem only arises because we have chosen 
to use ${\bar l}$ as our order parameter. We could use the
average plaquette instead since that also takes different
values in the confined and deconfined phases. However
because its value is the same in all $N$ deconfined vacua,
tunnellings between the latter will not populate
the range of plaquette values that signal tunnelling
between confined and deconfined phases. The reason we
do not analyse the plaquette probability distributions
is that the separation between the Gaussian peaks
is essentially non-existent for our range of $V$ and $N$.

Since our calculations are at $T\simeq T_c$ it is customary to 
define the dimensionless surface tension 
\begin{equation}
{\hat\sigma}_{cd}
\equiv
\frac{\sigma_{cd}}{T^3_c}
\label{eqn_sighat}
\end{equation}
in which case we see from eqn(\ref{eqn_pminmax}) that
\begin{equation}
{\hat\sigma}_{cd}
=
-\frac{L^2_t}{2L^2_s}
{\ln \frac{p_{min}}{p_{max}}}
+
\frac{L^2_t}{2L^2_s}
\{\frac{1}{2}\ln L_s - \ln N + c \}
\label{eqn_sig}
\end{equation}
where $c$ is a constant which may contain large $O(1/N^2)$ corrections.
Capillary waves along the interface
\cite{rough}
produce a constant that can be absorbed into the $c$ in
eqn(\ref{eqn_sig}). (The next correction
\cite{rough}
is $O(1/L^2_s)$ and we shall ignore it here.)
We now apply the above formalism, and calculate ${\hat\sigma}_{cd}$
from our calculated Polyakov loop probability distributions. 

In our calculation we choose $T_c$ as the temperature
at which the peak heights of the confined and deconfined Polyakov 
loop probability distributions are equal. This differs slightly 
from our above discussion but, since both peaks shrink with $N$ 
and $V$ in the same way, this does not affect the argument
in a material way. Defining
\begin{equation}
{\hat\Sigma}_{cd}
\equiv
-\frac{L^2_t}{2L^2_s}
{\ln \frac{p_{min}}{p_{max}}}
+
\frac{L^2_t}{4L^2_s}
\ln L_s
\label{eqn_Sig}
\end{equation}
we see from eqn(\ref{eqn_sig}) that 
\begin{eqnarray}
{\hat\Sigma}_{cd}
& = &
{\hat\sigma}_{cd}
-
\frac{L^2_t}{2L^2_s}
\Bigl\{c -\ln N \Bigr\} + \cdots \nonumber\\
& \stackrel{\frac{L_t}{L_s}\to 0}{\longrightarrow} &
{\hat\sigma}_{cd}.
\label{eqn_SigV}
\end{eqnarray}
In Fig.\ref{fig_sigma} we plot our calculated values of 
${\hat\Sigma}_{cd}$ against $(L_t/L_s)^2$ and show the best fits
obtained by including the leading large-$V$ correction
shown in eqn(\ref{eqn_SigV}).
We note that the slope of the best fit decreases with increasing 
$N$ whereas from  eqn(\ref{eqn_SigV}) one would expect the opposite.
It is clear that higher order corrections in $N$ or $V$ must be 
important here. For example, we find that the two Gaussian
distributions are not well separated for most of our volumes. 
Thus there must be some doubt about the reliability of our 
large-$V$ extrapolations.

Ignoring this uncertainty, we extract the interface tensions
from the large-$V$ extrapolations in Fig.\ref{fig_sigma} 
and list them in Table~\ref{table_sig}. Plotting them against $N$ 
in Fig.\ref{fig_sigmaN} we observe a very strong increase with 
$N$. The fact that $\sigma_{cd}$ increases by a factor
of $\sim 30$ when we go from SU(3) to SU(8) implies, using
eqn(\ref{eqn_probt}), that we must decrease $L_s$ by a factor 
$\sim 5.5$ if we want to maintain a comparable tunnelling rate 
when we go from $N=3$ to $N=8$.
  
To test the functional form of the increase of $\sigma_{cd}$
with $N$, we show in  Fig.\ref{fig_sigmaN} the
best fits  corresponding to both linear
\begin{equation}
\frac{\sigma_{cd}}{T^3_c}
=
0.118(3)N-0.333(9) \quad ; \quad \chi^2/n_{df} = 6.5
\label{eqn_sigN1}
\end{equation}
and quadratic variations
\begin{equation}
\frac{\sigma_{cd}}{T^3_c}
=
0.0138(3)N^2-0.104(3) \quad ; \quad \chi^2/n_{df} = 2.7.
\label{eqn_sigN2}
\end{equation}
While a cursory glance at  Fig.\ref{fig_sigmaN} might have
suggested that a linear behaviour looks better, we see that
this is not supported by the statistical analysis which strongly
prefers $\sigma_{cd}\propto N^2$. However the $\chi^2$ is driven by 
the very accurate smaller-$N$ values, which might be sub-asymptotic.
So it is clear that more accurate calculations with a better
control of the systematic errors are needed, 
particularly for $N\geq 8$, if we are to be able to answer this
question.

\section{SU(2)}
\label{section_su2}

The SU(2) gauge theory is linearly confining at low $T$
and deconfined at high $T$. Therefore there will be a
phase transition at some $T=T_c$. If the phase transition
were first order then, in the $V=\infty$ thermodynamic limit,
$\partial \ln Z/\partial \beta$ would be discontinuous at
$T=T_c$. This is the case, as we have seen, for $N\geq 3$.
In SU(2), however, it is well known that 
there is no observed discontinuity, and that
instead the effective string tension appears to vanish 
smoothly as $T\to T^-_c$. The corresponding correlation 
length diverges as $T\to T^-_c$ and the transition is thus
taken to be second order.

In Fig.\ref{fig_mtsu2} we provide an example of how
the correlation length varies with $\beta$
and the spatial volume on a set of $L^3 5$ lattices
for which $V=\infty$ deconfinement occurs at
$\beta_c = 2.3714(6)$ (see Table 4 of
\cite{oxtemp03}). 
What is actually shown here is $\xi_t = 1/am_t$ where
$m_t$ is the mass of the lightest flux loop 
that winds around the time-torus, and which is related
to $\sigma_{eff}(T)$ by eqn(\ref{eqn_sigT}) when $V\to\infty$. 
We see how on a fixed volume near $T_c$ the value of  $\xi_t$
is constrained by the spatial size, but that it is 
consistent with diverging at  $T_c$ as we take $V\to\infty$. 
This is to be contrasted with the values of $am_t$ 
at $T=T_c$ for $N\geq 3$ that are listed in 
Table~\ref{table_mtCTc} (for the same value of $a$ in 
units of $T_c$). 
   
The usual way to identify a first or second order transition 
at $\beta=\beta_c$ is to calculate the specific heat 
$C(\beta)$ 
\begin{equation}
\frac{1}{\beta^2} C(\beta)
=
\frac{1}{V}
\frac{\partial^2 \ln Z(\beta)}{\partial\beta^2}.
\label{eqn_Cheat}
\end{equation}
If $V$ is finite but
large, then  $C(\beta)$ should have a peak at some
$\beta=\beta_c(V)\stackrel{V\to\infty}\longrightarrow\beta_c$ 
and this peak should diverge as
\begin{equation}
C(\beta_c(V))
\propto 
V^\alpha
\label{eqn_CheatV}
\end{equation}
where $\alpha = 1$ for a
first order transition and $\alpha < 1$ for a second order
transition. The behaviour for a first order transition
follows trivially from the discontinuity in the action, 
when one notes that $C$ is proportional to the integrated
connected plaquette correlation function
\begin{equation}
C(\beta)
\propto 
\frac{1}{V}
\langle \sum_p u_p \sum_{p^\prime} u_{p^\prime} \rangle
- \langle \sum_p u_p\rangle^2
\propto
\sum_p
\langle (u_p-\bar{u}) (u_{p_0}-\bar{u}) \rangle .
\label{eqn_Cheatcor}
\end{equation}
Of course it will only be seen
if the Monte Carlo calculation is long enough to sample both 
phases. For a second order transition the divergence of
$C$ arises because of the vanishing mass, which implies
a divergence in the integrated correlator in
eqn(\ref{eqn_Cheatcor}). For the first order transition, 
for $N\geq 3$, a growing specific heat peak is precisely 
what we see. However for the
SU(2) transition $C(\beta)$ not only does not grow with $V$ 
in the neighbourhood of $\beta_c$, but shows no peak at all; 
despite the fact that, as we see in  Fig.\ref{fig_mtsu2},  
there is a rapidly growing correlation length for these
volumes. This is the implicit reason why earlier studies
have not used  $C(\beta)$ to study the SU(2) transition,
but have rather used the Polyakov loop susceptibility,
which shows the expected diverging behaviour.

We are already familiar with one example of a striking
phenomenon in the high $T$ Euclidean system which is visible 
in Polyakov loops but which is not directly reflected
in the real thermal properties of the gauge theory: the
$Z_N$ symmetry breaking and vacuum degeneracy discussed
in Section~\ref{subsection_deconf_artifact}. Here we have 
a diverging correlation length which has only been visible
in Polyakov loops. Is it possible that this divergence is also 
not reflected in the  thermal properties of the gauge theory,
so that the phase transition is not second but higher order?
This may seem very unlikely but provides us with a motivation
for understanding better what is going on.

The vanishing of the mass of the timelike flux loop at $\beta_c$ 
implies, from eqn(\ref{eqn_sigT}), the vanishing of the effective 
string tension, $\sigma_{eff}(T)$, at $T=T_c=1/a(\beta_c)L_t$.
When we calculate the overlap of our (unsmeared) Polyakov loop onto 
this lightest flux loop, at say $\beta = 2.3725\simeq\beta_c$,
we see that it remains very large, $\sim 0.95$, on all our 
volumes. This implies that the effective potential between 
static sources will show a linear behaviour 
$V(r) \simeq \sigma_{eff}(T) r$ from some relatively short
distance, and that the potential will flatten from this distance 
onwards as $T\to T_c$. Thus the vanishing of the SU(2) effective 
string tension at the transition is a robust, highly visible 
feature of the thermal gauge theory.

Nonetheless this is a property of the gauge theory probed with static
fundamental sources, and is not related in an obvious way to
the vanishing in the pure gauge theory of a screening length 
which couples to the plaquette. Indeed we know that in the
confining phase a non-contractible winding flux loop must have 
an exactly zero overlap on to a contractible Wilson loop such 
as the plaquette, and so cannot contribute directly to $C(\beta)$.
But this is not the end of the story. Scattering states composed of 
two such (mutually conjugate) flux loops will have 
energies all the way down to zero at $T_c$ and these can have
non-zero overlaps onto the plaquette. We now estimate this
overlap.

Consider a $20^3 5$ lattice at $\beta_c=2.3675$, which is
very close to $\beta_c$. We consider operators composed
of $\vec{p}=0$ sums of spatial (magnetic) and timelike 
(electric) plaquettes and also an operator that is composed 
of a product of two $\vec{p}=0$ Polyakov loops.
(Since we reserve the label $t$ for the short torus, we take
correlations in the $z$-direction and the momentum $\vec{p}$
is defined in the 3-space orthogonal to the latter.) 
Looking at the effective mass plot of the latter it is clear 
that at this $a$ this operator has a very good overlap onto 
the state composed of two of the lightest $\vec{p}=0$ timelike
flux loops. We will therefore make the approximation that
this operator is in fact the wavefunctional for this two loop
state, $| ll \rangle$. We label the latter operator by 
$\phi_{ll}$, and the plaquette sums by $u_s,u_t$ respectively. 
Then the normalised cross-correlations give us the matrix 
element which determines the desired overlap
\begin{equation}
\frac{\langle vac | u_a \phi_{ll} | vac \rangle}
{ \langle vac | u_a u_a | vac \rangle^\frac{1}{2}
\langle vac | \phi_{ll} \phi_{ll} | vac \rangle^\frac{1}{2} }
\simeq
\frac{\langle vac | u_a | ll \rangle}
{\langle vac | u_a u_a | vac \rangle^\frac{1}{2}}
\simeq  \begin{cases}
0.027(3) &:\quad a=s\\
0.042(3) &:\quad a=t .
\end{cases}
\label{eqn_llplaq}
\end{equation}
This is very small, and one naturally wonders if this smallness
does not indicate that the overlap actually vanishes 
either as one approaches $V = \infty$ or as one approaches  
$\beta=\beta_c$. This we shall now show is not the case. We have
performed calculations on $14^3 5$ and $12^3 5$ lattices
which show that the overlap in eqn(\ref{eqn_llplaq}) is
volume independent, within errors. We have also performed
calculations at $\beta=2.3600$ where $T-T_c$ is about
3 times larger than in the above example, and again we
find that the overlap is unchanged. We thus conclude that
at the lattice spacing $a \simeq 1/5T_c$ there is in fact 
a non-zero overlap between the plaquette and the double 
flux loop whose mass, we recall, vanishes as $V\to\infty$
and $T\to T_c$. That is to say, the thermal average of a 
plaquette-plaquette correlator must, at large enough
separations, display the contribution of this diverging
correlation length, and so the deconfining transition in the
SU(2) thermal gauge theory is indeed a conventional second 
order one.

In the normalised $\vec{p}=0$ plaquette-plaquette correlator, the 
contribution of the double flux loop will be
\begin{equation}
\frac{C_{uu}(n_z)}{C_{uu}(0)}
=
\frac{|{\langle vac | u_a | ll \rangle}|^2} 
{\langle vac | u_a u_a | vac \rangle}
e^{-2am_t n_z} + \ldots
\label{eqn_llplaqcor}
\end{equation}
(ignoring periodicity). We can use this to estimate how large
we have to make $L_s$ for the diverging correlation length to
start making a visible contribution. First we note that
we can replace $\sum_p u_p$ in eqn(\ref{eqn_Cheatcor}) by the 
sum over $z$ of the $\vec{p}=0$ sums of plaquettes. And we can
make a crude estimate for $am_t(L_s)$ using Fig.\ref{fig_mtsu2}.
Using the latter and the overlaps in eqn(\ref{eqn_llplaq}),
we can easily see, via a very approximate calculation, that for the
term in eqn(\ref{eqn_llplaqcor}) to contribute about half of the 
specific heat at $\beta=\beta_c$, we need to have a lattice
with $L_s \sim 2000$. It is now clear why even on our `very large'
$40^3 5$ lattices we saw no sign of a growing specific heat peak!

The above calculations have been for a specific small range of 
$a$ close to $a = 1/5T_c$. As $a\to 0$ all physical states
decouple from the plaquette, since wavelengths on physical
scales contribute only $\sim a^4\Lambda_{QCD}$ of its value. 
So in the continuum one would want to use a suitably regularised
form of the action when calculating the specific heat. Since
we have  no reason to expect any pathology as $a\to 0$ we
have not attempted a detailed study at smaller $a$.
We have however performed some calculations at  $a = 1/8T_c$ which 
confirm that the overlap, as defined in eqn(\ref{eqn_llplaq}), 
remains non-zero, and that it is smaller, roughly as one would expect.

Finally we comment that the above near-decoupling will obviously
also occur for $N>2$ where the transition is first order.
This means that the specific heat will give a clear, precocious
signal for the developing two peak structure in the plaquette
distribution, unobscured by an additional term coming from
an increasing correlation length  (albeit increasing only to
some finite value). This is to be contrasted with the 
Polyakov loop susceptibility. This effect is strongest
for the weakly first order transition in SU(3) -- compare
Fig.5 and Fig.3 in
\cite{oxtemp03}.
\section{The bulk phase transition}
\label{section_bulk}

It is well known that SU(2) and SU(3) lattice gauge theories
with the plaquette action exhibit a rapid crossover between 
the strong and weak coupling regions.
This crossover becomes more pronounced for SU(4)
\cite{blmt-glue01}
and, as we have found in the present work, 
transforms into a weakly first order transition for SU(5) and
becomes robustly first order for SU($N\geq 6$). The strong
coupling bulk phase is believed to contain $Z_N$ vortices
and monopoles which disorder Wilson loops down to the ultraviolet
length scale, so that the string tension in lattice units
is $a^2\sigma \sim O(1)$. On the weak coupling side, the physics 
in the ultraviolet is determined by asymptotic freedom and  
so $a^2\sigma \sim O(a^2) \ll 1$. Only calculations on the weak coupling
side are useful for the continuum extrapolation. 

The $N\to\infty$ limit has been studied within the twisted 
Eguchi-Kawai formalism, and the critical inverse bare 't Hooft 
coupling is
\cite{bulkN}
\begin{equation}
\lambda^{-1}_b(N=\infty)
=
\lim_{N\to\infty} \frac{\beta_b}{2N^2}
=
0.3596(2).
\label{eqn_lambdab}
\end{equation}
In Table~\ref{table_bulk} we list the values of $\beta_b$ that 
we have obtained  for $5\leq N \leq 12$ during the course of
our calculations. For $N=5$ we can estimate $\beta_b$ directly
but for $N\geq 6$ the interface tension becomes large enough 
that all we see is a hysteresis effect. Once the interface tension
is large the size of the hysteresis varies weakly with both the 
volume size, as we see in Table~\ref{table_bulk}, and also
with the length of Monte Carlo run at each value of $\beta$ 
(within practical limits).
This leads to a large uncertainty in $\beta_b$, which can
be eliminated by adding a suitable function of the order
parameter to the action, as in
\cite{bulkN}.
However our purpose here is not to study the bulk transition
{\it{per se}} but rather to determine its influence, if any, on our
calculations of $T_c$.
Nonetheless we note that our results for $N\geq 6$ are consistent
with eqn(\ref{eqn_lambdab}) together with a modest $O(1/N^2)$
correction. (Note that this will not be the same
$O(1/N^2)$ correction as determined in
\cite{bulkN}
since the Eguchi-Kawai equivalence only exists at the 
leading planar level.)

We compare the values of  $\beta_b$ in Table~\ref{table_bulk}
to the values of  $\beta_c$ in 
Tables~\ref{table_su4lt}-\ref{table_su8lt}.
We note that for SU(6) the deconfining transitions for $L_t\geq 5$
are safely beyond the bulk transition. Interpolating to SU(5)
we would expect the $L_t=5$ deconfining transition to be
at $\beta_c(L_t=5) \simeq 16.87$, which is again well beyond $\beta_b$.
For SU(8) on the other hand, the hysteresis has widened and now
encompasses both $\beta_c(L_t=5)$ and  $\beta_c(L_t=6)$. 
These calculations of $\beta_c$ have been performed by working on
the weak coupling branch of the hysteresis curve and clearly
we need to know if this introduces a bias.

A useful order parameter for the bulk transition is the average
plaquette $\bar{u}_p$. For larger $N$ its discontinuity is 
$\Delta \bar{u}_p \sim 0.13$. Thus, in contrast to the 
deconfining transition,
where the discontinuity is only $O(a^4)$, here it is $O(1)$ and 
so even partial tunnellings are easily visible on single
lattice fields. We have examined the plaquette distributions
at $\beta\simeq \beta_c(L_t=5)$ and we find no trace
of even a partial attempt at tunnelling. It is clear that the 
barrier between the bulk and weak coupling phases is so large 
that the deconfining phase transition in the weak coupling
phase at $a\simeq 1/5T_c$ is largely unaffected by the
fact that it occurs in only a local rather than in the absolute
minimum of the effective potential -- except perhaps for some minor
distortion of the  potential that we have earlier
hypothesised might be the reason for the unexpectedly large 
lattice spacing corrections to the latent heat at this $a$.

\section{Conclusions}
\label{section_conclusions}

In this paper we studied various properties of SU($N$)
gauge theories at and around the deconfining phase transition.
We checked that there is indeed only one transition, i.e.
that the centre symmetry is not broken in steps, with each step 
corresponding to the condensation of an appropriate $k$-string.
We showed that the transition is robustly first order at
large $N$ with a latent heat $L_h\propto N^2$ as expected,
and an interface tension $\sigma_{cd}$ that grows with
$N$, although our calculations were unable to determine 
whether this growth is $\propto N$ or $\propto N^2$. In the 
former case the transition would become a real phase transition  
even on a small volume as $N\to\infty$ (which is possible because
the number of degrees of freedom per unit volume diverges
in that limit).
In the latter case one expects a growing hysteresis,
even on small volumes. If this hysteresis is large enough
one might encounter the `Hagedorn' string condensation
transition along the confining branch, above $T_c$.

We calculated the spatial string tension and showed that
at larger $N$ its value in the confining phase appears to 
be entirely independent of $T$. However, when we compare the
spatial string tensions in the two phases at $T=T_c$ it is
clear that there is a significant discontinuity. This is
easy to see for larger $N$ because the transition is robustly
first order and the growing interface tension ensures that
even on moderately large lattices there is no tunnelling, or 
even attempted tunnelling. So one can study the properties of
the two phases separately at $T=T_c$. For example, on a
$12^3 5$ lattice for SU(8). By contrast, in the well-studied
case of SU(3), even on a $64^3 5$ lattices we still observe
tunnelling. Thus such a discontinuity has not been observed
in SU(3) although it undoubtedly exists there as well.

At higher $T$ we have seen that the spatial string tension 
grows $\propto T^2$ and that it does so essentially from 
$T=T_c$ for larger $N$. The spatial $k$-string tensions are 
consistent with Casimir Scaling. For high $T$ the 
Debye mass grows $\propto T$ as expected,
although there are large deviations near $T_c$,. This is
not surprising because the leading-order theoretical expression,
$m^2_D \propto g^2(T)T^2$ contains an additional `logarithmic'
dependence on $T$ in the coupling.

We also showed why the SU(2) transition has no peak in the 
specific heat, let alone one that grows with $V$, although there 
is a diverging correlation length. We explicitly showed that 
this correlation length will eventually appear as a
visible screening length in thermal averages, so that
the transition is a bona fide second order one.
For larger $N$ we discussed the bulk transition, in
the shadow of whose hysteresis we have performed some of our
deconfinement studies.

The main motivation for studying SU($N$) gauge theories
at large $N$ is the hope that SU(3) might be `close' to this
limit and the gauge theory is expected to be much simpler at
$N = \infty$ than at $N=3$. The former has been largely
confirmed by recent lattice calculations
\cite{blmt-glue,oxglue04,oxtemp03}. 
Signs of the latter
are found throughout our calculations in this paper; for example 
the reduction in finite volume effects at the deconfinement
transition, which naturally points
\cite{oxtemp03} 
to a form of Eguchi-Kawai space-time reduction
\cite{EK}.
On the other hand, while the suppression of `small' fluctuations
as $N\to\infty$ does indeed lead to the simple idea of a 
Master Field
\cite{Master},
what we have also seen is that there is actually a
plethora of Master Fields -- confining, deconfining,
bulk, one for each of the $N$ degenerate phases obtained if 
a spatial torus is reduced below a 
critical size, not to mention the $N$ $\theta$-vacua branches
\cite{Ntheta}
and conjectured non-analyticities in scale size
\cite{HNRN}. 
All this 
\cite{mtN04}
serves to emphasise  how much there is still to be understood
about the physics of this theory.

\section*{Acknowledgements}

Our lattice calculations were carried out on the JIF/PPARC funded 
Astrophysics Beowulf cluster in Oxford Physics, on PPARC 
and EPSRC funded Alpha Compaq workstations in Oxford Theoretical 
Physics,  on a desktop funded by All Souls College, and on the 
APE in Swansea Physics funded by PPARC under grant PPA/G/S/1999/00026.
During much of this research, UW was supported 
by a PPARC SPG fellowship, and BL by a
EU Marie Sk{\l}odowska-Curie postdoctoral fellowship. 
MT would like to thank the KITP for its hospitality
while this paper was being written up, and would like
to thank participants at the `QCD and String Theory' 
Workshop at the KITP, 
as well as at the `Large $N$' and `QCD and Strings' Workshops
at the ECT, Trento, for useful and interesting discussions.

\vfill\eject

\begin{table}
\begin{center}
\begin{tabular}{|c|c|c|c|c|}\hline
\multicolumn{5}{|c|}{SU(4)} \\ \hline
$L_t$  & $L_s$ & $h$ & $\beta_c(V=\infty)$ & $T_c/\surd\sigma$ \\ \hline
 5     & 12-20  & 0.090(17) & 10.63727(53) & 0.6148(13) \\
 6     & 16     & 0.100(22) & 10.7898(16) & 0.6166(24) \\
 8     & 24     & 0.123(27) & 11.0880(22) & 0.6310(30) \\ \hline
\end{tabular}
\caption{\label{table_su4lt} Critical values of $\beta$
calculated on $L^3_s L_t$ lattices and
extrapolated to $V=\infty$ using eqn(\ref{Tc1}) and the values of 
$h$ shown. The corresponding values of
$T_c/\surd\sigma$ are shown.}
\end{center}
\end{table}

\begin{table}
\begin{center}
\begin{tabular}{|c|c|c|c|c|}\hline
\multicolumn{5}{|c|}{SU(6)} \\ \hline
$L_t$  & $L_s$ & $h$ & $\beta_c(V=\infty)$ & $T_c/\surd\sigma$ \\ \hline
 5     & 8-14  & 0.112(19) & 24.5139(24) & 0.5894(36) \\
 6     & 16    & 0.128(29) & 24.8467(30) & 0.5956(29) \\
 8     & 16    & 0.168(39) & 25.4782(64) & 0.6024(29) \\ \hline
\end{tabular}
\caption{\label{table_su6lt} Critical values of $\beta$
calculated on $L^3_s L_t$ lattices and
extrapolated to $V=\infty$ using eqn(\ref{Tc1}) and the values of 
$h$ shown. The corresponding values of
$T_c/\surd\sigma$ are shown.}
\end{center}
\end{table}

\begin{table}
\begin{center}
\begin{tabular}{|c|c|c|c|c|}\hline
\multicolumn{5}{|c|}{SU(8)} \\ \hline
$L_t$  & $L_s$ & $h$ & $\beta_c(V=\infty)$ & $T_c/\surd\sigma$ \\ \hline
 5     & 8,10  & 0.134(73)  & 43.982(14) & 0.5819(41) \\
 6     & 8     & 0.155(84)  & 44.535(37) & 0.5850(70) \\
 8     & 10,12 & 0.218(123) & 45.654(32) & 0.5888(69) \\ \hline
\end{tabular}
\caption{\label{table_su8lt} Critical values of $\beta$
calculated on $L^3_s L_t$ lattices and
extrapolated to $V=\infty$ using eqn(\ref{Tc1}) and the values of 
$h$ shown. The corresponding values of
$T_c/\surd\sigma$ are shown.}
\end{center}
\end{table}

\begin{table}
\begin{center}
\begin{tabular}{|c|c|c|}\hline
\multicolumn{3}{|c|}{SU($N$) : continuum} \\ \hline
$N$  & $T_c/\surd\sigma$ & $\chi^2/n_{df}$ \\ \hline
 2        & 0.7091(36)  & 0.28 \\
 3        & 0.6462(30)  & 0.05 \\
 4        & 0.6344(42)[(81)]  & 3.7[1.0] \\
 6        & 0.6101(51)  & 0.02 \\
 8        & 0.5928(107) & 0.003 \\ \hline
\end{tabular}
\caption{\label{table_suN}  $T_c/\surd\sigma$ extrapolated to
the continuum limit for various SU($N$) gauge theories,
with the $\chi^2$ per degree of freedom of the best fits.
The SU(2) and SU(3) values are taken from \cite{oxtemp03}.}
\end{center}
\end{table}

\begin{table}
\begin{center}
\begin{tabular}{|c|c|c|c|c|c|c|c|c|}\hline
 $N$ & $\beta\geq$ &  $\beta\leq$ & $\beta_0$ & $c_0$  & $c_1$ & $c_2$ & $c_3$ & $\chi^2/n_{df}$ \\ \hline
 2  &  2.1768  & 2.5115 & 2.3726 & 0.2877 & 0.9195 & -3.2961 & --   & 0.55 \\
 3  &  5.6925  & 6.3380 & 5.8945 & 0.2607 & 2.6312 & 12.2494 & 24.0 & 0.54 \\
 3  &  5.6925  & 6.3380 & 5.8945 & 0.2610 & 2.7271 & 10.6846 & --   & 1.1 \\
 4  &  10.637  & 11.400 & 10.789 & 0.2702 & 3.5393 & 18.7900 & --   & 0.9 \\
 6  &  24.500  & 25.452 & 24.670 & 0.3082 & 4.9649 & 34.5288 & --   & 0.5 \\
 8  &  43.85   & 45.700 & 44.350 & 0.3022 & 5.0812 & 42.0732 & --   & 1.2 \\ \hline
\end{tabular}
\caption{\label{table_Kfit} Fit parameters to
the string tension as a function of $\beta$ using
the interpolation function in eqn(\ref{eqn_Kfit}).}
\end{center}
\end{table}

\begin{table}
\begin{center}
\begin{tabular}{|c|c|c|c|}\hline
 $N$ & $L_t$ &  $L_s\in$ & $\Delta\langle u_p\rangle$ \\ \hline
 3  &  5   & 20-40 & 0.00084(2) \\
 4  &  5   & 14-20 & 0.00198(5) \\
 6  &  5   & 8-14  & 0.00238(6) \\
 8  &  5   & 8-10  & 0.00251(34) \\ \hline
\end{tabular}
\caption{\label{table_lhCmax} The jump in the average plaquette,
$\Delta \langle u_p \rangle$, between the confined and deconfined
phases at $\beta_c(a=1/5T_c)$, as obtained from a $V\to\infty$ 
extrapolation of the specific heat peak from the range of lattice
sizes shown.}
\end{center}
\end{table}

\begin{table}
\begin{center}
\begin{tabular}{|c|c|c|c|c|c||c|}\hline
\multicolumn{7}{|c|}{ $L_h$ at $a=1/L_tT_c$ } \\ \hline
 $N$ & $L_t$ &  $L_s$ &  $\beta$ & $\Delta\langle u_p \rangle$ & 
$d\ln{a}/d\beta$ & $L_h/T_c^4$ \\ \hline
 3  &  5   & 64 & 5.799  & 0.00078(3)  & -2.070(9)   & 1.413(55) \\ \hline
 4  &  5   & 32 & 10.635 & 0.00187(4)  & -1.2878(90) & 5.39(13)  \\
    &  6   & 16 & 10.780 & 0.00064(5)  & -1.1666(80) & 4.06(34) \\
    &  8   & 24 & 11.085 & 0.000165(20) & -0.9400(90) & 4.30(53) \\ \hline
 6  &  5   & 16 & 24.515 & 0.00248(3)  & -0.6347(120) & 14.69(35) \\
    &  6   & 16 & 24.845 & 0.00084(2)  & -0.5472(80)  & 11.90(35) \\
    &  8   & 16 & 25.46  & 0.000202(7) & -0.4012(76)  & 12.00(49) \\ \hline
 8  &  5   & 12 &  43.965 & 0.00261(4)   & -0.3717(54) & 25.67(65) \\
    &  6   & 8  &  44.45  & 0.00075(5)   & -0.3240(26) & 16.0(1.4) \\
    &  8   & 12 &  45.50  & 0.000195(10) & -0.2304(48) & 17.8(1.1) \\ \hline
\end{tabular}
\caption{\label{table_lh} The jump in the average plaquette,
$\Delta \langle u_p \rangle$, and the latent heat, $L_h$, at $T=T_c$,
for the gauge groups and parameters shown.}
\end{center}
\end{table}

\begin{table}
\begin{center}
\begin{tabular}{|c|c|c|}\hline
\multicolumn{3}{|c|}{ $L_h$ : continuum } \\ \hline
 $N$ & $L_t\in$ & $L^\frac{1}{4}_h/T_c$ \\ \hline
  4  &  6,8 &  1.47(10) \\
  6  &  6,8 &  1.87(5) \\
  8  &  6,8 &  2.12(9) \\ \hline
\end{tabular}
\caption{\label{table_lhcont} The latent heat in physical units
after a continuum extrapolation from the $L_t=6,8$ values.}
\end{center}
\end{table}

\begin{table}
\begin{center}
\begin{tabular}{|c|c|c|c|c|c|}\hline
\multicolumn{6}{|c|}{ $am_k$ at $a\simeq 1/5T_c$ : confined} \\ \hline
 $N$ & $L_s$ & $\beta$ & $k=1$ & $k=2$ & $k=3$ \\ \hline
 3  & 64 & 5.799  & 0.105(8)   &           & \\
 4  & 32 & 10.635 & 0.1967(89) & 0.335(13) & \\
 6  & 16 & 24.515 & 0.2343(87) & 0.474(25) & 0.657(36) \\
 8  & 12 & 43.965 & 0.2835(103) & 0.485(43) & 0.69(10) \\ \hline
\end{tabular}
\caption{\label{table_mtCTc} The masses of the lightest time-like 
$k$-loops in the confined phase at $T=T_c$.}
\end{center}
\end{table}

\begin{table}
\begin{center}
\begin{tabular}{|c|c|c|c|c|c|}\hline
\multicolumn{6}{|c|}{ $am_k$ at $a\simeq 1/5T_c$ : deconfined } \\ \hline
 $N$ & $L_s$ & $\beta$ & $k=1$ & $k=2$ & $k=3$ \\ \hline
 3  & 64 & 5.799  & 0.080(6)  &            & \\
 4  & 32 & 10.635 & 0.191(34) &  0.185(32) & \\
 6  & 16 & 24.515 & 0.359(14) &  0.346(13) & 0.346(12) \\
 8  & 12 & 43.965 & 0.418(14) &  0.405(13) & 0.402(12) \\ \hline
\end{tabular}
\caption{\label{table_mtDTc} The lightest non-zero masses coupling 
to the time-like $k$-loop in the deconfined phase at $T=T_c$.}
\end{center}
\end{table}

\begin{table}
\begin{center}
\begin{tabular}{|c|c|c|c|c|}\hline
\multicolumn{5}{|c|}{ $am_t$ at $a\simeq 1/5T_c$ : SU(4)} \\ \hline
$L_s$ & $\beta$ & $C: k=1$ & $C: k=2$ & $D$ \\ \hline
32 & 10.635 & 0.1967(89) & 0.335(13) & 0.185(32) \\
20 & 10.635 & 0.203(10)  & 0.312(20) & 0.215(20) \\
16 & 10.635 & 0.195(20)  & 0.24(4)   & 0.25(3) \\ \hline
20 & 10.642 &        --  &        -- & 0.240(20) \\
20 & 10.633 & 0.185(10)  & 0.255(20) & -- \\
16 & 10.630 & 0.204(15)  & 0.27(4)   & -- \\ \hline
\end{tabular}
\caption{\label{table_mtTcN4} The lightest non-zero masses coupling 
to the time-like $k$-loop in the confined ($C$) and deconfined ($D$)
phases at $T\simeq T_c$ on various volumes.}
\end{center}
\end{table}

\begin{table}
\begin{center}
\begin{tabular}{|c|c|c|c|c|c|c|}\hline
\multicolumn{7}{|c|}{ $2am_D$ for $T\geq T_c$} \\ \hline
 $N$ & $\beta$ & $T=T_c$ & $T=5T_c/4$ & $T=5T_c/3$ 
& $T=5T_c/2$  & $T=5T_c/2$; $L_s=6$ \\ \hline
3 & 5.80 & --   & 0.615(21) & 1.040(15) & 1.596(40) & -- \\
4 & 10.635 & -- & 0.660(8)  & 1.053(19) & 1.643(40) & 1.578(30)  \\
8 & 44.0 & 0.527(11) & 0.670(12) & 1.054(20) & 1.599(46) & 1.62(5) \\ \hline
\end{tabular}
\caption{\label{table_mtDT} The lightest non-zero mass that couples 
to the time-like flux loop in the deconfined phase for $a \simeq 1/5T_c$ 
on $L_s=8$ lattices (unless otherwise indicated).}
\end{center}
\end{table}

\begin{table}
\begin{center}
\begin{tabular}{|c|c|c|}\hline
\multicolumn{3}{|c|}{ SU(8) : $am_s(t=an_t)$ } \\ \hline
 $n_t$ & $D$ & $C$  \\ \hline
1 & 1.326(7)  & 1.527(11) \\
2 & 1.187(24) & 1.377(40) \\
3 & 1.173(81) & 1.16(15) \\
4 & 1.44(39)  & 1.20(42) \\ \hline
\end{tabular}
\caption{\label{table_msCD} Effective masses extracted at $t=an_t$
from correlators of spacelike Polyakov loops on $12^3 5$ lattices
at $T=T_c$ in the confined, $C$, and deconfined, $D$, phases.}
\end{center}
\end{table}

\begin{table}
\begin{center}
\begin{tabular}{|c|c|c|c|c|c|c|}\hline
\multicolumn{7}{|c|}{ $am_s(k)$ for $T\geq T_c$} \\ \hline
 $N$ & $k$ & $T=T_c$ & $T=5T_c/4$ & $T=5T_c/3$ 
& $T=5T_c/2$ & $T=5T_c/2$; $L_s=6$ \\ \hline
3 & 1 & -- & 0.753(7) & 1.119(11) & 2.027(42) & -- \\ \hline
4 & 1 & -- & 0.835(7) & 1.192(12) & 2.211(66) & 1.607(23) \\
  & 2 & -- & 1.126(17) & 1.59(4)  & 2.52(32)  & 2.09(8) \\ \hline
8 & 1 & 0.695(9)  & 0.910(11) & 1.311(19) & 2.39(13) & 1.805(32) \\
  & 2 & 1.171(26) & 1.610(34) & 2.12(12)  & -- & 3.1(7) \\ 
  & 3 & 1.507(41) & 1.933(81) & 2.29(32)  & -- & --   \\
  & 4 & 1.603(72) & 2.41(23)  & 2.75(99)  & -- & --   \\ \hline
\end{tabular}
\caption{\label{table_msDT} The spacelike $k$-loop masses
on $8^3 5$ lattices (unless otherwise indicated) in the
deconfined phase.}
\end{center}
\end{table}

\begin{table}
\begin{center}
\begin{tabular}{|c|c|c|c|c|}\hline
\multicolumn{5}{|c|}{ interface tension } \\ \hline
 $N$ & $\sigma_{cd}/T^3_c$ & fit range & $\chi^2/n_{df}$ & $n_{df}$ \\ \hline
 3  & 0.0200(6)  & 16-40  & 1.51 & 3 \\
 4  & 0.1208(56) & 14-20  & 0.37 & 2 \\ 
 6  & 0.394(11)  & 8-14   & 6.02 & 2 \\ 
 8  & 0.56(10)   & 8-10   & -    & 0 \\ \hline
\end{tabular}
\caption{\label{table_sig} The tension of the confining-deconfining
interface, in units of $T_c$, for various SU($N$) gauge theories,
extracted from the $V\to\infty$ extrapolations in Fig.\ref{fig_sigma}.
All for fixed $a$: $aT_c = 1/5$.
The best fit $\chi^2$ and the number of degrees of freedom, $n_{df}$,
are shown. }
\end{center}
\end{table}

\begin{table}
\begin{center}
\begin{tabular}{|c|c|c|c|}\hline
\multicolumn{4}{|c|}{ bulk transition } \\ \hline
 $N$ & lattice & $\beta^\downarrow_b$ &  $\beta^\uparrow_b$ \\ \hline
5 &  $8^4$  & 16.6565(5)  & 16.6565(5) \\
5 &  $10^4$ & 16.6550(5)  & 16.6550(5) \\
5 &  $12^4$ & 16.6550(10) & 16.6550(10) \\
6 &  $4^4$  & 24.100(25) & 24.475(25) \\
6 &  $6^4$  & 24.290(5)  & 24.480(5)  \\
8 &  $4^4$  & 43.25(5)   & 44.85(5) \\
8 &  $6^4$  & 43.60(5)   & 44.90(5) \\ 
12 &  $4^4$ & 98.5(1)    & 103.9(1) \\
12 &  $8^35$ & 98.7(1)    &  --   \\ \hline
\end{tabular}
\caption{\label{table_bulk} The location of the strong-weak 
coupling bulk transition when decreasing or increasing $\beta$
from large or small values respectively. For various SU($N$)
groups on the lattices shown.}
\end{center}
\end{table}

\clearpage

\clearpage

\begin	{figure}[p]
\begin	{center}
\leavevmode
\setlength{\unitlength}{0.240900pt}
\ifx\plotpoint\undefined\newsavebox{\plotpoint}\fi
\sbox{\plotpoint}{\rule[-0.200pt]{0.400pt}{0.400pt}}%
\begin{picture}(1500,1800)(0,0)
\font\gnuplot=cmr10 at 12pt
\gnuplot
\sbox{\plotpoint}{\rule[-0.200pt]{0.400pt}{0.400pt}}%
\put(375.0,250.0){\rule[-0.200pt]{4.818pt}{0.400pt}}
\put(350,250){\makebox(0,0)[r]{\ \ {$0$}}}
\put(1405.0,250.0){\rule[-0.200pt]{4.818pt}{0.400pt}}
\put(375.0,550.0){\rule[-0.200pt]{4.818pt}{0.400pt}}
\put(350,550){\makebox(0,0)[r]{\ \ {$500$}}}
\put(1405.0,550.0){\rule[-0.200pt]{4.818pt}{0.400pt}}
\put(375.0,850.0){\rule[-0.200pt]{4.818pt}{0.400pt}}
\put(350,850){\makebox(0,0)[r]{\ \ {$1000$}}}
\put(1405.0,850.0){\rule[-0.200pt]{4.818pt}{0.400pt}}
\put(375.0,1150.0){\rule[-0.200pt]{4.818pt}{0.400pt}}
\put(350,1150){\makebox(0,0)[r]{\ \ {$1500$}}}
\put(1405.0,1150.0){\rule[-0.200pt]{4.818pt}{0.400pt}}
\put(375.0,1450.0){\rule[-0.200pt]{4.818pt}{0.400pt}}
\put(350,1450){\makebox(0,0)[r]{\ \ {$2000$}}}
\put(1405.0,1450.0){\rule[-0.200pt]{4.818pt}{0.400pt}}
\put(375.0,1750.0){\rule[-0.200pt]{4.818pt}{0.400pt}}
\put(350,1750){\makebox(0,0)[r]{\ \ {$2500$}}}
\put(1405.0,1750.0){\rule[-0.200pt]{4.818pt}{0.400pt}}
\put(375.0,250.0){\rule[-0.200pt]{0.400pt}{4.818pt}}
\put(375,200){\makebox(0,0){\ {$0$}}}
\put(375.0,1730.0){\rule[-0.200pt]{0.400pt}{4.818pt}}
\put(480.0,250.0){\rule[-0.200pt]{0.400pt}{4.818pt}}
\put(480,200){\makebox(0,0){\ {$0.01$}}}
\put(480.0,1730.0){\rule[-0.200pt]{0.400pt}{4.818pt}}
\put(585.0,250.0){\rule[-0.200pt]{0.400pt}{4.818pt}}
\put(585,200){\makebox(0,0){\ {$0.02$}}}
\put(585.0,1730.0){\rule[-0.200pt]{0.400pt}{4.818pt}}
\put(690.0,250.0){\rule[-0.200pt]{0.400pt}{4.818pt}}
\put(690,200){\makebox(0,0){\ {$0.03$}}}
\put(690.0,1730.0){\rule[-0.200pt]{0.400pt}{4.818pt}}
\put(795.0,250.0){\rule[-0.200pt]{0.400pt}{4.818pt}}
\put(795,200){\makebox(0,0){\ {$0.04$}}}
\put(795.0,1730.0){\rule[-0.200pt]{0.400pt}{4.818pt}}
\put(900.0,250.0){\rule[-0.200pt]{0.400pt}{4.818pt}}
\put(900,200){\makebox(0,0){\ {$0.05$}}}
\put(900.0,1730.0){\rule[-0.200pt]{0.400pt}{4.818pt}}
\put(1005.0,250.0){\rule[-0.200pt]{0.400pt}{4.818pt}}
\put(1005,200){\makebox(0,0){\ {$0.06$}}}
\put(1005.0,1730.0){\rule[-0.200pt]{0.400pt}{4.818pt}}
\put(1110.0,250.0){\rule[-0.200pt]{0.400pt}{4.818pt}}
\put(1110,200){\makebox(0,0){\ {$0.07$}}}
\put(1110.0,1730.0){\rule[-0.200pt]{0.400pt}{4.818pt}}
\put(1215.0,250.0){\rule[-0.200pt]{0.400pt}{4.818pt}}
\put(1215,200){\makebox(0,0){\ {$0.08$}}}
\put(1215.0,1730.0){\rule[-0.200pt]{0.400pt}{4.818pt}}
\put(1320.0,250.0){\rule[-0.200pt]{0.400pt}{4.818pt}}
\put(1320,200){\makebox(0,0){\ {$0.09$}}}
\put(1320.0,1730.0){\rule[-0.200pt]{0.400pt}{4.818pt}}
\put(1425.0,250.0){\rule[-0.200pt]{0.400pt}{4.818pt}}
\put(1425,200){\makebox(0,0){\ {$0.1$}}}
\put(1425.0,1730.0){\rule[-0.200pt]{0.400pt}{4.818pt}}
\put(375.0,250.0){\rule[-0.200pt]{252.945pt}{0.400pt}}
\put(1425.0,250.0){\rule[-0.200pt]{0.400pt}{361.350pt}}
\put(375.0,1750.0){\rule[-0.200pt]{252.945pt}{0.400pt}}
\put(150,1300){\makebox(0,0){$N$}}
\put(875,25){\makebox(0,0){\large{$|{\bar l}_{k=1}|$}}}
\put(375.0,250.0){\rule[-0.200pt]{0.400pt}{361.350pt}}
\put(386,524){\circle*{18}}
\put(407,1013){\circle*{18}}
\put(428,1290){\circle*{18}}
\put(449,1574){\circle*{18}}
\put(470,1595){\circle*{18}}
\put(491,1564){\circle*{18}}
\put(512,1456){\circle*{18}}
\put(533,1236){\circle*{18}}
\put(554,1072){\circle*{18}}
\put(575,886){\circle*{18}}
\put(596,722){\circle*{18}}
\put(617,607){\circle*{18}}
\put(638,525){\circle*{18}}
\put(659,459){\circle*{18}}
\put(680,416){\circle*{18}}
\put(701,392){\circle*{18}}
\put(722,363){\circle*{18}}
\put(743,327){\circle*{18}}
\put(764,321){\circle*{18}}
\put(785,311){\circle*{18}}
\put(806,303){\circle*{18}}
\put(827,296){\circle*{18}}
\put(848,299){\circle*{18}}
\put(869,299){\circle*{18}}
\put(890,300){\circle*{18}}
\put(911,305){\circle*{18}}
\put(932,300){\circle*{18}}
\put(953,289){\circle*{18}}
\put(974,296){\circle*{18}}
\put(995,290){\circle*{18}}
\put(1016,295){\circle*{18}}
\put(1037,312){\circle*{18}}
\put(1058,321){\circle*{18}}
\put(1079,327){\circle*{18}}
\put(1100,341){\circle*{18}}
\put(1121,355){\circle*{18}}
\put(1142,358){\circle*{18}}
\put(1163,363){\circle*{18}}
\put(1184,353){\circle*{18}}
\put(1205,338){\circle*{18}}
\put(1226,311){\circle*{18}}
\put(1247,310){\circle*{18}}
\put(1268,291){\circle*{18}}
\put(1289,277){\circle*{18}}
\put(1310,261){\circle*{18}}
\put(1331,254){\circle*{18}}
\put(1352,251){\circle*{18}}
\put(1373,251){\circle*{18}}
\put(1394,251){\circle*{18}}
\put(1415,250){\circle*{18}}
\end{picture}
\end	{center}
\vskip 0.15in
\caption{Histogram of timelike $k=1$ Polyakov loops averaged 
over single $20^3 5$ lattice fields in SU(4) at $T\simeq T_c$.
Separate confined and deconfined peaks are visible.}  
\label{fig_histk1}
\end 	{figure}
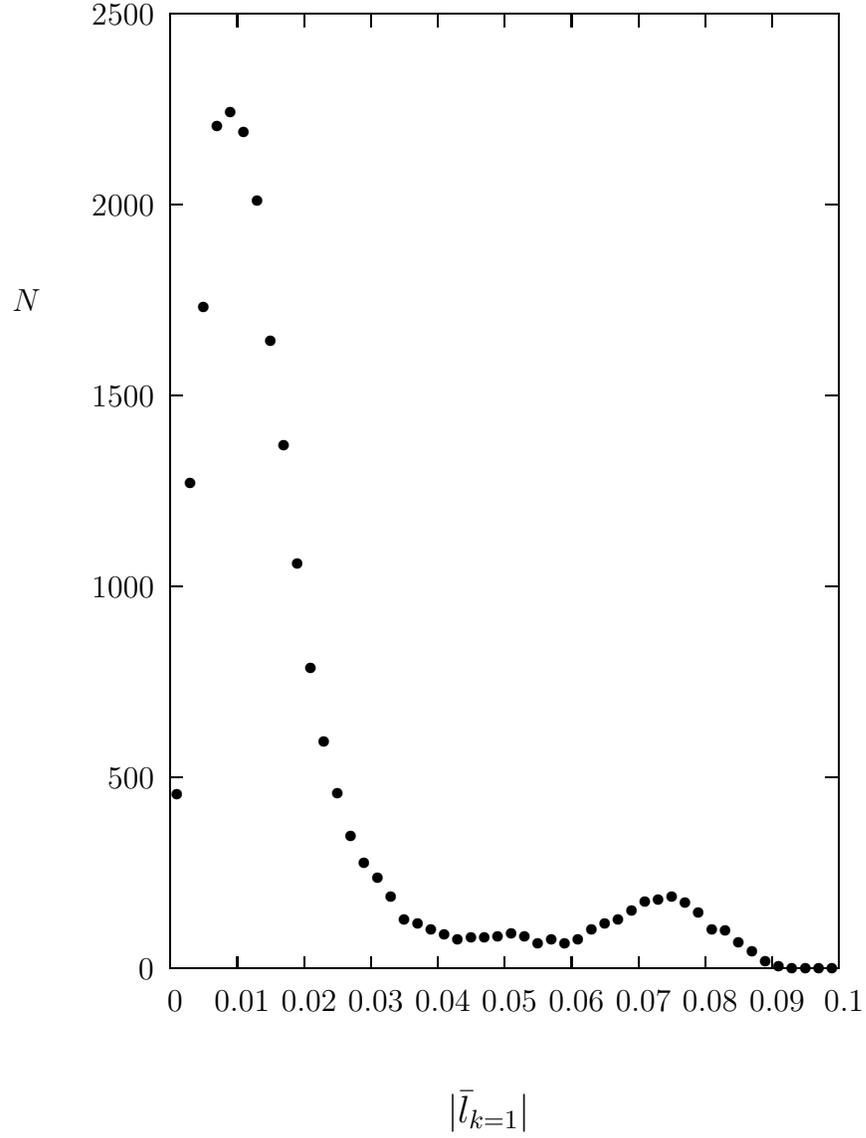

\begin	{figure}[p]
\begin	{center}
\leavevmode
\setlength{\unitlength}{0.240900pt}
\ifx\plotpoint\undefined\newsavebox{\plotpoint}\fi
\sbox{\plotpoint}{\rule[-0.200pt]{0.400pt}{0.400pt}}%
\begin{picture}(1500,1800)(0,0)
\font\gnuplot=cmr10 at 12pt
\gnuplot
\sbox{\plotpoint}{\rule[-0.200pt]{0.400pt}{0.400pt}}%
\put(375.0,250.0){\rule[-0.200pt]{4.818pt}{0.400pt}}
\put(350,250){\makebox(0,0)[r]{\ \ {$0$}}}
\put(1405.0,250.0){\rule[-0.200pt]{4.818pt}{0.400pt}}
\put(375.0,464.0){\rule[-0.200pt]{4.818pt}{0.400pt}}
\put(350,464){\makebox(0,0)[r]{\ \ {$500$}}}
\put(1405.0,464.0){\rule[-0.200pt]{4.818pt}{0.400pt}}
\put(375.0,679.0){\rule[-0.200pt]{4.818pt}{0.400pt}}
\put(350,679){\makebox(0,0)[r]{\ \ {$1000$}}}
\put(1405.0,679.0){\rule[-0.200pt]{4.818pt}{0.400pt}}
\put(375.0,893.0){\rule[-0.200pt]{4.818pt}{0.400pt}}
\put(350,893){\makebox(0,0)[r]{\ \ {$1500$}}}
\put(1405.0,893.0){\rule[-0.200pt]{4.818pt}{0.400pt}}
\put(375.0,1107.0){\rule[-0.200pt]{4.818pt}{0.400pt}}
\put(350,1107){\makebox(0,0)[r]{\ \ {$2000$}}}
\put(1405.0,1107.0){\rule[-0.200pt]{4.818pt}{0.400pt}}
\put(375.0,1321.0){\rule[-0.200pt]{4.818pt}{0.400pt}}
\put(350,1321){\makebox(0,0)[r]{\ \ {$2500$}}}
\put(1405.0,1321.0){\rule[-0.200pt]{4.818pt}{0.400pt}}
\put(375.0,1536.0){\rule[-0.200pt]{4.818pt}{0.400pt}}
\put(350,1536){\makebox(0,0)[r]{\ \ {$3000$}}}
\put(1405.0,1536.0){\rule[-0.200pt]{4.818pt}{0.400pt}}
\put(375.0,250.0){\rule[-0.200pt]{0.400pt}{4.818pt}}
\put(375,200){\makebox(0,0){\ {$0$}}}
\put(375.0,1730.0){\rule[-0.200pt]{0.400pt}{4.818pt}}
\put(537.0,250.0){\rule[-0.200pt]{0.400pt}{4.818pt}}
\put(537,200){\makebox(0,0){\ {$0.1$}}}
\put(537.0,1730.0){\rule[-0.200pt]{0.400pt}{4.818pt}}
\put(698.0,250.0){\rule[-0.200pt]{0.400pt}{4.818pt}}
\put(698,200){\makebox(0,0){\ {$0.2$}}}
\put(698.0,1730.0){\rule[-0.200pt]{0.400pt}{4.818pt}}
\put(860.0,250.0){\rule[-0.200pt]{0.400pt}{4.818pt}}
\put(860,200){\makebox(0,0){\ {$0.3$}}}
\put(860.0,1730.0){\rule[-0.200pt]{0.400pt}{4.818pt}}
\put(1021.0,250.0){\rule[-0.200pt]{0.400pt}{4.818pt}}
\put(1021,200){\makebox(0,0){\ {$0.4$}}}
\put(1021.0,1730.0){\rule[-0.200pt]{0.400pt}{4.818pt}}
\put(1183.0,250.0){\rule[-0.200pt]{0.400pt}{4.818pt}}
\put(1183,200){\makebox(0,0){\ {$0.5$}}}
\put(1183.0,1730.0){\rule[-0.200pt]{0.400pt}{4.818pt}}
\put(1344.0,250.0){\rule[-0.200pt]{0.400pt}{4.818pt}}
\put(1344,200){\makebox(0,0){\ {$0.6$}}}
\put(1344.0,1730.0){\rule[-0.200pt]{0.400pt}{4.818pt}}
\put(375.0,250.0){\rule[-0.200pt]{252.945pt}{0.400pt}}
\put(1425.0,250.0){\rule[-0.200pt]{0.400pt}{361.350pt}}
\put(375.0,1750.0){\rule[-0.200pt]{252.945pt}{0.400pt}}
\put(150,1300){\makebox(0,0){$N$}}
\put(875,25){\makebox(0,0){\large{$|{\bar l}_{k=2}|$}}}
\put(375.0,250.0){\rule[-0.200pt]{0.400pt}{361.350pt}}
\put(383,1679){\circle*{18}}
\put(399,1609){\circle*{18}}
\put(415,1494){\circle*{18}}
\put(432,1285){\circle*{18}}
\put(448,1079){\circle*{18}}
\put(464,884){\circle*{18}}
\put(480,752){\circle*{18}}
\put(496,603){\circle*{18}}
\put(512,509){\circle*{18}}
\put(528,442){\circle*{18}}
\put(545,380){\circle*{18}}
\put(561,361){\circle*{18}}
\put(577,334){\circle*{18}}
\put(593,313){\circle*{18}}
\put(609,297){\circle*{18}}
\put(625,296){\circle*{18}}
\put(642,282){\circle*{18}}
\put(658,283){\circle*{18}}
\put(674,282){\circle*{18}}
\put(690,275){\circle*{18}}
\put(706,270){\circle*{18}}
\put(722,266){\circle*{18}}
\put(738,260){\circle*{18}}
\put(755,256){\circle*{18}}
\put(771,255){\circle*{18}}
\put(787,250){\circle*{18}}
\put(803,251){\circle*{18}}
\put(819,250){\circle*{18}}
\put(835,250){\circle*{18}}
\put(852,250){\circle*{18}}
\put(609,250){\circle{18}}
\put(625,250){\circle{18}}
\put(642,252){\circle{18}}
\put(658,251){\circle{18}}
\put(674,253){\circle{18}}
\put(690,259){\circle{18}}
\put(706,263){\circle{18}}
\put(722,263){\circle{18}}
\put(738,267){\circle{18}}
\put(755,267){\circle{18}}
\put(771,274){\circle{18}}
\put(787,271){\circle{18}}
\put(803,275){\circle{18}}
\put(819,285){\circle{18}}
\put(835,284){\circle{18}}
\put(852,278){\circle{18}}
\put(868,287){\circle{18}}
\put(884,283){\circle{18}}
\put(900,296){\circle{18}}
\put(916,296){\circle{18}}
\put(932,298){\circle{18}}
\put(948,304){\circle{18}}
\put(965,320){\circle{18}}
\put(981,301){\circle{18}}
\put(997,297){\circle{18}}
\put(1013,296){\circle{18}}
\put(1029,290){\circle{18}}
\put(1045,289){\circle{18}}
\put(1062,286){\circle{18}}
\put(1078,276){\circle{18}}
\put(1094,270){\circle{18}}
\put(1110,267){\circle{18}}
\put(1126,259){\circle{18}}
\put(1142,257){\circle{18}}
\put(1158,256){\circle{18}}
\put(1175,253){\circle{18}}
\put(1191,252){\circle{18}}
\put(1207,250){\circle{18}}
\put(1223,250){\circle{18}}
\put(1239,250){\circle{18}}
\put(1255,250){\circle{18}}
\put(1272,250){\circle{18}}
\end{picture}
\end	{center}
\vskip 0.15in
\caption{As in Fig.\ref{fig_histk1} but for (antisymmetric) $k=2$.
From lattice fields with $|\bar{l}_{k=1}| \leq 0.05$,
$\bullet$, and with $|\bar{l}_{k=1}| > 0.05$, $\circ$.}
\label{fig_histk2n}
\end 	{figure}
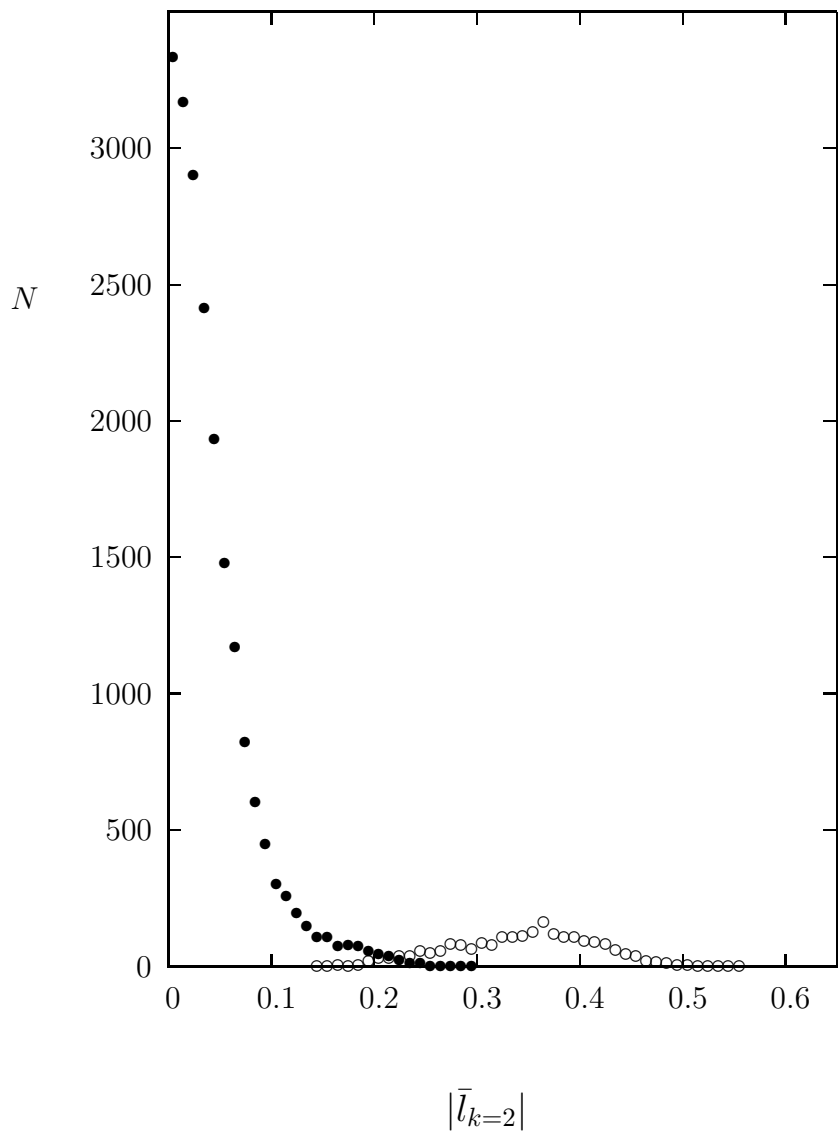

\begin	{figure}[p]
\begin	{center}
\leavevmode
\input	{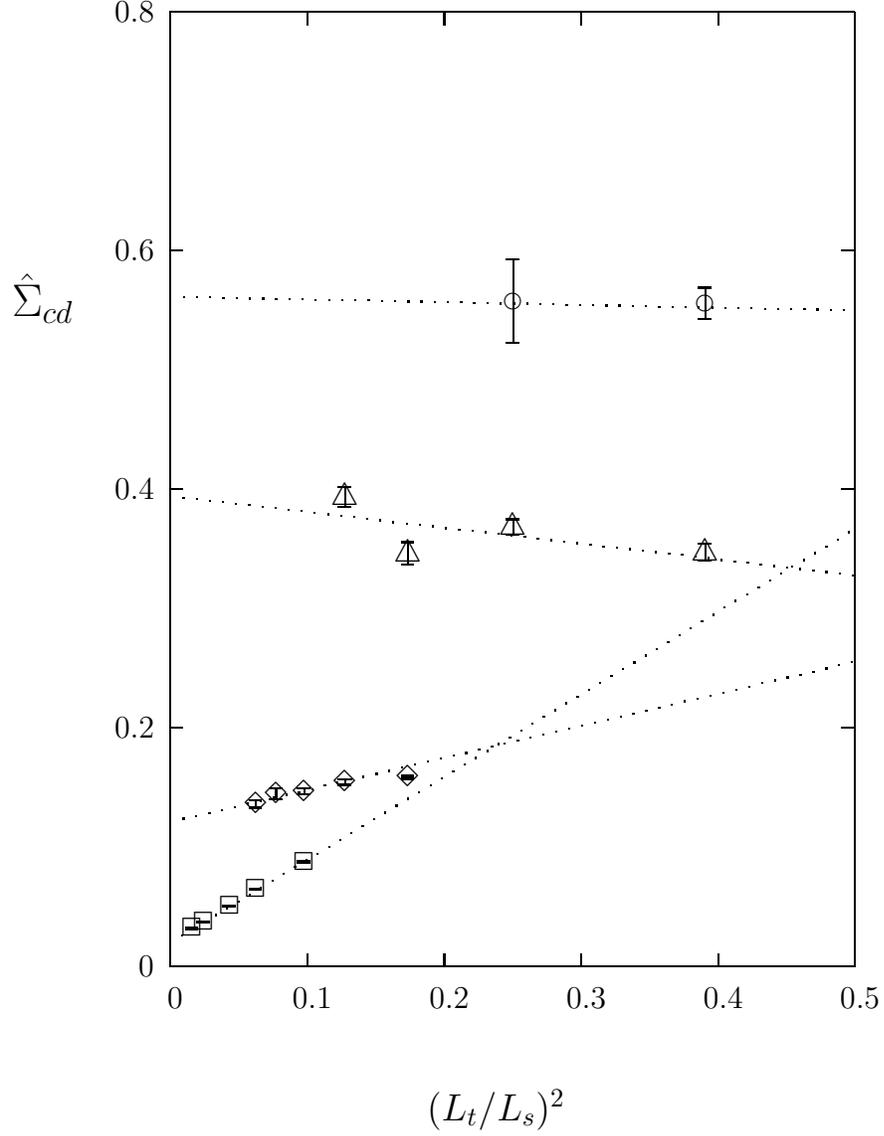}
\end	{center}
\vskip 0.15in
\caption{The quantity ${\hat\Sigma}_{cd}$ defined in 
eqn(\ref{eqn_SigV}) whose $L_s=\infty$ intercept
is the interface tension, $\sigma_{cd}/T^3_c$. 
Calculations for SU(3), $\Box$, SU(4), $\Diamond$, 
SU(6), $\bigtriangleup$ and SU(8), $\circ$.
All  are for $a\simeq 1/5T_c$. Linear extrapolations to
infinite $V$ are shown.}
\label{fig_sigma}
\end 	{figure}

\begin	{figure}[p]
\begin	{center}
\leavevmode
\input	{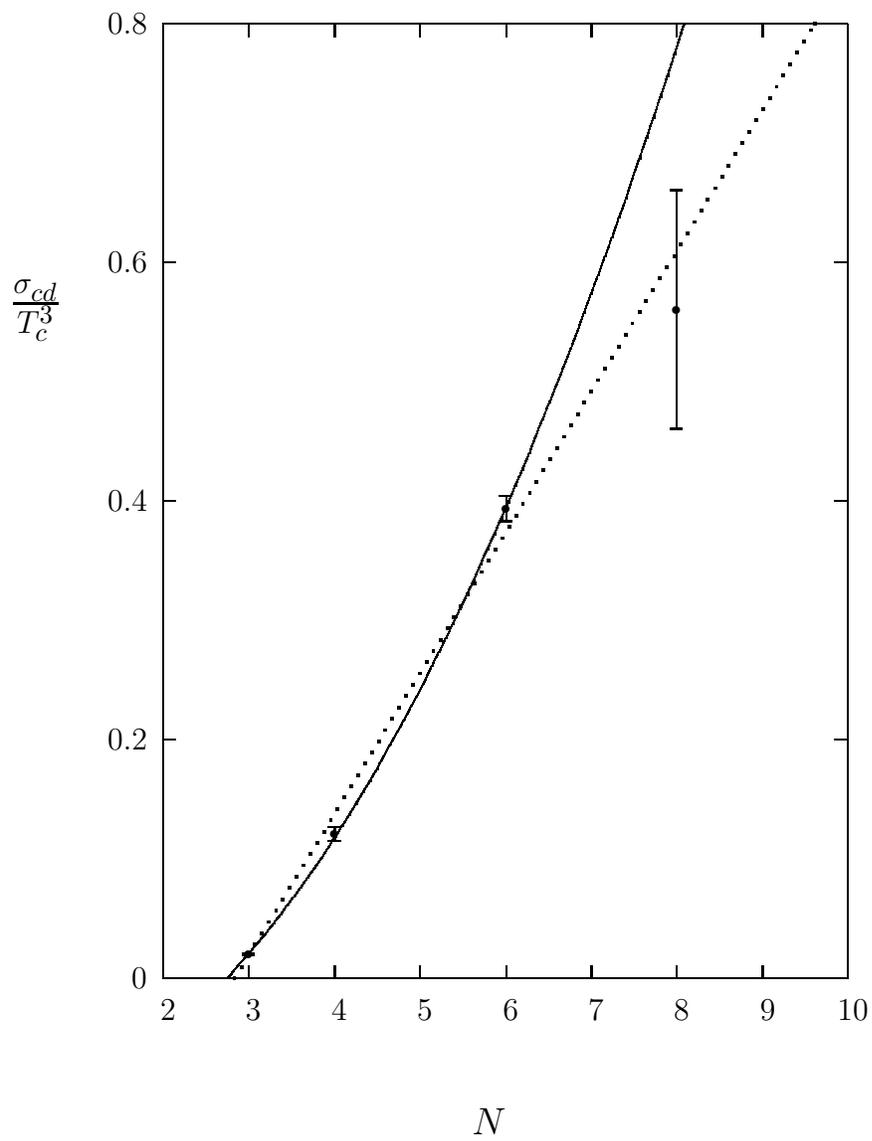}
\end	{center}
\vskip 0.15in
\caption{The interface tension at the fixed lattice spacing 
$a\simeq 1/5T_c$ plotted against $N$. 
Also shown are the best linear and quadratic fits.}
\label{fig_sigmaN}
\end 	{figure}

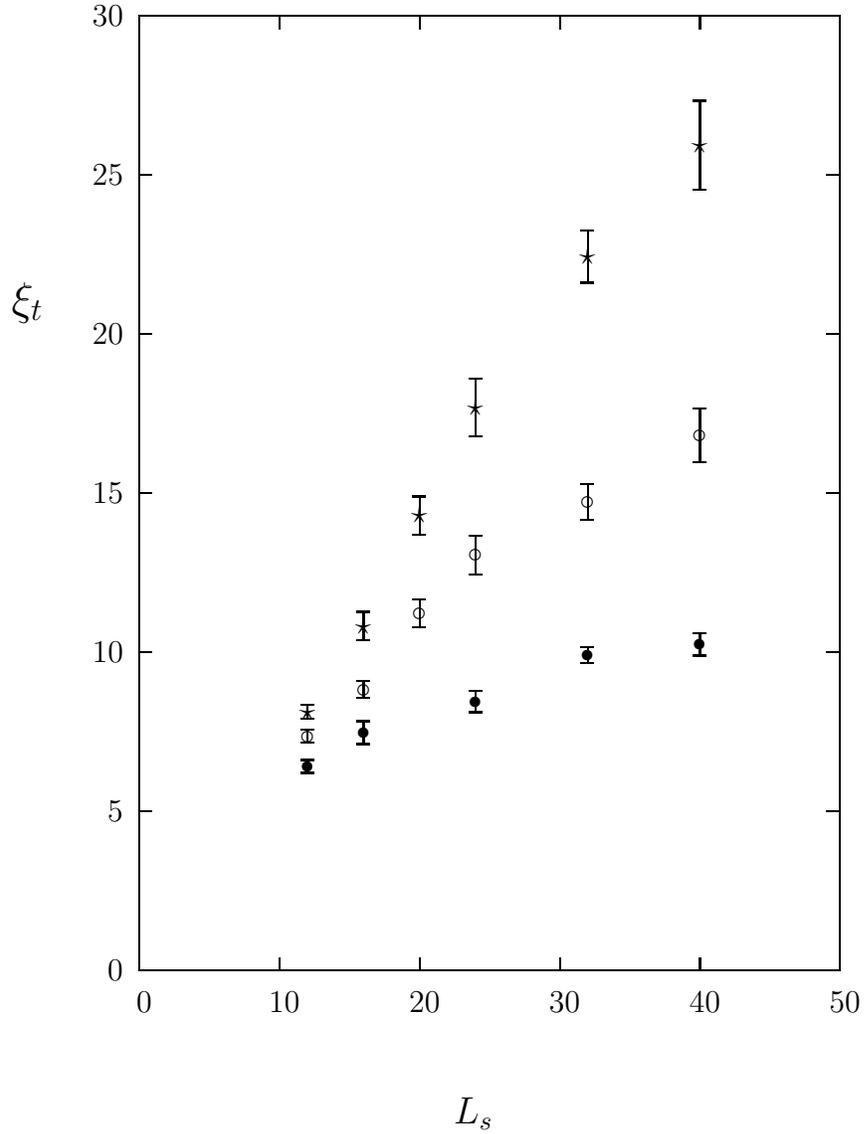
\begin	{figure}[p]
\begin	{center}
\leavevmode
\setlength{\unitlength}{0.240900pt}
\ifx\plotpoint\undefined\newsavebox{\plotpoint}\fi
\sbox{\plotpoint}{\rule[-0.200pt]{0.400pt}{0.400pt}}%
\begin{picture}(1500,1800)(0,0)
\font\gnuplot=cmr10 at 12pt
\gnuplot
\sbox{\plotpoint}{\rule[-0.200pt]{0.400pt}{0.400pt}}%
\put(325.0,250.0){\rule[-0.200pt]{4.818pt}{0.400pt}}
\put(300,250){\makebox(0,0)[r]{\ \ {$0$}}}
\put(1405.0,250.0){\rule[-0.200pt]{4.818pt}{0.400pt}}
\put(325.0,500.0){\rule[-0.200pt]{4.818pt}{0.400pt}}
\put(300,500){\makebox(0,0)[r]{\ \ {$5$}}}
\put(1405.0,500.0){\rule[-0.200pt]{4.818pt}{0.400pt}}
\put(325.0,750.0){\rule[-0.200pt]{4.818pt}{0.400pt}}
\put(300,750){\makebox(0,0)[r]{\ \ {$10$}}}
\put(1405.0,750.0){\rule[-0.200pt]{4.818pt}{0.400pt}}
\put(325.0,1000.0){\rule[-0.200pt]{4.818pt}{0.400pt}}
\put(300,1000){\makebox(0,0)[r]{\ \ {$15$}}}
\put(1405.0,1000.0){\rule[-0.200pt]{4.818pt}{0.400pt}}
\put(325.0,1250.0){\rule[-0.200pt]{4.818pt}{0.400pt}}
\put(300,1250){\makebox(0,0)[r]{\ \ {$20$}}}
\put(1405.0,1250.0){\rule[-0.200pt]{4.818pt}{0.400pt}}
\put(325.0,1500.0){\rule[-0.200pt]{4.818pt}{0.400pt}}
\put(300,1500){\makebox(0,0)[r]{\ \ {$25$}}}
\put(1405.0,1500.0){\rule[-0.200pt]{4.818pt}{0.400pt}}
\put(325.0,1750.0){\rule[-0.200pt]{4.818pt}{0.400pt}}
\put(300,1750){\makebox(0,0)[r]{\ \ {$30$}}}
\put(1405.0,1750.0){\rule[-0.200pt]{4.818pt}{0.400pt}}
\put(325.0,250.0){\rule[-0.200pt]{0.400pt}{4.818pt}}
\put(325,200){\makebox(0,0){\ {$0$}}}
\put(325.0,1730.0){\rule[-0.200pt]{0.400pt}{4.818pt}}
\put(545.0,250.0){\rule[-0.200pt]{0.400pt}{4.818pt}}
\put(545,200){\makebox(0,0){\ {$10$}}}
\put(545.0,1730.0){\rule[-0.200pt]{0.400pt}{4.818pt}}
\put(765.0,250.0){\rule[-0.200pt]{0.400pt}{4.818pt}}
\put(765,200){\makebox(0,0){\ {$20$}}}
\put(765.0,1730.0){\rule[-0.200pt]{0.400pt}{4.818pt}}
\put(985.0,250.0){\rule[-0.200pt]{0.400pt}{4.818pt}}
\put(985,200){\makebox(0,0){\ {$30$}}}
\put(985.0,1730.0){\rule[-0.200pt]{0.400pt}{4.818pt}}
\put(1205.0,250.0){\rule[-0.200pt]{0.400pt}{4.818pt}}
\put(1205,200){\makebox(0,0){\ {$40$}}}
\put(1205.0,1730.0){\rule[-0.200pt]{0.400pt}{4.818pt}}
\put(1425.0,250.0){\rule[-0.200pt]{0.400pt}{4.818pt}}
\put(1425,200){\makebox(0,0){\ {$50$}}}
\put(1425.0,1730.0){\rule[-0.200pt]{0.400pt}{4.818pt}}
\put(325.0,250.0){\rule[-0.200pt]{264.990pt}{0.400pt}}
\put(1425.0,250.0){\rule[-0.200pt]{0.400pt}{361.350pt}}
\put(325.0,1750.0){\rule[-0.200pt]{264.990pt}{0.400pt}}
\put(150,1300){\makebox(0,0){\Large{$\xi_t$}}}
\put(850,25){\makebox(0,0){\large{$L_s$}}}
\put(325.0,250.0){\rule[-0.200pt]{0.400pt}{361.350pt}}
\put(1205.0,744.0){\rule[-0.200pt]{0.400pt}{8.431pt}}
\put(1195.0,744.0){\rule[-0.200pt]{4.818pt}{0.400pt}}
\put(1195.0,779.0){\rule[-0.200pt]{4.818pt}{0.400pt}}
\put(1029.0,733.0){\rule[-0.200pt]{0.400pt}{6.022pt}}
\put(1019.0,733.0){\rule[-0.200pt]{4.818pt}{0.400pt}}
\put(1019.0,758.0){\rule[-0.200pt]{4.818pt}{0.400pt}}
\put(853.0,655.0){\rule[-0.200pt]{0.400pt}{8.191pt}}
\put(843.0,655.0){\rule[-0.200pt]{4.818pt}{0.400pt}}
\put(843.0,689.0){\rule[-0.200pt]{4.818pt}{0.400pt}}
\put(677.0,605.0){\rule[-0.200pt]{0.400pt}{8.672pt}}
\put(667.0,605.0){\rule[-0.200pt]{4.818pt}{0.400pt}}
\put(667.0,641.0){\rule[-0.200pt]{4.818pt}{0.400pt}}
\put(589.0,560.0){\rule[-0.200pt]{0.400pt}{4.818pt}}
\put(579.0,560.0){\rule[-0.200pt]{4.818pt}{0.400pt}}
\put(1205,762){\circle*{18}}
\put(1029,745){\circle*{18}}
\put(853,672){\circle*{18}}
\put(677,623){\circle*{18}}
\put(589,570){\circle*{18}}
\put(579.0,580.0){\rule[-0.200pt]{4.818pt}{0.400pt}}
\put(1205.0,1048.0){\rule[-0.200pt]{0.400pt}{20.476pt}}
\put(1195.0,1048.0){\rule[-0.200pt]{4.818pt}{0.400pt}}
\put(1195.0,1133.0){\rule[-0.200pt]{4.818pt}{0.400pt}}
\put(1029.0,957.0){\rule[-0.200pt]{0.400pt}{13.731pt}}
\put(1019.0,957.0){\rule[-0.200pt]{4.818pt}{0.400pt}}
\put(1019.0,1014.0){\rule[-0.200pt]{4.818pt}{0.400pt}}
\put(853.0,872.0){\rule[-0.200pt]{0.400pt}{14.695pt}}
\put(843.0,872.0){\rule[-0.200pt]{4.818pt}{0.400pt}}
\put(843.0,933.0){\rule[-0.200pt]{4.818pt}{0.400pt}}
\put(765.0,789.0){\rule[-0.200pt]{0.400pt}{10.600pt}}
\put(755.0,789.0){\rule[-0.200pt]{4.818pt}{0.400pt}}
\put(755.0,833.0){\rule[-0.200pt]{4.818pt}{0.400pt}}
\put(677.0,678.0){\rule[-0.200pt]{0.400pt}{6.263pt}}
\put(667.0,678.0){\rule[-0.200pt]{4.818pt}{0.400pt}}
\put(667.0,704.0){\rule[-0.200pt]{4.818pt}{0.400pt}}
\put(589.0,608.0){\rule[-0.200pt]{0.400pt}{4.818pt}}
\put(579.0,608.0){\rule[-0.200pt]{4.818pt}{0.400pt}}
\put(1205,1091){\circle{18}}
\put(1029,986){\circle{18}}
\put(853,903){\circle{18}}
\put(765,811){\circle{18}}
\put(677,691){\circle{18}}
\put(589,618){\circle{18}}
\put(579.0,628.0){\rule[-0.200pt]{4.818pt}{0.400pt}}
\put(1205.0,1476.0){\rule[-0.200pt]{0.400pt}{33.726pt}}
\put(1195.0,1476.0){\rule[-0.200pt]{4.818pt}{0.400pt}}
\put(1195.0,1616.0){\rule[-0.200pt]{4.818pt}{0.400pt}}
\put(1029.0,1330.0){\rule[-0.200pt]{0.400pt}{19.754pt}}
\put(1019.0,1330.0){\rule[-0.200pt]{4.818pt}{0.400pt}}
\put(1019.0,1412.0){\rule[-0.200pt]{4.818pt}{0.400pt}}
\put(853.0,1089.0){\rule[-0.200pt]{0.400pt}{21.681pt}}
\put(843.0,1089.0){\rule[-0.200pt]{4.818pt}{0.400pt}}
\put(843.0,1179.0){\rule[-0.200pt]{4.818pt}{0.400pt}}
\put(765.0,934.0){\rule[-0.200pt]{0.400pt}{14.454pt}}
\put(755.0,934.0){\rule[-0.200pt]{4.818pt}{0.400pt}}
\put(755.0,994.0){\rule[-0.200pt]{4.818pt}{0.400pt}}
\put(677.0,768.0){\rule[-0.200pt]{0.400pt}{10.840pt}}
\put(667.0,768.0){\rule[-0.200pt]{4.818pt}{0.400pt}}
\put(667.0,813.0){\rule[-0.200pt]{4.818pt}{0.400pt}}
\put(589.0,645.0){\rule[-0.200pt]{0.400pt}{5.300pt}}
\put(579.0,645.0){\rule[-0.200pt]{4.818pt}{0.400pt}}
\put(1205,1546){\makebox(0,0){$\star$}}
\put(1029,1371){\makebox(0,0){$\star$}}
\put(853,1134){\makebox(0,0){$\star$}}
\put(765,964){\makebox(0,0){$\star$}}
\put(677,790){\makebox(0,0){$\star$}}
\put(589,656){\makebox(0,0){$\star$}}
\put(579.0,667.0){\rule[-0.200pt]{4.818pt}{0.400pt}}
\end{picture}
\end	{center}
\vskip 0.15in
\caption{The correlation length near the SU(2) deconfinement
transition, as a function of the spatial lattice size.
For $\beta=2.360$ ($\bullet$), $\beta=2.3675$ ($\circ$) , and 
$\beta=2.3725$  ($\star$) on $L^3_s 5$ lattices where
$\beta_c(V=\infty) = 2.3714(6)$ \cite{oxtemp03}.}
\label{fig_mtsu2}
\end 	{figure}


\begin{thebibliography}{99}


\bibitem{oxtemp02}
B. Lucini, M. Teper and U. Wenger,
Phys. Lett. B545 (2002) 197 (hep-lat/0206029).

\bibitem{oxtemp03}
B. Lucini, M. Teper and U. Wenger,
JHEP 0401 (2004) 061 (hep-lat/0307017).

\bibitem{oxtempq04}
B. Lucini, M. Teper and U. Wenger,
hep-lat/0401028.

\bibitem{polyTc}
A. Polyakov, Phys. Lett. B72 (1978) 477.

\bibitem{fluxtube}
N. Isgur and J. Paton, Phys. Rev. D31 (1985) 2910. \\
T. Moretto and M. Teper, hep-lat/9312035. \\
R. Johnson and M. Teper, 
Phys. Rev. D66 (2002) 036006 
(hep-ph/0012287).

\bibitem{Tcpuzzle}
N. Ishii and H. Suganuma,
hep-ph/0210158.

\bibitem{ArvisLW}
J. Arvis, Phys. Lett. 127B (1983) 106. \\
M. Luscher and P. Weisz,
JHEP 0407 (2004) 014 (hep-th/0406205).

\bibitem{mtd3su2}
M. Teper, Phys. Lett. B313 (1993) 417
and unpublished.

\bibitem{legeland}
J. Engels, F. Karsch, E. Laermann, C. Legeland, M. Lutgemeier,
B. Petersson and T. Scheideler,
Nucl. Phys. Proc. Suppl. 53 (1997) 420
(hep-lat/9608099).

\bibitem{CKA}
C. Korthals Altes, private communication.

\bibitem{blmt-string}
B. Lucini and M. Teper,
Phys. Lett. B501 (2001) 128
(hep-lat/0012025);
Phys. Rev. D64 (2001) 105019
(hep-lat/0107007).

\bibitem{CS}
J. Ambjorn, P. Olesen and C. Peterson,
Nucl. Phys. B240 (1984) 189, 533; B244 (1984) 262;
Phys. Lett. B142 (1984) 410.


\bibitem{MQCD}
A. Hanany, M. Strassler and A. Zaffaroni,
Nucl. Phys. B513 (1998) 87 (hep-th/9707244). \\ 
M. Strassler,
Nucl. Phys. Proc. Suppl. 73 (1999) 120 (hep-lat/9810059). \\
M. Strassler,
Prog. Theor. Phys. Suppl. 131 (1998) 439 (hep-lat/9803009).

\bibitem{oxglue04}
B. Lucini, M. Teper and U. Wenger,
JHEP 0406 (2004) 012
(hep-lat/0404008).

\bibitem{bagTH}
T. H. Hansson, 
Phys. Lett. B166 (1986) 343. \\
K. Johnson and C. B. Thorn, 
Phys. Rev. D13 (1976) 1934.

\bibitem{GPY}
D. Gross, R. Pisarski and L. Yaffe,
Rev. Mod. Phys. 53 (1981) 43.

\bibitem{ckaDW}
P. Giovannangeli and C. P. Korthals Altes,
Nucl. Phys. B608 (2001) 203 
(hep-ph/0102022); hep-ph/0412322.

\bibitem{oxDW96}
C. Korthals Altes, A. Michels, M. Stephanov and M. Teper,
Phys. Rev. D55 (1997) 1047 
(hep-lat/9606021). 


\bibitem{blmt-glue}
B. Lucini and M. Teper,
JHEP 0106 (2001) 050
(hep-lat/0103027).

\bibitem{largeN}
G. 't Hooft, Nucl. Phys. B72 (1974) 461. \\
E. Witten, Nucl. Phys. B160 (1979) 57. \\
S. Coleman, 1979 Erice Lectures. \\
A. Manohar, 1997 Les Houches Lectures, hep-ph/9802419. \\ 
S.R. Das, Rev. Mod. Phys. 59(1987)235.  \\ 
Y. Makeenko, hep-th/0001047. 


\bibitem{lattice-thooft}
G. 't Hooft,
in {\it Large $N$ QCD} 
(Ed. R.F. Lebed, World Scientific 2002)
(hep-th/0204069); hep-th/0408183.

\bibitem{gross-witten}
D. Gross and E. Witten,
Phys. Rev. D21 (1980) 446.

\bibitem{mtN04}
M. Teper,  Talk at `Large N QCD', ECT, Trento July 2004
(hep-th/0412005).

\bibitem{hmmt-string04}
H. Meyer and M. Teper, hep-lat/0411039.

\bibitem{CKAmon}
C. Korthals Altes, 
hep-ph/0406138; hep-ph/0408301.

\bibitem{interface}
Y. Iwasaki, K. Kanaya, L. Karkainnen, K. Rummukainen and T. Yoshie,
Phys. Rev. D49 (1994) 3540
(hep-lat/9309003).

\bibitem{phdf04}
Ph. de Forcrand, B. Lucini and M. Vettorazzo, hep-lat/0409148 and in
preparation.

\bibitem{rough}
M. Caselle, R. Fiori, F. Gliozzi, M. Hasenbusch, K. Pinn
and S. Vinti,
Nucl. Phys. B432 (1994) 590
(hep-lat/9407002).

\bibitem{blmt-glue01}
B. Lucini and M. Teper,
JHEP 0106 (2001) 050
(hep-lat/0103027).

\bibitem{bulkN}
M. Campostrini,
Nucl. Phys. Proc. Suppl. 73 (1999) 724
(hep-lat/9809072).

\bibitem{EK}
T. Eguchi and H. Kawai,
Phys. Rev. Lett. 48 (1982) 1063. \\
A. Gonzales-Arroyo and M. Okawa,
Phys. Lett. 120B (1983) 174.

\bibitem{Master}
E. Witten, in: ``Recent Developments in Gauge Theories'',
Ed. G. 't Hooft et al. (Plenum Press, 1980).

\bibitem{Ntheta}
E. Witten,
Phys. Rev. Lett. 81 (1998) 2862 (hep-th/9807109). \\
G. Gabadadze and M. Shifman,
Int. J. Mod. Phys. A17 (2002) 3689 (hep-ph/0206123).\\
A. Armoni and M. Shifman,
Nucl. Phys. B 664 (2003) 233 (hep-th/0304127).

\bibitem{HNRN}
R. Narayanan and H. Neuberger, Talks at `Large N QCD',
ECT, Trento July 2004 (hep-lat/0501031).


\end{thebibliography}
\end{document}